\documentclass[a4paper,11pt]{article}
\pdfoutput=1 
\usepackage{ae,aecompl}
\usepackage{jheppub} 

\usepackage[T1]{fontenc}
\usepackage[none]{hyphenat}

\usepackage[italian,english]{babel}
\usepackage{hyperref}
\usepackage{changepage}
\usepackage{ifpdf}
\usepackage{subfigure}
\usepackage{amssymb}
\usepackage{amsfonts}
\usepackage{epsf}
\usepackage{rotating}
\usepackage{graphicx}
\usepackage{amsmath}
\usepackage{fancyhdr}
\usepackage{lineno}
\usepackage{babel}
\usepackage{graphics}
\usepackage{pstricks}
\usepackage{color}
\usepackage{multirow}
\usepackage{slashed}
\usepackage{physics}
\usepackage{caption}
\usepackage{tabularx, array,booktabs, makecell}
\usepackage[toc,page]{appendix}
\usepackage{pifont}

\usepackage{tikz}
\usepackage{tikz-feynman}
\tikzfeynmanset{compat=1.0.0}

\definecolor{pantoneCB}{rgb}{0.0588235, 0.298039, 0.505882}

\usepackage{listings}
\hypersetup{
	colorlinks=true,
	linkcolor=purple,
	filecolor=magenta,
	urlcolor=[rgb]{0.0,0.42,0.59},
	citecolor=[rgb]{0.64,0.0,0.0},
}

\newcommand{\nn}{\nonumber}
\newcommand{\lsim}{\mathrel{\mathop{\kern 0pt \rlap
			{\raise.2ex\hbox{$<$}}}
		\lower.9ex\hbox{\kern-.190em $\sim$}}}
\newcommand{\gsim}{\mathrel{\mathop{\kern 0pt \rlap
			{\raise.2ex\hbox{$>$}}}
		\lower.9ex\hbox{\kern-.190em $\sim$}}}

\newcommand{\be}{\begin{equation}}
	\newcommand{\ee}{\end{equation}}
\newcommand{\bea}{\begin{eqnarray}}
	\newcommand{\eea}{\end{eqnarray}}
\newcommand{\sigmav}{\langle \sigma v \rangle}

\def\ptmiss{\not\!\!{p_T}}

\newcommand{\half}{\frac{1}{2}}

\newcommand{\sarah}{\texttt{SARAH-4.14.2} }
\newcommand{\chep}{\texttt{CalcHEP-3.8.10} } 

\newcommand{\py}{\texttt{Pythia8.2} }
\newcommand{\fj}{\texttt{fastjet-3.3.4} }
\newcommand{\mic}{\texttt{micrOMEGAs} }

\newcommand{\fbi}{fb$^{-1}$}
\newcommand{\y}{\mathcal{Y}}

\newcolumntype{P}[1]{>{\centering\arraybackslash}p{#1}}
\newcolumntype{M}[1]{>{\centering\arraybackslash}m{#1}}

\title{\boldmath Interplay of inert  doublet  and vector-like lepton triplet with displaced vertices at the LHC/FCC  and MATHUSLA}

\author[a,b]{Priyotosh Bandyopadhyay,}
\author[c]{Mariana Frank,}
\author[a]{Snehashis Parashar,}
\author[a]{and Chandrima Sen}

\affiliation[a]{Indian Institute of Technology Hyderabad, Kandi,  Sangareddy-502285, Telangana, India}
\affiliation[b]{Korea Institute for Advanced Study, Seoul, 02455, Republic of Korea}
\affiliation[c]{Department of Physics, Concordia University, 7141 Sherbrooke St. West, Montreal, Quebec, Canada H4B 1R6}

\emailAdd{bpriyo@phy.iith.ac.in}
\emailAdd{mariana.frank@concordia.ca}
\emailAdd{ph20resch11006@iith.ac.in}
\emailAdd{ph19resch11014@iith.ac.in}

\preprint{IITH-PH-0001/23 \\ \hspace*{14cm} KIAS-Q23017}

\abstract{We study the interaction between the inert Higgs doublet (IDM) dark matter and  a vector-like $SU(2)$ triplet lepton (VLL), both of which are $Z_2$-odd. The  vector current of the VLL with the $Z$-boson rules out a fermionic or two-component dark matter scenario. However, a compressed mass spectrum and a sufficiently small Yukawa coupling allows co-annihilation and late decay of the VLL into the IDM sector, affecting the relic density of the pseudoscalar dark matter. The same two factors enable displaced decay of the VLL states, providing novel signatures involving hadronically quiet displaced multi-lepton final states. Such signatures to probe the model are studied at the 14 and 27 TeV  LHC, as well as the 100 TeV FCC-hh. In addition to  being detectable at the CMS/ATLAS experiments,  if the new particles have sub-100 GeV masses, signals can also be seen at the proposed MATHUSLA detector. 
 }

\begin{document}

	\maketitle
	\flushbottom
	
\section{Introduction}
The Large Hadron Collider (LHC) was successful in discovering  the last keystone of the Standard Model (SM), which is the Higgs boson \cite{CMS:2012qbp,ATLAS:2012yve}. However, since then, different Beyond the Standard Model (BSM) scenarios were looked for, but no such signs have  been found out in the standard search modes.  The motivation for such searches comes from the experimental and theoretical shortcomings of the SM. One  of the  illusive particles frequently looked for at colliders or through various direct and indirect searches  is the dark matter (DM) candidate, a particle missing from the SM, but for which there is considerable experimental evidence.  The origin of dark matter, and whether or not it is composed of one or more particles, is also unknown. Most BSM scenarios consider a variation of the SM augmented by a dark matter particle or a dark sector.

Here, we propose a scenario with two possible candidates for dark matter, where only one of them can be the actual stable DM of the current universe. Our model introduces a dark $SU(2)$ triplet fermion within the inert two Higgs doublet model (IDM), which itself contains a bosonic dark matter candidate \cite{Dolle:2009fn,LopezHonorez:2006gr,Goudelis:2013uca}. This provides an interplay between the bosonic and fermionic dark sectors, and we show that only one of them survives as the true dark matter, while the other one plays a significant role in attaining the correct dark matter relic today. To achieve this relic abundance, the model predicts very tiny Yukawa couplings, which can be probed at the LHC/FCC-hh via the displaced decays of doubly-charged, singly-charged, and the neutral triplet fermions.

Normally, a chiral $Y=0$ inert $SU(2)$ triplet fermionic dark matter requires a Majorana mass  term, and this possibility  has been investigated in  \cite{Ma:2008cu,Abada:2008ea,vonderPahlen:2016cbw,Chun:2009mh,Suematsu:2019kst,Choubey:2017yyn,Belanger:2022gqc}. However, no such Majorana mass term can be written for a $Y=1$ $SU(2)$ triplet fermionic dark matter, forcing  only a  Dirac type mass term, and thus requiring a vector-like lepton (VLL), which is what we propose to study here. 

Vector-like leptons have been investigated in the literature for a long time in various scenarios \cite{Graham:2009gy, Bernreuther:2023uxh,delAguila:2008pw,Endo:2011xq,Ellis:2014dza,Falkowski:2013jya,Kumar:2015tna,Freitas:2020ttd,Moroi:1991mg,Bhattiprolu:2019vdu,Xu:2018pnq,Poh:2017tfo,Bahrami:2016has,Dermisek:2015oja,Dermisek:2014qca,Dermisek:2013gta,Ishiwata:2013gma,Thomas:1998wy,DeJesus:2020yqx,Moroi:1992zk, Babu:2004xg, Martin:2009bg,Camargo:2019ukv}. Conventional searches for VLLs at the LHC involve prompt searches with hadronic $\tau$ \cite{CMS:2019hsm, CMS:2022pvz, ATLAS:2023sbu}. Searches for VLLs decaying into a prompt electron and a $W$-boson provide a lower bound of $~100$ GeV from the LEP data \cite{L3:2001xsz}, which is evaded by long-lived VLLs. Future collider search strategies for singlet or doublet VLLs are studied in \cite{Shang:2021mgn, Bhattiprolu:2019vdu}. The phenomenology of a triplet VLL with $Y=-1$ is studied in refs \cite{Delgado:2011iz, Ma:2013tda}, where decay signatures with prompt multileptons are predicted at the LHC. The possibility of a compressed mass spectrum leading to displaced decay of a VLL was explored in \cite{Thomas:1998wy}, with soft displaced pions. Introducing VLLs with a discrete $Z_2$-odd symmetry for various purposes is also frequently found in literature \cite{Ghosh:2023xbj, Chakrabarty:2021kmr, Bahrami:2016has}, including establishing a DM candidate mixed with an extra Majorana fermion \cite{Bhattacharya:2015qpa, Bhattacharya:2018fus, Dutta:2020xwn}. An extension with a scalar singlet DM of the feebly interacting massive particle (FIMP) kind, along with a dark VLL singlet, is studied in \cite{Belanger:2018sti}, involving recasting of existing LHC searches for the long-lived VLL. In comparison, our work sticks to the weakly interacting massive particle (WIMP)-type DM, with the aim to study the interplay and the subsequent displaced multilepton signatures at CMS/ATLAS and the proposed MATHUSLA detector \cite{MATHUSLA:2022sze}, of a $Z_2$-odd triplet VLL that exists alongside a $Z_2$-odd scalar doublet.

In this article, we concentrate on the effects of an anomaly-free $Y=-1$ vector-like lepton, which includes a doubly-charged fermion, along with singly-charged and neutral ones.  Unlike the $Y=0$ inert triplet fermion case \cite{Ma:2008cu,vonderPahlen:2016cbw,Suematsu:2019kst,Choubey:2017yyn,Belanger:2022gqc},  here the $Y=-1$ vector-like lepton couples with the $Z$-boson and thus, in such a scenario the  neutral component of the triplet fermion is prohibited to be the dark matter particle. On the other hand, the inert Higgs  doublet  model, where one of the Higgs boson doublets is  also  $Z_2$-odd, provides the much needed  dark matter in the form of a scalar or a pseudoscalar \cite{Datta:2016nfz, Belyaev:2016lok, Jangid:2020qgo, Chakrabarty:2021kmr,Choubey:2017hsq,Banerjee:2019luv,Bhattacharya:2019fgs}, while guaranteeing  the  stability of the electroweak vacuum \cite{Chakrabarty:2016smc, Chakrabarty:2015yia, Swiezewska:2015paa, Khan:2015ipa, Jangid:2020dqh, Bandyopadhyay:2020djh}. To study the  interplay between this vector-like lepton and the inert  Higgs doublet model,  we assume both the inert Higgs doublet and the triplet fermion to be  $Z_2$-odd, even  though  IDM alone provides the necessary  scalar dark matter.  Due to the  $Z_2$-odd nature  of  both states, they can decay into each other via a dark  Yukawa coupling.  For  the choice of the  vector-like lepton being heavier than the inert doublet, the doubly- and singly-charged leptons decay into an SM charged lepton and missing energy, with their decay widths proportional to the square of the Yukawa coupling. 

Another interesting  aspect of this  scenario is that, the vector-like lepton that decays into the  dark scalar now plays  a crucial role in attaining the correct dark  matter relic density. This kind of interplay between scalar and fermionic dark sector has been studied in \cite{Bandyopadhyay:2018qcv,Bandyopadhyay:2017bgh,Bandyopadhyay:2011qm,Ghosh:2023dhj}. Here, we discuss in detail how a lower Yukawa  coupling and/or a lower  mass gap between these two $Z_2$-odd fields play a significant role in the the freeze-out of the scalar dark matter.  This  dark matter  analysis  predicts the dark Yukawa coupling to be  $\mathcal{O}(10^{-7}-10^{-9})$, and also  predicts  displaced vertex signatures when produced at the collider. The $Y=-1$ VLL can be pair produced at the LHC, as well as in association with different states, via processes which are mediated by $Z,\, \gamma$ or $W^\pm$ bosons. Even though the production modes are electroweak in nature, their cross-sections can be non-negligible for a lower mass spectrum and higher centre of mass energies of the LHC, as well as the Future Circular Collider (FCC). Specially, the displaced decays of the doubly-charged vector-like lepton plus missing energy are quite unique signatures for this scenario, which can be  searched  at the  LHC via CMS, ATLAS and the proposed MATHUSLA experiment. 

This article is organized as follows: In  \autoref{model}  we  briefly describe the model, and in \autoref{ModelC} we isolate parameter regions allowed by electroweak precession data (EWPD), $h \to \gamma \gamma$, vacuum stability and perturbativity, for further analysis. The dark matter constraints including  relic abundance, direct and indirect bounds are detailed in \autoref{DM}, with a special emphasis on the interplay between the fermionic and scalar dark sectors. The production and decays of the VLLs are discussed in \autoref{dcprod}. The collider simulations at the LHC/FCC-hh  are carried out in \autoref{csim},  with the results from the analysis of displaced multi-lepton final states given in \autoref{results}. Finally, we summarize the results and conclude the work in \autoref{concl}.

\section{Brief description of the model}\label{model}

Our model extends the particle content of the Standard Model (SM) with an additional $SU(2)_L$ scalar doublet $\Phi_2$, and one generation of a vector-like $SU(2)_L$ triplet lepton $N$ with $U(1)_Y$ hypercharge $Y= - 1$, following the convention $Q=T_3 + Y$. Being vector-like, both the left- and right-chiral components of $N$ i.e. $N_{L,R}$ transform the same way under $SU(2)_L$. Additionally, we impose a discrete $Z_2$ symmetry on the particle content, under which all the SM particles are even, while the new particles i.e. $\Phi_2$ and $N$ are odd.  \autoref{tab:partcon} summarizes the matter particle content and their quantum numbers under the gauge groups.
\begin{table}[ht]
	\centering
	\renewcommand{\arraystretch}{1.5}
	\begin{tabular}{|c|c|c|c|c|c|}
		\hline
		\multirow{2}{*}{Description}&\multirow{2}{*}{Field definition} & \multicolumn{4}{c|}{Quantum Numbers}\\
		\cline{3-6}
		&&$SU(3)_C$&$SU(2)_L$&$U(1)_Y$&$Z_2$\\
		\hline
		\makecell{vector-like\\ lepton (VLL)}&$N  = \begin{pmatrix}
		\frac{N^-}{\sqrt{2}} & N^0 \\ N^{--} & -\frac{N^-} {\sqrt{2}}
		\end{pmatrix}$ & 1 & 3 & $-1$ & $-$ \\
		\hline
		\multirow{2}{*}{\makecell{Scalars}}&$\Phi_1 = (\phi_1^+ \,\, \phi_1^0)^T$ & 1 & 2 & 1/2 & $+$\\
		\cline{2-6}
		&$\Phi_2 = (\phi_2^+ \,\, \phi_2^0)^T$ & 1 & 2 & 1/2 & $-$\\
		\hline
		\multirow{2}{*}{Leptons}&$L_L^i = (\nu_L^i\,\,\ell_L^i)^T$ & 1 & 2 & $-1/2$ & $+$ \\
		\cline{2-6}
		& $\ell_R^i$ & 1 & 1 & $-1$ & $+$\\
		\hline
		\multirow{3}{*}{Quarks} & $Q_L^i  = (u_L^i,d_L^i)$ & 3 & 2& 1/6& $+$ \\
		\cline{2-6} 
		& $u_R^i$ & 3 & 1 & 2/3 & $+$ \\
		\cline{2-6}
		& $d_R^i$ & 3 & 1 & $-1/3$ & $+$ \\
		\hline
	\end{tabular}
\caption{Matter fields in IDM+VLL and their corresponding gauge charges.}
\label{tab:partcon}
\end{table}

The scalar potential of the model is governed  by the regular Inert Doublet $\Phi_2$, and the SM-like  Higgs  doublet  $\Phi_1$ and  can be written as follows:
\begin{align}
\begin{split}
V_{\text{scalar}} &= -m_{\Phi_1}^2 \Phi_1^\dagger \Phi_1 - m_{\Phi_2}^2 \Phi_2^\dagger \Phi_2 + \lambda_1 (\Phi_1^\dagger \Phi_1)^2 + \lambda_2 (\Phi_2^\dagger \Phi_2)^2 + \lambda_{3} (\Phi_1^\dagger \Phi_1)(\Phi_2^\dagger \Phi_2) \\  {}& + \lambda_{4} (\Phi_1^\dagger \Phi_2)(\Phi_2^\dagger \Phi_1) + \left[\lambda_5 (\Phi_1^\dagger \Phi_2)^2 + h.c\right],
\end{split}
\label{eq:scalpot}
\end{align}
where $m_{\Phi_1},   m_{\Phi_2}$ are  the   respective mass terms for the Higgs  doublets  and  $\lambda_i, i=1,2,3,4,5$ are the dimensionless quartic couplings.   Due to the $Z_2$-odd nature, $\Phi_2$ does not couple to SM fermions, whereas $\Phi_1$ couples to quarks  and  leptons.  Nevertheless, $\Phi_2$ can couple to  the $Z_2$-odd vector-like lepton via the  dark Yukawa  coupling $\y_N$ as shown  below, where  $M_N$ is  the Dirac mass for the vector-like lepton.  
\begin{align}
\begin{split}
\mathcal{L}_{VLL} &\supset \left[-\frac{M_N}{2} \overline{N_L}N_R + \y_N \overline{L_L^e} N_R \Phi_2 \right] + h.c.
\end{split}
\label{eq:Nlag}
\end{align}


The Yukawa coupling $\y_N$ facilitates that interaction between the two $Z_2$-odd fields. As we are considering only one generation of the VLL, we are only keeping the coupling of $N$ to the first generation of the SM lepton doublet, in order to avoid unwanted complications from flavour-changing neutral current (FCNC) and lepton flavour violating (LFV) decays. One can expand the Yukawa term in \autoref{eq:Nlag} to see the possible interactions:
\begin{align}
\begin{split}
 \y_N \overline{L_L^e} N_R \Phi_2={}& \frac{1}{\sqrt{2}} \left(\y_N\bar{\nu_e} N^- \phi_2^+ -\y_N\bar{e}N^- \phi_2^0  \right) + \y_N\bar{\nu_e}N^0 \phi_2^0 + \y_N\bar{e} N^{--} \phi_2^+  \label{eq:yuk}
\end{split}
\end{align}

As $\Phi_2$ is odd under the discrete $Z_2$ symmetry, its neutral component does not obtain a vacuum expectation value (VEV), and in turn does not participate in the electroweak symmetry breaking (EWSB). After $\phi_1^0$ acquires the SM vacuum expectation values (VEV) $v = 246.2$ GeV, we can write the scalar fields as fluctuations around the minima: 
\begin{equation}
\Phi_1 = \begin{pmatrix}
G^+ \\ \frac{v + h + iG^0}{\sqrt{2}} 
\end{pmatrix}, \quad \Phi_2 = \begin{pmatrix}
H^+ \\ \frac{H^0 + iA^0}{\sqrt{2}}
\end{pmatrix}
\end{equation}

After EWSB, $G^\pm$ and $G^0$ become the Goldstone bosons, which are then absorbed by the $W$- and $Z$- bosons, and so we have the following particles as scalar mass eigenstates: the SM Higgs boson $h$, the inert neutral CP-even Higgs boson $H^0$, the inert CP-odd pseudoscalar boson $A^0$, and a pair of inert charged Higgs bosons $H^\pm$. There is no mass mixing between the two doublets as $\Phi_2$ is  $Z_2$-odd. The  physical  scalar masses after electroweak symmetry breaking are given  below:
\begin{align}
M_{h}^2 &= 2\lambda_1 v^2, \\
M_{H^0 / A^0}^2 &= m_{\Phi_2}^2 + \half v^2 \lambda_{L/S},\\
M_{H^\pm}^2 &= m_{\Phi_2}^2 + \half v^2 \lambda_{3},
\end{align}

where we have defined 
\begin{equation}
\lambda_{L/S} = \lambda_{3}+\lambda_{4}\pm 2\lambda_5.
\end{equation}
Depending on the sign of $\lambda_{L/S}$, one can have either $A^0$ or $H^0$ as the lightest  pseudoscalar  or scalar, which is a viable dark matter candidate. Moreover, as the VLL only couples to the inert doublet, which receives no VEV, there is no mass mixing between the SM leptons and $N$. Hence, at the tree level, the masses of each of the components of $N$ are degenerate, and they   are  equal to $M_N$. However, at one-loop level, mass splitting occurs between each of the differently charged components, as given below  \cite{Cirelli:2005uq}:

\begin{align}
\Delta M_{N^\pm N^0} ={}& \frac{\alpha_2 M_N}{4 \pi} \left[ (s_W^2 + 1) \mathcal{G} \left( \frac{M_Z}{M_N}\right) - \mathcal{G} \left( \frac{M_W}{M_N}\right)  \right], \label{eq:delm10} \\
\Delta M_{N^{\pm\pm} N^0} ={}& \frac{\alpha_2 M_N}{4 \pi} \left[ 4\,s_W^2 \mathcal{G} \left( \frac{M_Z}{M_N}\right)\right]. \label{eq:delm20}
\end{align}

Here, $\alpha_2$ is the weak coupling constant, $s_W = \sin \theta_W$ with $\theta_W$ being the Weinberg angle, and the function $\mathcal{G}(x)$ for fermions defined as:

\begin{equation}
\mathcal{G}(x) = \frac{x}{2} \left[ 2x^3 \, \ln x - 2x +  (x^2 + 2) \sqrt{x^2 - 4}\,\ln(\frac{x^2 - 2 - x\sqrt{x^2-4}}{2})  \right]. 
\end{equation}

Now, it is important to note that, $N$ and $H^\pm/H^0$ can contribute to the experimental excess of muon $(g-2)$, if one considers $\y_N$ values of $\mathcal{O}(0.1)$ or higher \cite{Dermisek:2021ajd,Ghosh:2022vpb,Poh:2017tfo}. Similar contributions can also be expected for the magnetic and electric dipole moment of the electron, which again require $\y_N \gsim \mathcal{O}(10^{-2})$ \cite{Jana:2020joi}. However, such large couplings will not lead to  displaced decays of the odd particles, which is one of the main interests of this paper.  Having sufficiently small Yukawa couplings for displaced decays will not affect the magnetic/electric dipole moment of the leptons, and they will also have cosmological implications, where late decay of $N^0$ to the DM scalar/pseudoscalar can alter its relic abundance, as compared to the simple IDM scenario. We will discuss this in detail in our DM analysis section. Before moving to the analysis, we summarize the other theoretical and experimental constraints on the model, especially for the lower mass  values range.

\section{Model Constraints}\label{ModelC}

In this section we briefly summarize the basic theoretical and experimental constraints that are imposed upon the model. 

\begin{itemize}
	\item \textbf{Tree-level vacuum stability:} For the  requirement that the scalar potential of the IDM part of our model be bounded from below, the following inequalities must be satisfied by the scalar quartic couplings \cite{Belyaev:2016lok}:
	\begin{equation}
	\lambda_1, \, \lambda_2 > 0; \quad \lambda_{3} + 2\sqrt{\lambda_1 \lambda_2} > 0; \quad \lambda_L + 2\sqrt{\lambda_1 \lambda_2} > 0. 
	\end{equation}
	
	\item \textbf{Perturbative unitarity:} The perturbativity requirement leads to the conditions of $\abs{\lambda_i} \leq 4\pi$. Additionally, the unitarity conditions as described in ref. \cite{Belyaev:2016lok} imply $\lambda_2 \leq 4\pi/3$. Imposing the 125 GeV SM Higgs boson mass requirement of $\lambda_1 = 0.129$, we then have a combined limit on $\lambda_{3}$ and $\lambda_{L}$ from vacuum stability and unitarity as:
	\begin{equation}
	\lambda_{3}, \, \lambda_{L}  \geq -1.47.
	\end{equation}
	\item \textbf{Constraints from $h\to\gamma\gamma$ at the LHC:} The charged scalar in the IDM contributes to the $h\to\gamma\gamma$ one-loop decay, with the partial decay width estimated as \cite{Posch:2010hx, Krawczyk:2013jta}
	\begin{equation}
	\Gamma(h\to\gamma\gamma) = \frac{G_F \alpha^2 M_h^3}{128\sqrt{2}\pi^3}\abs{\sum_{f\in \{t,b,c,\tau\}}N_c Q_f^2 \mathcal{A}_{1/2}(\tau_f) + \mathcal{A}_1(\tau_W) + \frac{\lambda_{3} v^2}{2M_{H^\pm}^2}\mathcal{A}_0(\tau_{H^\pm})}^2.
	\label{eq:hgg}
	\end{equation}
	
	Here, $N_c$ is the colour factor, $Q_f$ is the charge of the fermion in the loop, and $\tau_x = \frac{4M_x^2}{M_h^2}$. The loop functions $\mathcal{A}_i$, where $i\in\{0,1/2,1\}$ representing the spin of the particle in the loop, are defined as \cite{Swiezewska:2012eh}:
	\begin{align}
	\mathcal{A}_0(\tau) &= -\tau\left[1-\tau f(\tau)\right] \label{eq:a0}\\
	\mathcal{A}_{1/2}(\tau) &= 2\tau\left[1+(1-\tau)f(\tau)\right] \label{eq:a12}\\
	\mathcal{A}_1(\tau) &= -\left[2+3\tau+3\tau(2-\tau)f(\tau)\right]\label{eq:a1},
	\end{align}
	where we also have
	\begin{equation}
	f(\tau) = 
	\begin{cases}
	\arcsin[2](\tau^{-1/2}) & \text{if } \tau \geq 1, \\
	-\frac{1}{4}\left[\ln(\frac{1+\sqrt{1-\tau}}{1-\sqrt{1-\tau}} - i\pi)\right]^2 & \text{if } \tau < 1.
	\end{cases}
	\end{equation}

	\autoref{eq:hgg} clearly indicates that $\lambda_{3}$ is the coupling that can be immediately constrained from this bound, with respect to the charged scalar mass. Additionally, while our VLL triplet also contains charged components, the $Z_2$-odd nature prevents them from coupling to the SM Higgs boson, hence keeping it safe from the $h\to\gamma \gamma$ constraint. The constraint is interpreted in terms of the ratio defined a
	\begin{equation}
	\mu_{\gamma\gamma} = \frac{\Gamma(h\to\gamma\gamma)}{\Gamma(h\to\gamma\gamma)_{SM}}.
	\end{equation}

The recent ATLAS measurement\cite{ATLAS:2022tnm} restricts this parameter as $\mu_{\gamma\gamma} =1.04_{-0.09}^{+0.10}$ in the $1\sigma$ limit. Hence, based on vacuum stability, perturbativity, unitarity and $h\to\gamma\gamma$ bounds, we can draw up a comprehensive exclusion plot for the $M_{H^\pm}-\lambda_{3}$ parameter space, as depicted in \autoref{fig:mhclam3}. The coloured regions of red, brown, and green depict the regions excluded by the $h\to \gamma\gamma$ bound, the perturbativity bound, and the combined bound from vacuum stability and unitarity, respectively. In the plot, we also notice a very thin funnel-like region in the $\lambda_{3} > 0$ region, where the parameter space is not restricted by the $h\to\gamma\gamma$ bound. This is accounted for by the positive but small contribution of $H^\pm$, which gives the desired cancellation required to satisfy  $\Gamma(h\to\gamma\gamma)$ for those particular points\cite{Jueid:2020yfj}.
	
\begin{figure}[ht]
	\centering
	\includegraphics[width=0.5\linewidth]{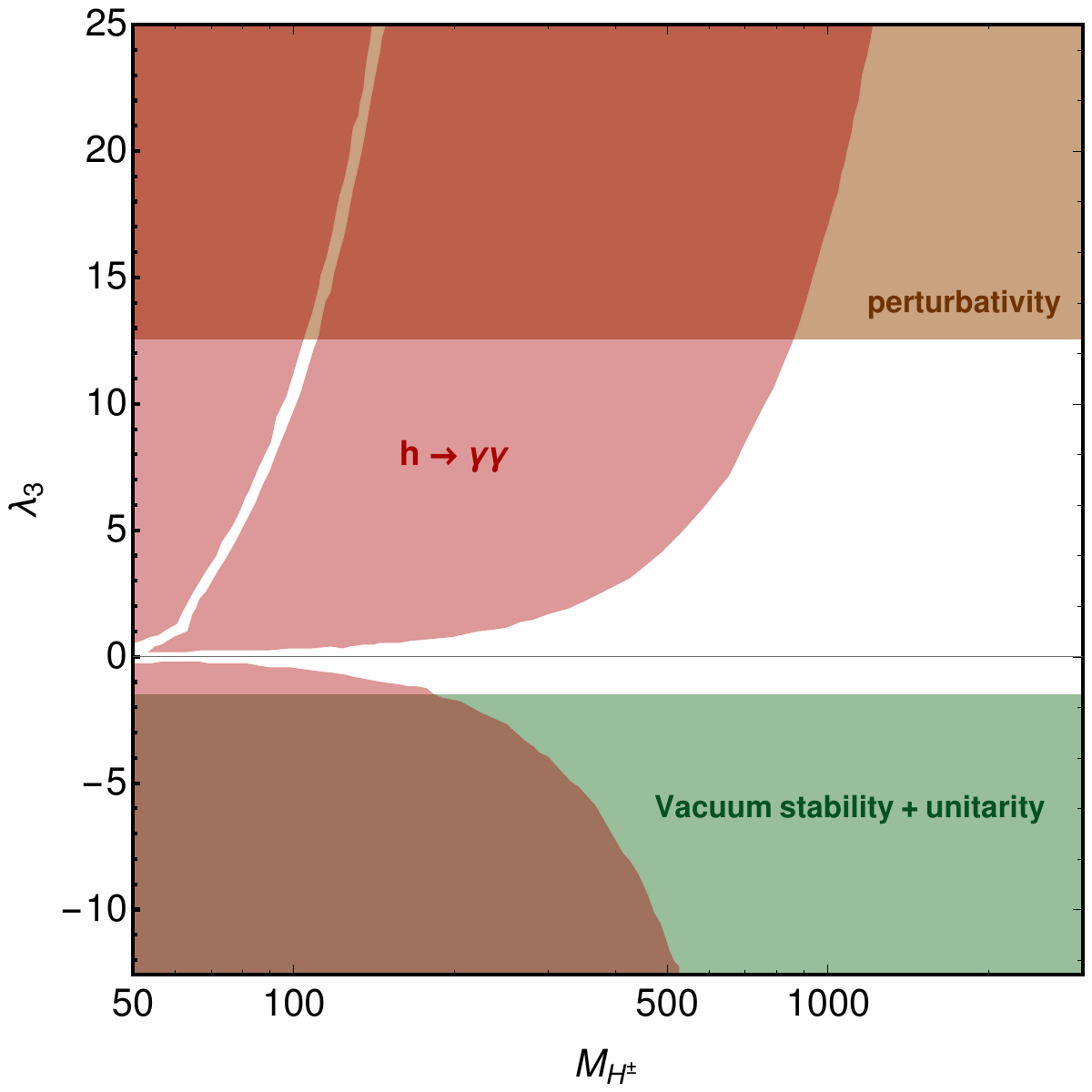}
	\caption{Excluded regions of the $M_{H^\pm}-\lambda_{3}$ parameter space from $h\to\gamma\gamma$ (red), perturbativity (brown), and vacuum stability + unitarity (green) constraints. For the $h\to\gamma\gamma$ region, the latest ATLAS limit of $\mu_{\gamma\gamma} =1.04_{-0.09}^{+0.10}$\cite{ATLAS:2022tnm} is considered.}
	\label{fig:mhclam3}
\end{figure}

\item \textbf{Electroweak precision bounds on IDM and VLL:}

The electroweak precision data (EWPD) establishes three oblique parameters viz. $S$, $T$, $U$, which get contribution from the extra scalars present in our model. The most recent measurements from ref. \cite{ParticleDataGroup:2022pth} are given as:
\begin{align}
S = - 0.02 \pm 0.10, \quad T = 0.03 \pm 0.12 \quad \text{and} \quad U = 0.01 \pm 0.11. 
\end{align}

The additional contribution from IDM can be given by \cite{Chakrabarty:2021kmr},
\begin{align}
S &= \frac{1}{2\pi}\left[\frac{1}{6}\ln(\frac{M_{H^0}^2}{M_{H^\pm}^2})-\frac{5}{36}+\frac{M_{H^0}^2 M_{A^0}^2}{3\left(M_{A^0}^2-M_{H^0}^2\right)^2}+\frac{M_{A^0}^4\left(M_{A^0}^2-3M_{H^0}^2\right)}{6\left(M_{A^0}^2-M_{H^0}^2\right)^3} \ln(\frac{M_{A^0}^2}{M_{H^0}^2})\right],\\
T&= \frac{1}{16 \pi\, s_W^2\, M_W^2}\left[\mathcal{F}\left(M_{H^\pm}^2, M_{H^0}^2\right)+\mathcal{F}\left(M_{H^\pm}^2, M_{A^0}^2\right)-\mathcal{F}\left(M_{H^0}^2, M_{A^0}^2\right)\right],\\
U&=0,
\end{align}
where, 

\begin{equation}
\mathcal{F}(a,b)= 
\begin{cases}
 \frac{a+b}{2}-\frac{ab}{a-b} \ln(\frac{a}{b}) & \text{for } a \neq b, \\
0 & \text{for } a=b.
\end{cases}
\end{equation}

Since, in our case the VLL is odd under $Z_2$ symmetry and does not have any interaction with the SM Higgs boson, it will not contribute to the $S$, $T$, $U$ parameters \cite{vonderPahlen:2016cbw}. In our further analysis, we will be working within the parameter space allowed by the EWPD observables, alongside the bounds presented in \autoref{fig:mhclam3}.
	
\end{itemize}
\section{Dark Matter constraints}\label{DM}
In this  section  we  shall  discuss  the various possibilities of dark matter in the  framework where there  are two  $Z_2$-odd DM candidates $N^0$ and $A^0$, in the fermionic and the scalar dark sectors, respectively. Due to its Dirac  nature, $N^0$ has a  non-zero vector current coupling with the $Z$-boson, leading to high values of DM-nucleon scattering cross-section. Hence, $N^0$ as a DM candidate is conclusively excluded by the DM direct detection experiments \cite{Essig:2007az}. For the same reason,  the  multi-component scenario is also ruled out,  leaving  the only possibility of  $A^0$ being the  dark  matter. Nonetheless, if one considers a compressed mass spectrum between the $Z_2$-odd particles, as well as small Yukawa couplings $\y_N$, they can ensure some significant interplay via co-annihilation and late decay effects, influencing the relic density of $A^0$.

\subsection{DM relic density}\label{relic}
The interplay between the scalar dark sector ($\Phi_2$) and the fermionic dark sector ($N$) can lead to a pair of coupled Boltzmann equations, which can then be solved using the \texttt{darkOmegaNTR} routine in \mic 5.3.41 \cite{Belanger:2001fz, Alguero:2022inz}, to obtain the yields of the two sectors. The traditional \texttt{darkOmega} function of \mic solves only one Boltzmann equation, keeping all the $Z_2$-odd particles in the same sector. This approach does not correctly take into account the late decay of one species to the other, facilitated by small Yukawa couplings, which takes place out of equilibrium. Hence, following the prescription of the \texttt{darkOmegaNTR} routine in ref. \cite{Alguero:2022inz}, we denote the scalar dark sector with $1 \equiv [A^0, H^0, H^\pm]$ and the fermionic dark sector with $2 \equiv [N^0, N^\pm, N^{\pm\pm}]$, while assigning $0 \equiv$ all SM particles. With this notation, the number density evolution equations of the dark sectors, denoted as $Y_{\Phi_2}$ and $Y_N$, can be written as follows:

\begin{align}
	\begin{split}
		\frac{dY_{\Phi_2}}{dx} {}= & -\frac{1}{x^2}\frac{s(M_{A^0})}{H(M_{A^0})} \Biggl[ \sigmav_{1100} \left( Y_{\Phi_2}^2 - (Y_{\Phi_2}^{eq})^2\right) - \sigmav_{2211} \left( Y_{\Phi_2}^2 - Y_{N}^2 \frac{(Y_{\Phi_2}^{eq})^2}{(Y_{N}^{eq})^2} \right) \\ & + \sigmav_{1200} \left( Y_{\Phi_2} Y_{N} - Y_{\Phi_2}^{eq} Y_{N}^{eq} \right)  \Biggr] + \frac{x\Gamma_{N\to \Phi_2 X  }}{H(M_{A^0})} \left(Y_{N} - Y_{\Phi_2} \frac{Y_{N}^{eq}}{Y_{\Phi_2}^{eq}}\right),
	\end{split} \label{eq:beq1}
\end{align}
\begin{align}
	\begin{split}
		\frac{dY_{N}}{dx} {}= & -\frac{1}{x^2}\frac{s(M_{A^0})}{H(M_{A^0})} \Biggl[ \sigmav_{2200} \left( Y_{N}^2 - (Y_{N}^{eq})^2\right) + \sigmav_{2211} \left( Y_{\Phi_2}^2 - Y_{N}^2 \frac{(Y_{\Phi_2}^{eq})^2}{(Y_{N}^{eq})^2} \right) \\ & + \sigmav_{1200} \left( Y_{\Phi_2} Y_{N} - Y_{\Phi_2}^{eq} Y_{N}^{eq} \right)  \Biggr] - \frac{x\Gamma_{N\to \Phi_2 X  }}{H(M_{A^0})} \left(Y_{N} - Y_{\Phi_2} \frac{Y_{N}^{eq}}{Y_{\Phi_2}^{eq}}\right).
	\end{split} \label{eq:beq2}
\end{align}

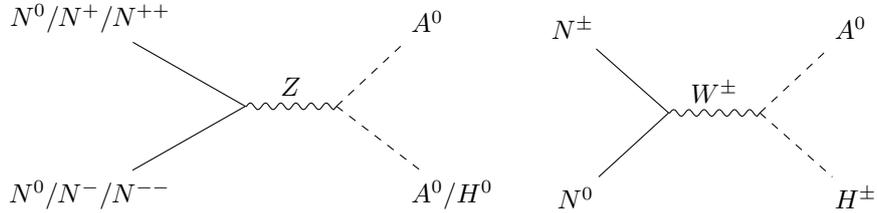
\begin{figure}[ht]
	\centering
	\textbf{Diagrams for 1100:}\\
	\begin{tikzpicture}
	\begin{feynman}
	\vertex (a2);
	\vertex [above left=1.1cm of a2] (a0){\small $A^0$};
	\vertex [left =0.1cm of a2] (d1);
	\vertex [below left=1.1cm of a2] (a1){\small $A^0$};
	\vertex [right =1.1cm of a2] (d2);
	\vertex [above right=1.1cm of a2] (f2){\small $W^+/Z$};
	\vertex [below right=1.1cm of a2] (f3){\small $W^-/Z$};
	
	\diagram {(a2)--[scalar](a1), (a2)--[scalar](a0), (a2)--[boson](f2), (a2)--[boson](f3),
	};
	\end{feynman}
	\end{tikzpicture}
	\begin{tikzpicture}
	\begin{feynman}
	\vertex (a2);
	\vertex [above left=1.2cm of a2] (a0){\small $A^0$};
	\vertex [left =0.1cm of a2] (d1);
	\vertex [below left=1.2cm of a2] (a1){\small $A^0$};
	\vertex [right =1.3cm of a2] (d2);
	\vertex [right=1cm of a2] (f1);
	\vertex [above right=1.2cm of f1] (f2){\small $W^+/Z$};
	\vertex [below right=1.2cm of f1] (f3){\small $W^-/Z$};
	
	\diagram {(a2)--[scalar](a1), (a2)--[scalar](a0), (a2)--[scalar, edge label=\small$h$](f1), (f1)--[boson](f2), (f1)--[boson](f3),
	};
	\end{feynman}
	\end{tikzpicture}
	\begin{tikzpicture}
	\begin{feynman}
	\vertex (a2);
	\vertex [left=1.1cm of a2] (a0){\small $A^0$};
	\vertex [above =0.05cm of a2] (d1);
	\vertex [below=1.9cm of a2] (a1);
	\vertex [below =0.05cm of a1] (d2);
	\vertex [left=1.0cm of a1] (i1){\small $A^0$};
	\vertex [right=1.0cm of a1] (i2){\small $W^+/Z$};
	\vertex [right=1.0cm of a2] (f1){\small $W^-/Z$};
	
	\diagram {(a2)--[boson](f1), (a2)--[scalar](a0), (a2)--[scalar, edge label=\small $H^+/H^0/A^0$](a1), (i1)--[scalar](a1), (a1)--[boson](i2)
	};
	\end{feynman}
	\end{tikzpicture}
	\begin{tikzpicture}
	\begin{feynman}
	\vertex (a2);
	\vertex [left=1.1cm of a2] (a0){\small $A^0$};
	\vertex [above =0.05cm of a2] (d1);
	\vertex [below=1.9cm of a2] (a1);
	\vertex [below =0.05cm of a1] (d2);
	\vertex [left=1.0cm of a1] (i1){\small $H^\pm$};
	\vertex [right=1.0cm of a1] (i2){\small $\gamma/Z$};
	\vertex [right=1.0cm of a2] (f1){\small $W^\pm$};
	
	\diagram {(a2)--[boson](f1), (a2)--[scalar](a0), (a2)--[scalar, edge label=\small $H^\pm$](a1), (i1)--[scalar](a1), (a1)--[boson](i2)
	};
	\end{feynman}
	\end{tikzpicture}
	
	\textbf{Diagrams for 2200:}\\
	\centering\begin{tikzpicture}
	\begin{feynman}
	\vertex (a2);
	\vertex [left=1.1cm of a2] (a0){\small $N^0$};
	\vertex [above =0.05cm of a2] (d1);
	\vertex [below=1.9cm of a2] (a1);
	\vertex [below =0.05cm of a1] (d2);
	\vertex [left=1.0cm of a1] (i1){\small $N^0$};
	\vertex [right=1.0cm of a1] (i2){\small $W^+/Z$};
	\vertex [right=1.0cm of a2] (f1){\small $W^-/Z$};
	
	\diagram {(a2)--[boson](f1), (a2)--(a0), (a2)--[edge label=\small $N^+/N^0$](a1), (i1)--[](a1), (a1)--[boson](i2)
	};
	\end{feynman}
	\end{tikzpicture}
	\quad 
	\centering\begin{tikzpicture}
	\begin{feynman}
	\vertex (a2);
	\vertex [left=1.1cm of a2] (a0){\small $N^0$};
	\vertex [above =0.05cm of a2] (d1);
	\vertex [below=1.9cm of a2] (a1);
	\vertex [below =0.05cm of a1] (d2);
	\vertex [left=1.0cm of a1] (i1){\small $N^\pm$};
	\vertex [right=1.0cm of a1] (i2){\small $\gamma/Z$};
	\vertex [right=1.0cm of a2] (f1){\small $W^\pm$};
	
	\diagram {(a2)--[boson](f1), (a2)--[](a0), (a2)--[ edge label=\small $N^\pm$](a1), (i1)--[](a1), (a1)--[boson](i2)
	};
	\end{feynman}
	\end{tikzpicture}
	\quad 
	\centering\begin{tikzpicture}
	\begin{feynman}
	\vertex (a2);
	\vertex [left=1.1cm of a2] (a0){\small $N^0$};
	\vertex [above =0.05cm of a2] (d1);
	\vertex [below=1.9cm of a2] (a1);
	\vertex [below =0.05cm of a1] (d2);
	\vertex [left=1.0cm of a1] (i1){\small $N^{\pm\pm}$};
	\vertex [right=1.0cm of a1] (i2){\small $W^\pm$};
	\vertex [right=1.0cm of a2] (f1){\small $W^\pm$};
	
	\diagram {(a2)--[boson](f1), (a2)--[](a0), (a2)--[ edge label=\small $N^\mp$](a1), (i1)--[](a1), (a1)--[boson](i2)
	};
	\end{feynman}
	\end{tikzpicture}
	
	\textbf{Diagrams for 1200:}\\
	\centering\begin{tikzpicture}
	\begin{feynman}
	\vertex (a2);
	\vertex [left=1.1cm of a2] (a0){\small $A^0$};
	\vertex [above =0.05cm of a2] (d1);
	\vertex [below=1.9cm of a2] (a1);
	\vertex [below =0.05cm of a1] (d2);
	\vertex [left=1.0cm of a1] (i1){\small $N^0$};
	\vertex [right=1.0cm of a1] (i2){\small $W^+/Z$};
	\vertex [right=1.0cm of a2] (f1){\small $\ell^-/\nu$};
	
	\diagram {(a2)--[](f1), (a2)--[scalar](a0), (a2)--[edge label=\small $N^+/N^0$](a1), (i1)--[](a1), (a1)--[boson](i2)
	};
	\end{feynman}
	\end{tikzpicture}
	\quad 
	\centering\begin{tikzpicture}
	\begin{feynman}
	\vertex (a2);
	\vertex [left=1.1cm of a2] (a0){\small $A^0$};
	\vertex [above =0.05cm of a2] (d1);
	\vertex [below=1.9cm of a2] (a1);
	\vertex [below =0.05cm of a1] (d2);
	\vertex [left=1.0cm of a1] (i1){\small $N^\pm$};
	\vertex [right=1.0cm of a1] (i2){\small $W^\pm$};
	\vertex [right=1.0cm of a2] (f1){\small $\nu$};
	
	\diagram {(a2)--[](f1), (a2)--[scalar](a0), (a2)--[ edge label=\small $N^0$](a1), (i1)--[](a1), (a1)--[boson](i2)
	};
	\end{feynman}
	\end{tikzpicture}
	\quad 
	\centering\begin{tikzpicture}
	\begin{feynman}
	\vertex (a2);
	\vertex [left=1.1cm of a2] (a0){\small $A^0$};
	\vertex [above =0.05cm of a2] (d1);
	\vertex [below=1.9cm of a2] (a1);
	\vertex [below =0.05cm of a1] (d2);
	\vertex [left=1.0cm of a1] (i1){\small $N^{\pm\pm}$};
	\vertex [right=1.0cm of a1] (i2){\small $W^\pm$};
	\vertex [right=1.0cm of a2] (f1){\small $\ell^\pm$};
	
	\diagram {(a2)--[](f1), (a2)--[scalar](a0), (a2)--[ edge label=\small $N^\mp$](a1), (i1)--[](a1), (a1)--[boson](i2)
	};
	\end{feynman}
	\end{tikzpicture}
	
	\textbf{Diagrams for 2211:}\\
	
	\begin{tikzpicture}
	\begin{feynman}
	\vertex (a2);
	\vertex [above left=1.2cm of a2] (a0){\small  $N^0/N^+/N^{++}$ };
	\vertex [below left=1.2cm of a2] (a1){\small $N^0/N^-/N^{--}$};
	\vertex [right=1.2cm of a2] (f1);
	\vertex [above right=1.2cm of f1] (f2){\small $A^0$};
	\vertex [below right=1.2cm of f1] (f3){\small $A^0/H^0$};
	
	\diagram {(a2)--[](a1), (a2)--[](a0), (a2)--[boson, edge label=\small $Z$](f1), (f1)--[scalar](f2), (f1)--[scalar](f3)
	};
	\end{feynman}
	\end{tikzpicture}
	\quad
	\centering\begin{tikzpicture}
	\begin{feynman}
	\vertex (a2);
	\vertex [above left=1.2cm of a2] (a0){\small $N^\pm$};
	\vertex [below left=1.2cm of a2] (a1){\small $N^0$};
	\vertex [right=1.2cm of a2] (f1);
	\vertex [above right=1.2cm of f1] (f2){\small  $A^0$};
	\vertex [below right=1.2cm of f1] (f3){\small  $H^\pm$};
	
	\diagram {(a2)--[](a1), (a2)--[](a0), (a2)--[boson, edge label=\small $W^\pm$](f1), (f1)--[scalar](f2), (f1)--[scalar](f3)
	};
	\end{feynman}
	\end{tikzpicture}
	\caption{Dominant annihilation and co-annihilation diagrams for $A^0$ and $N^0$, considering each possible combination of dark and SM sectors.}
	\label{fig:anndiag}
\end{figure}

Here, $Y_{\Phi_2, N}^{eq}$ are the equilibrium abundances, $x = \frac{M_{DM}}{T}$ with $T$ being the temperature of the universe, $H(M_{A^0})$ is the Hubble parameter, $s(M_{A^0})$ is the  entropy density, and $\Gamma_{N\to\Phi_2 X}$ is the total decay width of the fermionic dark sector particles into scalar dark sector particles alongside some SM particles $X$. The individual decay widths of the three fermionic states are written as follows:
\begin{equation}
\Gamma_{N^{0}\to A^0/H^0 \nu  } = \sum_{i\in[A^0,H^0]} \frac{\mathcal{Y}_N^2 M_{N^0}}{64\pi} \left( 1-\frac{M_{i}^2}{M_{N^0}^2}\right)^2 , \label{eq:n0dec}
\end{equation}
\begin{equation}
\Gamma_{N^{\pm}\to A^0/H^0 \ell^\pm  } = \sum_{i\in[A^0,H^0]} \frac{\mathcal{Y}_N^2 M_{N^\pm}}{128\pi} \left( 1-\frac{M_{i}^2}{M_{N^\pm}^2}\right)^2 , \label{eq:npdec}
\end{equation}
\begin{equation}
\Gamma_{N^{\pm}\to H^\pm \nu  } =  \frac{\mathcal{Y}_N^2 M_{N^\pm}}{64\pi} \left( 1-\frac{M_{H^\pm}^2}{M_{N^\pm}^2}\right)^2 , \label{eq:npdec2}
\end{equation}
\begin{equation}
\Gamma_{N^{\pm\pm}\to H^\pm \ell^\pm  } =  \frac{\mathcal{Y}_N^2 M_{N^{\pm\pm}}}{32\pi} \left( 1-\frac{M_{H^\pm}^2}{M_{N^{\pm\pm}}^2}\right)^2 . \label{eq:nppdec2}
\end{equation}



In \autoref{eq:beq1} and \autoref{eq:beq2}, the subscripts $1100, 2200$ signify the annihilation of the dark sector particles into SM particles, $1200$ represents the co-annihilation of the two dark sectors, and $2211$ is assigned to the annihilation of the fermionic dark sector into the scalar dark sector. A few dominant Feynman diagrams for each of these processes are shown in \autoref{fig:anndiag}. 

\begin{figure}[ht]
	\centering
	\includegraphics[width=0.5\linewidth]{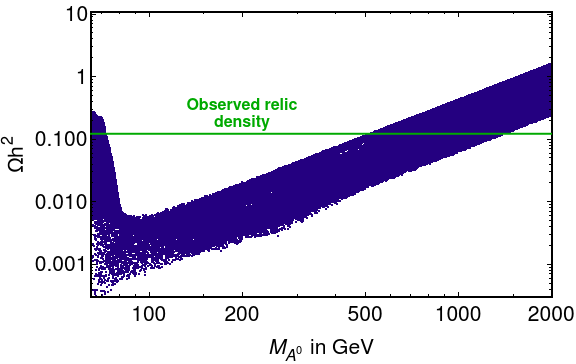}
	\caption{Mass of $A^0$ versus the DM relic density, from the parameter scan discussed in the text. Here, all the points are randomly generated such that they are allowed by the bounds from $h\to\gamma\gamma$, EWPD, stability and perturbativity.  }
	\label{fig:rel}
\end{figure}

As mentioned previously, we utilize the \texttt{darkOmegaNTR} routine in \mic 5.3.41 to solve these coupled Boltzmann equations including the decay effect, and obtain a plot of $M_{A^0}$ vs DM relic density $\Omega h^2$, as shown in \autoref{fig:rel}.  All the blue points are randomly generated such that they are allowed by the bounds from $h\to\gamma\gamma$, EWPD, stability and perturbativity  as discussed  in  \autoref{ModelC}.  The  green band shows the  observed relic  of $\Omega h^2=0.1199\pm  0.0027 $  \cite{Planck:2018vyg}.  We  can see  a  lower mass  region satisfying  relic around  $70$ GeV,  which is due to the $s$-channel resonance through the SM Higgs boson, as  can  be seen in \autoref{fig:anndiag}. Such annihilation in a pure IDM case i.e. without the VLL involvement is also reported in \cite{Datta:2016nfz}.

It is important to note that, from the \autoref{fig:rel} itself, the interplay between the VLL and the IDM is not very apparent, due to the large parameter space that has been scanned over. On closer inspection, one can indeed witness the effects of the co-annihilation, and the out-of-equilibrium decay of $N$, in obtaining the final relic density of $A^0$. Such effects, in essence, reorganize the parameter space of a pure IDM scenario (without the VLL), for fixed values of $M_{A^0}$ and $\lambda_{S}$. This reorganization can include bringing back parameter points which are ruled out in pure IDM due to overabundance, or with very late decay of the VLL rendering previously relic-satisfying (or underabundant) points as overabundant. For obvious reasons, we are interested in the former type of reorganization. A comprehensive study of the dynamics of the co-annihilation and decay effects from the VLL are detailed in \autoref{interplay}, after we establish a set of benchmark points.

\subsection{DM direct detection}

\begin{figure}[ht]
	\centering
	\includegraphics[width=0.7\linewidth]{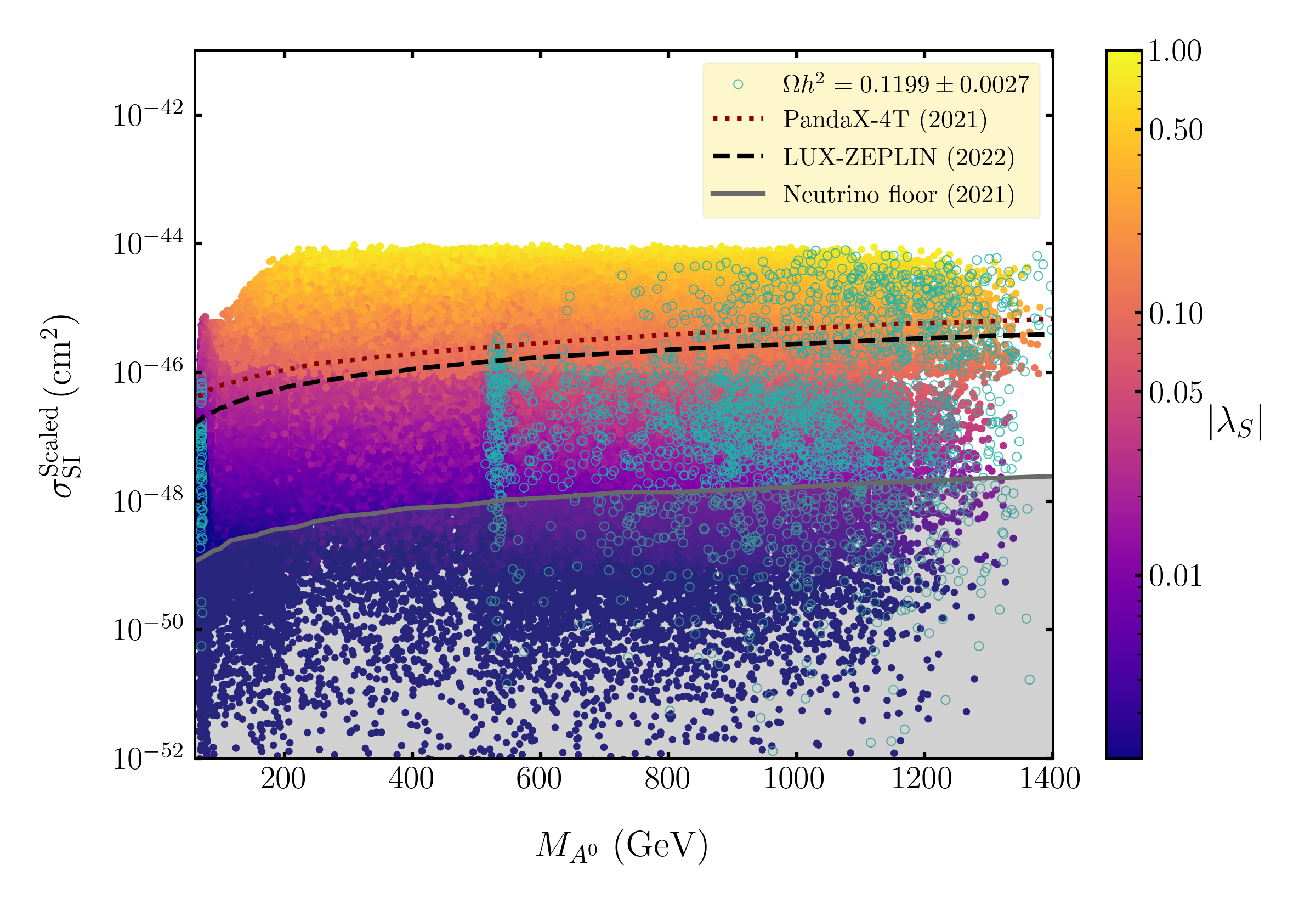}
	\caption{Scaled SI cross-section versus DM mass, with bounds from LZ and PandaX-4T experiments. A colour gradient is provided for the different values of $\lambda_S$. The points that satisfy the observed relic abundance are shown in cyan-coloured bubbles. All other points are underabundant in relic density.}
	\label{fig:sics}
\end{figure}

With an understanding of the mass region permitted by the Planck data, we can now move to the strongest constraints on the weakly interacting massive particles (WIMP), namely the DM direct detection bounds. In particular, we consider the two experiments with the most stringent upper limits on the  spin-independent (SI) DM-nucleon scattering cross-section $\sigma_{\rm SI}$, namely the PandaX-4T \cite{PandaX-4T:2021bab}, and the LUX-ZEPLIN (LZ) \cite{LZ:2022ufs} experiments. For the pseudoscalar DM in the inert doublet part of our model, the spin-independent (SI) DM-nucleon scattering cross-section is facilitated by the Higgs portal coupling $\lambda_{S}$, and the expression is given as:
\begin{equation}
\label{eq:sigma:SI}
\sigma_{\rm SI} \simeq \frac{\lambda_{S}^2 f_n^2}{4\pi M_h^4} \frac{M_n^4}{(M_n + M_{A^0})^2},
\end{equation}
where $M_n$ is the mass of the nucleon, and $f_n$ is the nucleon form factor.  This is  also the same as the pure IDM case  \cite{Belyaev:2016lok, Jangid:2020qgo}. We obtain the values of $\sigma_{\rm SI}$ for the points from \autoref{fig:rel} that either satisfy the correct relic or are underabundant, and plot them against the mass range in \autoref{fig:sics}. The SI cross-sections in the plot are scaled as follows, to account for the proportional decrease of DM-nucleon scattering rate for underabundant points:
\begin{equation}
\sigma_{\rm SI}^{\text{scaled}} = \sigma_{\rm SI} \times \frac{\Omega h^2}{\Omega_{\text{Planck}}h^2}.
\end{equation}

In \autoref{fig:sics}, a colour gradient corresponding to the absolute values of $\lambda_{S}$ has been presented, which is the only coupling that affects the SI cross-section. The cyan-coloured circles that are overlaid represent the ones that satisfy the observed relic density of $\Omega h^2 = 0.1199 \pm 0.0027$. One does not see any regular pattern on these cyan points, because in our model the relic density is governed by many factors such as the VLL masses, the Yukawa coupling $\y_N$, as well as the mass gap between the scalars themselves. The upper bounds from LZ (black dashed line) and PandaX-4T (red dotted line) are also shown, which rules out a significant chunk of the parameter space. The strong upper bound from the LZ experiment excludes $\abs{\lambda_{S}} \gsim 0.5$ values in the relic-satisfying region of $M_{A^0} \in [500, 1400]$ GeV. As far as the region near the SM Higgs boson resonance, with $M_{A^0} \in [65, 75]$ GeV, $\abs{\lambda_{S}} \gsim 0.05$ values are disallowed by the LZ data.  It is also important to consider the neutrino floor \cite{Billard:2021uyg}, a region of $\sigma_{\rm SI}$ below which the DM-nucleon recoil is mimicked by the neutrino-nucleon recoils, as depicted with the grey line and the shaded region below it. Because of this floor, for very low values of $\abs{\lambda_{S}}$, the direct detection experiment loses sensitivity. We wish to restrict ourselves to the parameter region between the LZ upper bound and the neutrino floor, within which a direct detection of DM can still be feasible. We extract the points satisfying this criteria, and proceed to discuss the constraints on these points coming from the DM indirect detection experiments.

%
\subsection{DM Indirect detection}\label{DDM}

The DM relic in our universe is distributed across galaxies in the form of DM haloes. If in principle they annihilate to SM particles, one can obtain a gamma ray flux coming from the regions in the galaxies where the DM are rich in abundance. Such gamma ray flux can be detected and translated into the $\sigmav$ of the DM in a particular channel of SM final states.

\begin{figure}[ht]
	\centering
	\subfigure[]{\includegraphics[width=0.45\linewidth]{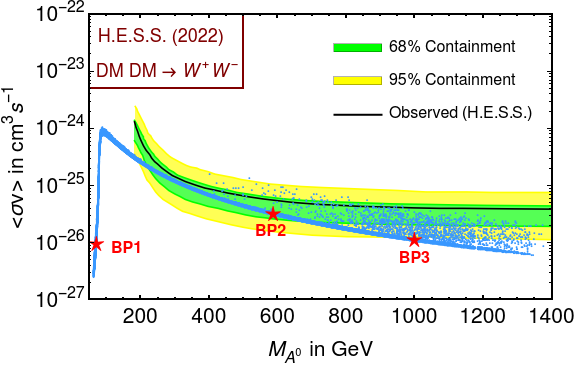}}
	\subfigure[]{\includegraphics[width=0.45\linewidth]{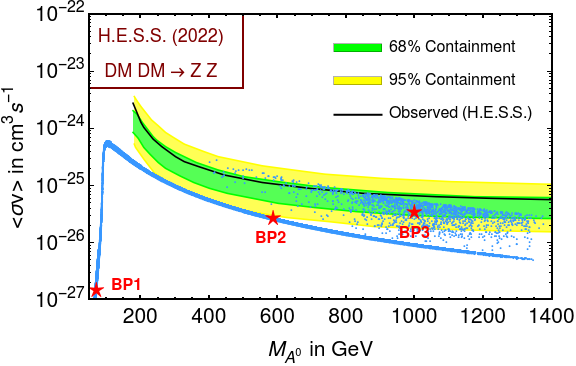}}
	\subfigure[]{\includegraphics[width=0.45\linewidth]{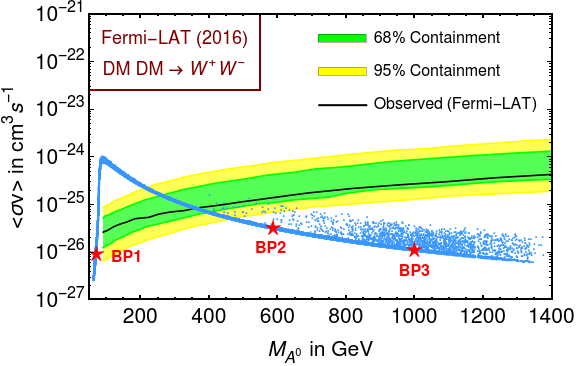}}
	\caption{DM indirect detection bounds on the $\sigmav$ of $A^0$ in the $W^+W^-, \, ZZ$ annihilation channel, for the mass range of $M_{A^0} \in [65, 1400]$ GeV, from (a), (b) H.E.S.S experiment \cite{HESS:2022ygk} and (c) Fermi-LAT experiment \cite{MAGIC:2016xys}. }
	\label{fig:indd}
\end{figure}

The  dominant  modes of $A^0$ annihilation are $W^\pm W^\mp$ and $ZZ$, which directly or via their decay products give rise to  the energetic photons that can be detected by the telescope experiments like H.E.S.S. \cite{HESS:2022ygk} and Fermi-LAT \cite{MAGIC:2016xys}. In \autoref{fig:indd} we present the $\sigmav$ values of the points allowed by the direct detection experiments and the neutrino floor, scaled with the percentage contribution of the $W^+ W^-, \, ZZ$ annihilation channels via the blue points, for the mass range of $65-1400$ GeV. \autoref{fig:indd}(a) shows the upper bound from the H.E.S.S. experiment \cite{HESS:2022ygk} in the  $W^+ W^-$ channel, which provides the strongest bound out of all channels, in the higher mass region. Most of the points with $W^+ W^-$ as the dominant annihilation mode form a narrow band in the lower $\sigmav$ region. A few widely scattered points, where the mass gaps in the scalar sector are $\gsim 2$ GeV, are observed for $M_A^0 \gsim 400$ GeV, having larger values of $\sigmav$. For those points, the $ZZ$ annihilation channel is dominant, which in general has higher cross-section, owing to the $A^0 H^0 Z$ vertex being larger than the $A^0 H^\pm W^\mp$ vertex. This effect is visibly strengthened in \autoref{fig:indd}(b), where we show the H.E.S.S upper bound from the $ZZ$ annihilation channel, which evidently is more relaxed. The scattered points above $M_A^0 \sim 400$ GeV are the ones where the $ZZ$ mode contributes more than the $W^+ W^-$ mode, and the number of possibilities increases as we go up in mass. The narrow band of blue points with lower cross-section contains the same points where the  $W^+ W^-$ mode has a dominant contribution.  \autoref{fig:indd}(c) presents the observed upper bound from the Fermi-LAT experiment \cite{MAGIC:2016xys}, which is more stringent in the lower mass region of the DM. The $W^+ W^-$ annihilation channel bound starts from $M_{DM} \sim 90$ GeV, and hence the low-mass region of our pseudoscalar DM candidate remains unconstrained.  We choose three benchmark points which are allowed by  these  indirect bounds, the direct detection bounds and the DM observed relic, density , which are also  motivated by the displaced decay features of this model. These are marked with red stars in the plot. The details of the benchmark points are discussed in the next section.

\subsection{VLL and IDM interplay}\label{interplay}

For the collider study, we identify three benchmark points (BPs) with different values of the DM mass, which we present here along with the masses of the other $Z_2$-odd particles. The BPs all satisfy the correct DM relic, are allowed by the direct and indirect detection bounds, and are in accordance with the constraints from $h\to\gamma\gamma$, EWPD  as well as the collider bounds for VLL \cite{L3:2001xsz}. In the \autoref{tab:bp} below, we list the BPs in detail, with the masses and relevant couplings of the additional scalars and fermions in the model.

\begin{table}[ht]
	\centering
	\renewcommand{\arraystretch}{1.4}
	\begin{tabular}{|c|c|c|c|c|c|c|c|c|c|c|}
		\hline
		BP & \makecell{$M_{A^0}$ \\ (GeV)} & \makecell{$M_{H^0}$\\ (GeV)}&  \makecell{$M_{H^\pm}$\\ (GeV)} & \makecell{$M_{N^0}$\\ (GeV)}& \makecell{$M_{N^-}$\\ (GeV)} & \makecell{$M_{N^{--}}$\\ (GeV)} &$\lambda_{3}$ & $\lambda_{S}$ & $\y_N$ & $\Omega h^2$\\
		\hline
		{\bf BP1} & 71.57 & 117.16 & 84.76 & 98.25 & 98.61 & 99.28 & 0.07 & 0.002 & 4.2 $\times 10^{-9}$ & 0.119 \\
		\hline
		{\bf BP2} & 587.6 & 589.4 & 588.2 & 595.5 & 595.9 & 596.8 & 0.05 & 0.03 & 1.1 $\times10^{-7}$ & 0.121 \\
		\hline
		{\bf BP3} & 1000.0 & 1010.5 & 1001.0 & 1010.6 & 1011.0 & 1011.9 & 0.02 & -0.04 & 5.4$\times 10^{-7}$ & 0.121 \\
		\hline
	\end{tabular}
	\caption{Benchmark points {\bf BP1}, {\bf BP2}, and {\bf BP3}, with the masses of the dark sector particles, the relevant couplings, and DM relic density values.}
	\label{tab:bp}
\end{table}

\begin{figure}[ht]
	\centering
	\subfigure[]{\includegraphics[width=0.42\linewidth]{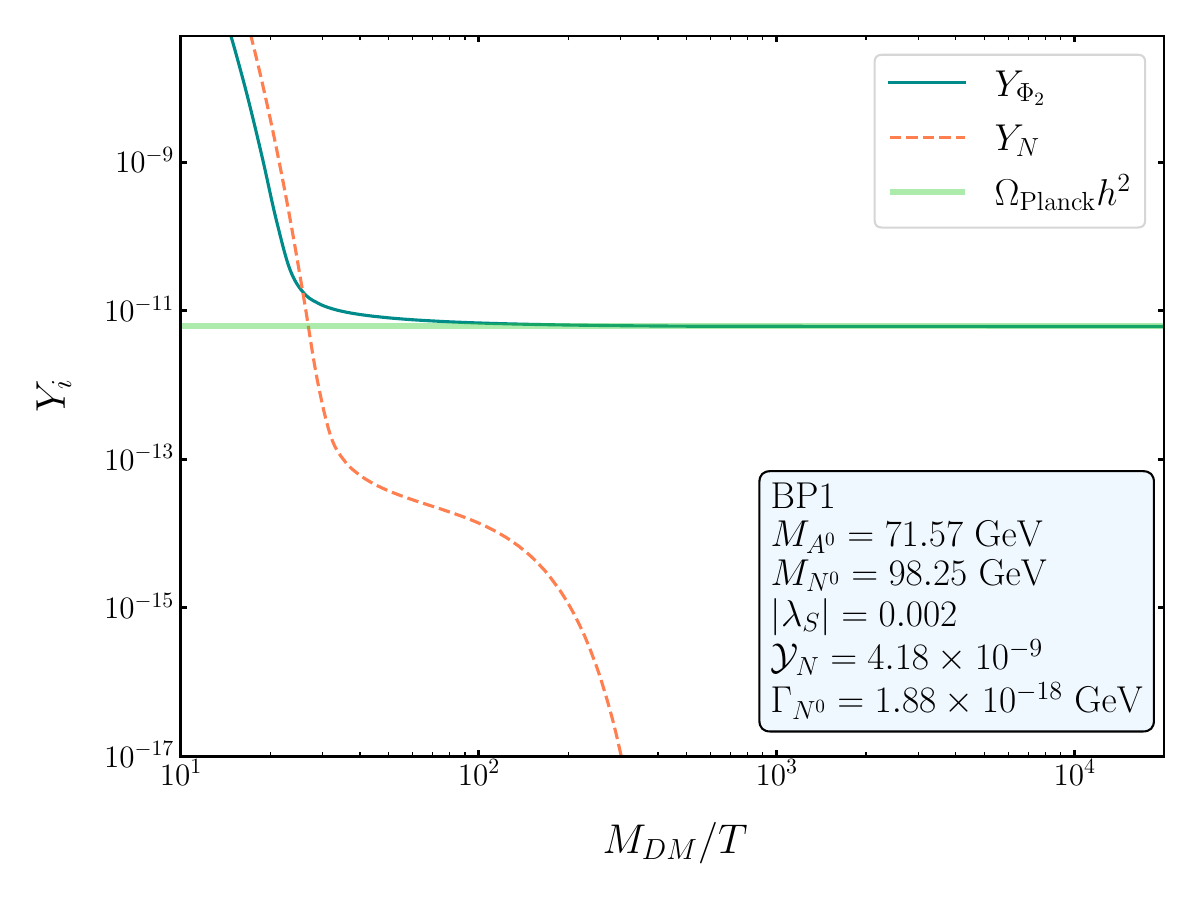}}
	\subfigure[]{\includegraphics[width=0.42\linewidth]{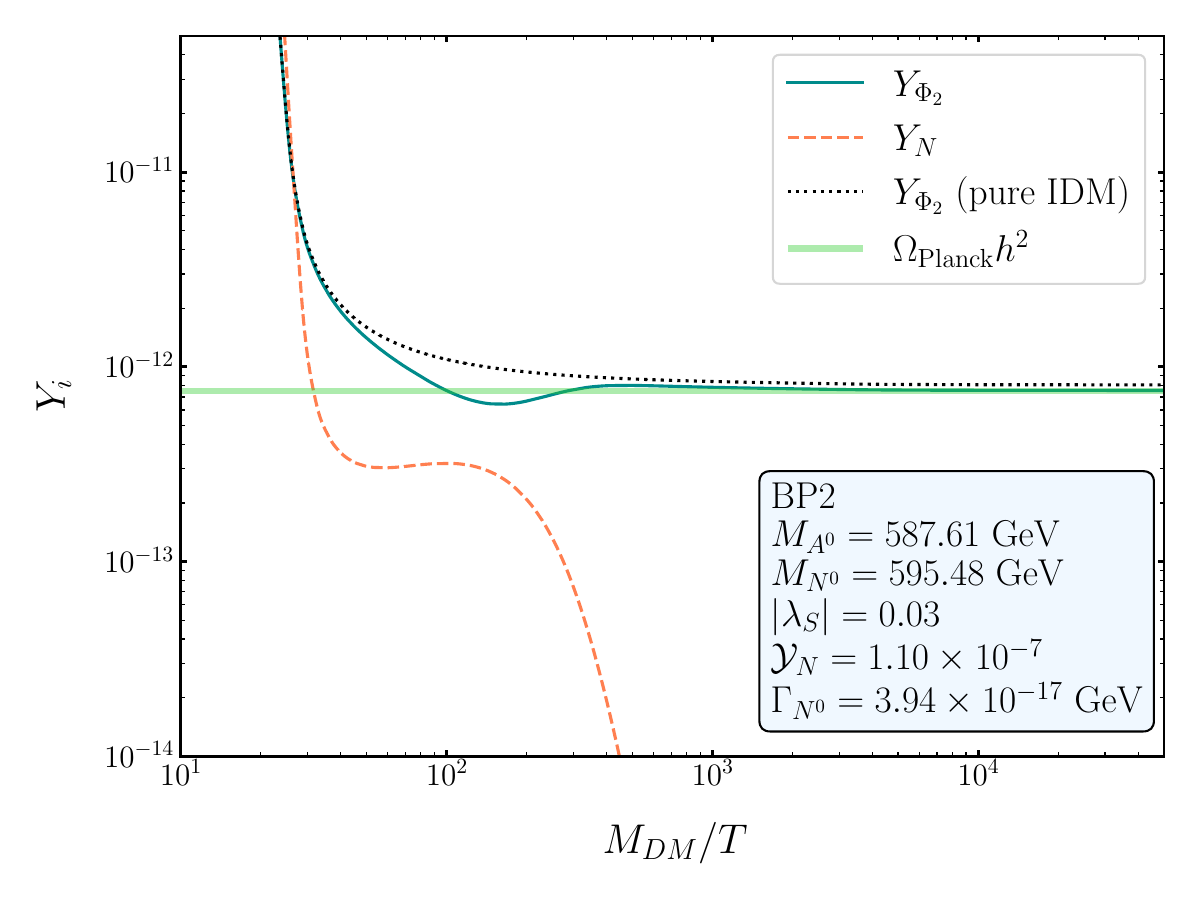}}
	\subfigure[]{\includegraphics[width=0.42\linewidth]{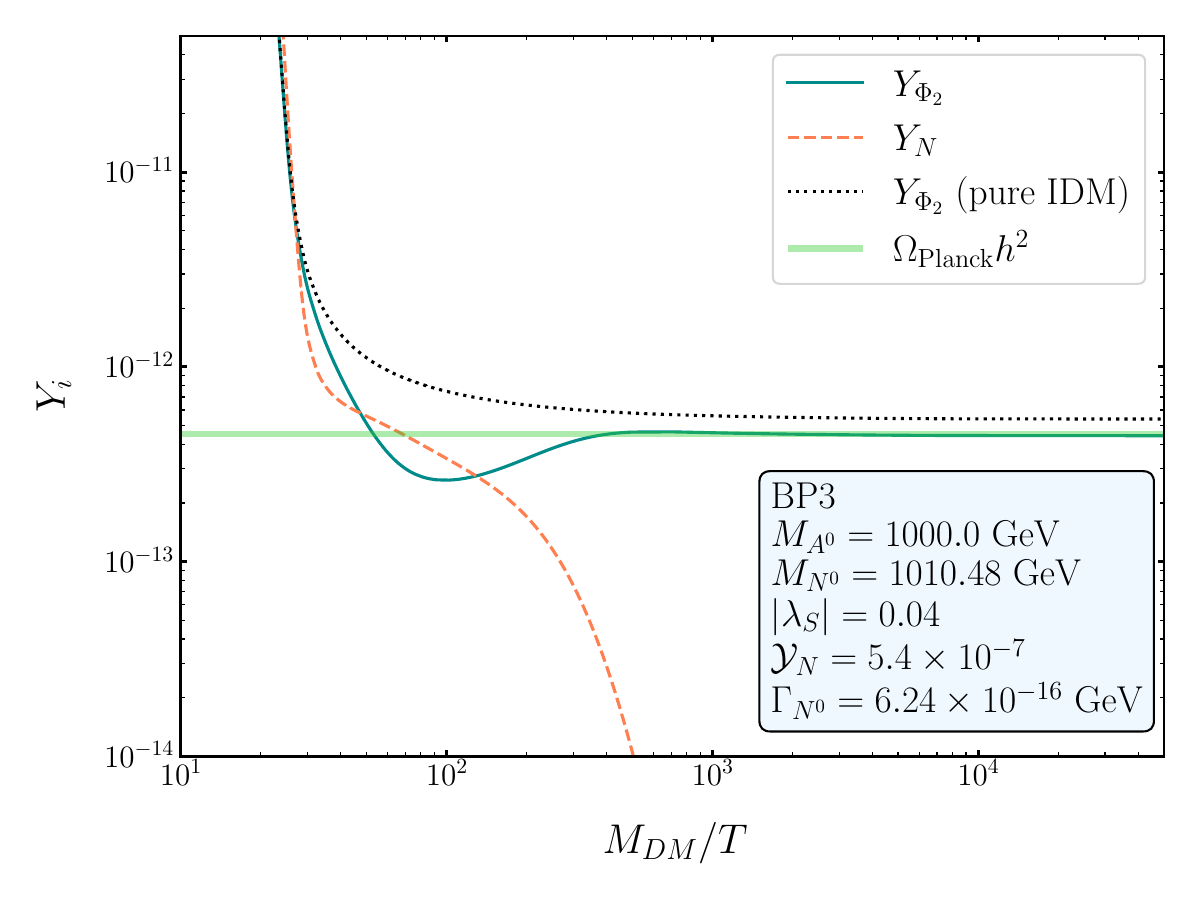}}
	\caption{The DM yields $Y_i$ as a function of $x = M_{DM}/T$ for (a) {\bf BP1}, (b) {\bf BP2}, (c) {\bf BP3}. The $Y_{\Phi_2}$ and $Y_{N}$ are shown with blue solid line and orange dashed line respectively. For {\bf BP2} and {\bf BP3}, the black dotted lines represent the yield for the same scalar masses for a pure IDM case, with no effect from the VLLs. The light green band represents the observed relic $\Omega_{\rm Planck}h^2 = 0.1199 \pm 0.0027$.}
	\label{fig:bpyield}
\end{figure}
Based on the mass difference between the scalar and fermionic dark sectors, as well as the Yukawa coupling values, the yields of the two neutral dark components show different features. The \autoref{fig:bpyield} describes the interplay for the three chosen benchmark points, with the yields of the two dark sectors, $Y_{\Phi_2}$ (blue solid line) and $Y_N$ (orange dashed line), being plotted as a function of $x = M_{DM}/T$ as per \autoref{eq:beq1} and \autoref{eq:beq2}.  As can be seen from \autoref{fig:bpyield}(a), in {\bf BP1} the decay effects of $N$ are almost negligible and $A^0$ decouples with higher yield compared to $N$. Even $N$ remains  in the equilibrium a little longer owing to relatively larger annihilation cross-section at lower mass  before  decoupling,  it decays at later time with decay width $\Gamma_{N}\sim 1.9\times 10^{-18}$  GeV. In case of {\bf BP2},  due to larger mass, $N$ now decouples with higher yield compared to {\bf BP1} and decays with $\Gamma_{N}\sim 3.9\times 10^{-17}$  GeV.  The $A^0$  yield sees a dip due to the enhancement of the co-annihilation attributed to the compressed   spectrum, which  later  gets  contribution from $N$ decay.  In {\bf BP3}, as the mass scale for both $N$ and $\Phi_2$ are $\mathcal{O}(1)$ TeV, the $N$ decouples relatively earlier compared to other BPs, due to reduction of the annihilation cross-section of $N$, with decay width $\Gamma_{N}\sim 6.2\times 10^{-16}$  GeV.  Whereas the compressed spectrum makes co-annihilation dominant, which causes a dip of the $A^0$ yield again. $A^0$ finally attains correct relic with some contribution coming from the $N$ decay. For {\bf BP2} and {\bf BP3} we also show the yields for a pure IDM case (with black dotted lines), where the fermionic sector has no effect on the scalar/pseudoscalar DM evolution. One can note how without the co-annihilation and decay of the VLL, a pure inert doublet DM shows overabundance for these two BPs. However, with the inclusion of the fermionic dark sector, the interplay can bring the scalar parameter points back to the region allowed by the Planck data. Thus, the late decay of  $N$ becomes  more important  as we get to higher DM masses in the observed relic regions.	

\begin{figure}[ht]
	\centering
	\subfigure[]{\includegraphics[width=0.42\linewidth]{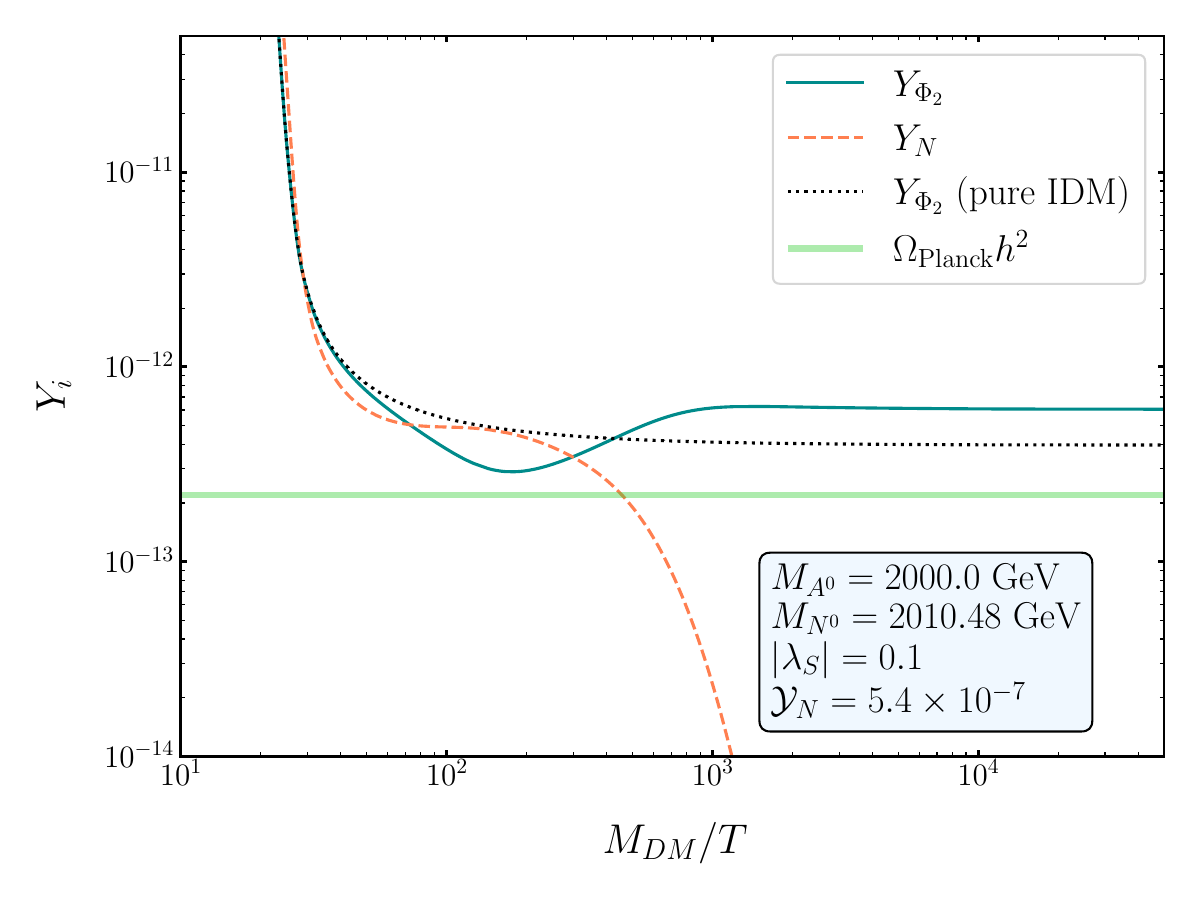}}
	\subfigure[]{\includegraphics[width=0.42\linewidth]{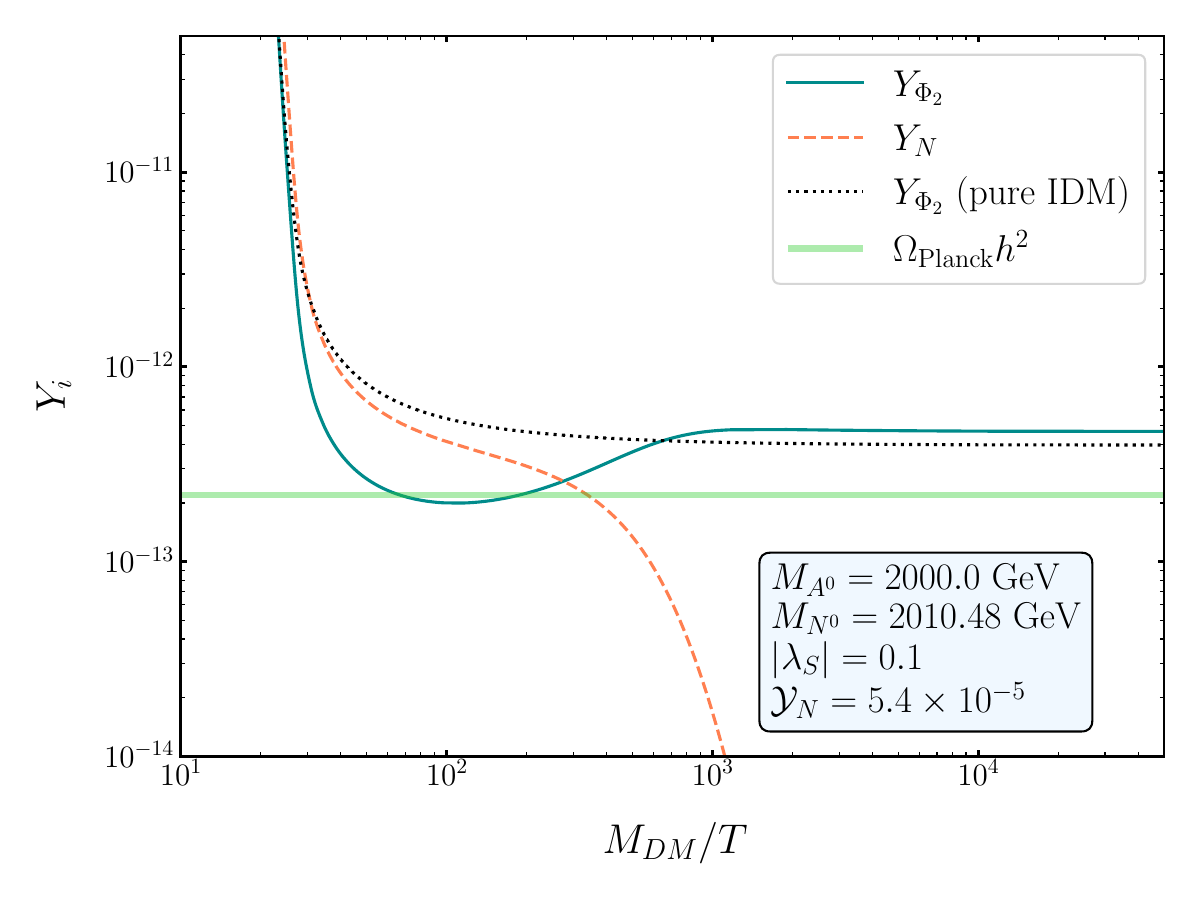}}
	\subfigure[]{\includegraphics[width=0.42\linewidth]{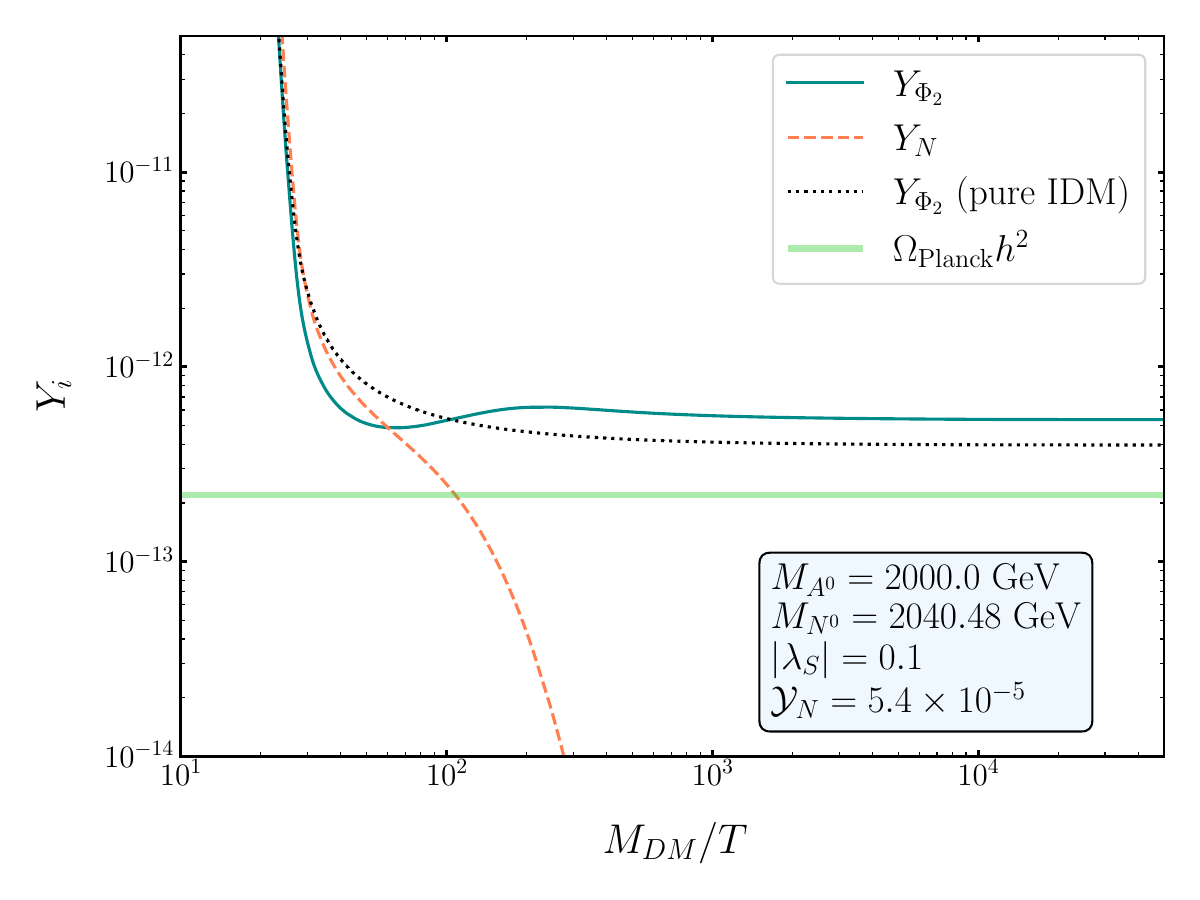}}
	\caption{The DM yields $Y_i$ as a function of $x = M_{DM}/T$ considering a value of $M_{A^0} = 2$ TeV, with (a) $M_{N^0} = 2010.5$ GeV and $\y_N = 5.4 \times 10^{-7}$, (b) $M_{N^0} = 2010.5$ GeV and $\y_N = 5.4 \times 10^{-5}$, (c) $M_{N^0} = 2040.5$ GeV and $\y_N = 5.4 \times 10^{-5}$.  The $Y_{\Phi_2}$ and $Y_{N}$ are shown with blue solid line and orange dashed line respectively. The black dotted lines represent the yield for the same scalar masses for a pure IDM case, with no effect from the VLLs. The light green band represents the observed relic $\Omega_{\rm Planck}h^2 = 0.1199 \pm 0.0027$.}
	\label{fig:2tev}
\end{figure}

It is evident from \autoref{fig:bpyield}(b)-(c) that for the BPs with higher masses, one can revitalize some parameter points ruled out for the pure IDM from the overabundance of the DM, due to the presence of the additional VLLs as we have presented. However, this sort of recovery of the parameter points becomes more and more improbable when  increasing the mass. To illustrate this with an example, we present three scenarios for a point with $M_{A^0}$ = 2.0 TeV, in \autoref{fig:2tev}. In \autoref{fig:2tev}(a), the mass gap $M_{N^0} - M_{A^0} \sim 10$ GeV case is shown, with a Yukawa coupling of $\y_N = 5.4\times 10^{-7}$. As one can see, the heavy mass of the dark sector does not allow enough annihilation, and the small Yukawa coupling suppresses the co-annihilation as well. Hence the number density $Y_{\Phi_2}$ (solid blue line) does not come down to the yield that corresponds to the correct relic (light green band), and the late decay of $Y_N$ (orange dashed line) enhances the yield even further, keeping it even more overabundant than in the pure IDM scenario in this case (black dotted line). \autoref{fig:2tev}(b) keeps the same mass gap as before, but shows the effect of the Yukawa coupling via increasing the value by two orders of magnitude i.e. with $\y_N = 5.4\times 10^{-5}$. This increases the co-annihilation, initially keeping the $Y_{\Phi_2}$ lower compared to \autoref{fig:2tev}(a) and crossing the light green band of correct relic. However, the decay effect inevitably kicks in and eventually results into overabundance. Lastly, in \autoref{fig:2tev}(c) we keep $\y_N = 5.4\times 10^{-5}$ and increase the mass gap to $M_{N^0} - M_{A^0} \sim 40$ GeV. Higher mass gap means two things: first, less amount of co-annihilation that keeps the $Y_{\Phi_2}$ line well above the required yield for the observed relic, and second, increased decay width that brings in the decay effect earlier compared to the previous case, resulting into an inevitable overabundance again. Hence, the lack of enough (co-)annihilation and the more pronounced decay effect combine to forbid the salvation of the ruled out pure IDM parameter points, for the higher mass range of $M_A^0 \geq 1.8$ TeV, as indicated originally in \autoref{fig:rel}.

\begin{figure}[ht]
	\centering
	\subfigure[]{\includegraphics[width=0.49\linewidth]{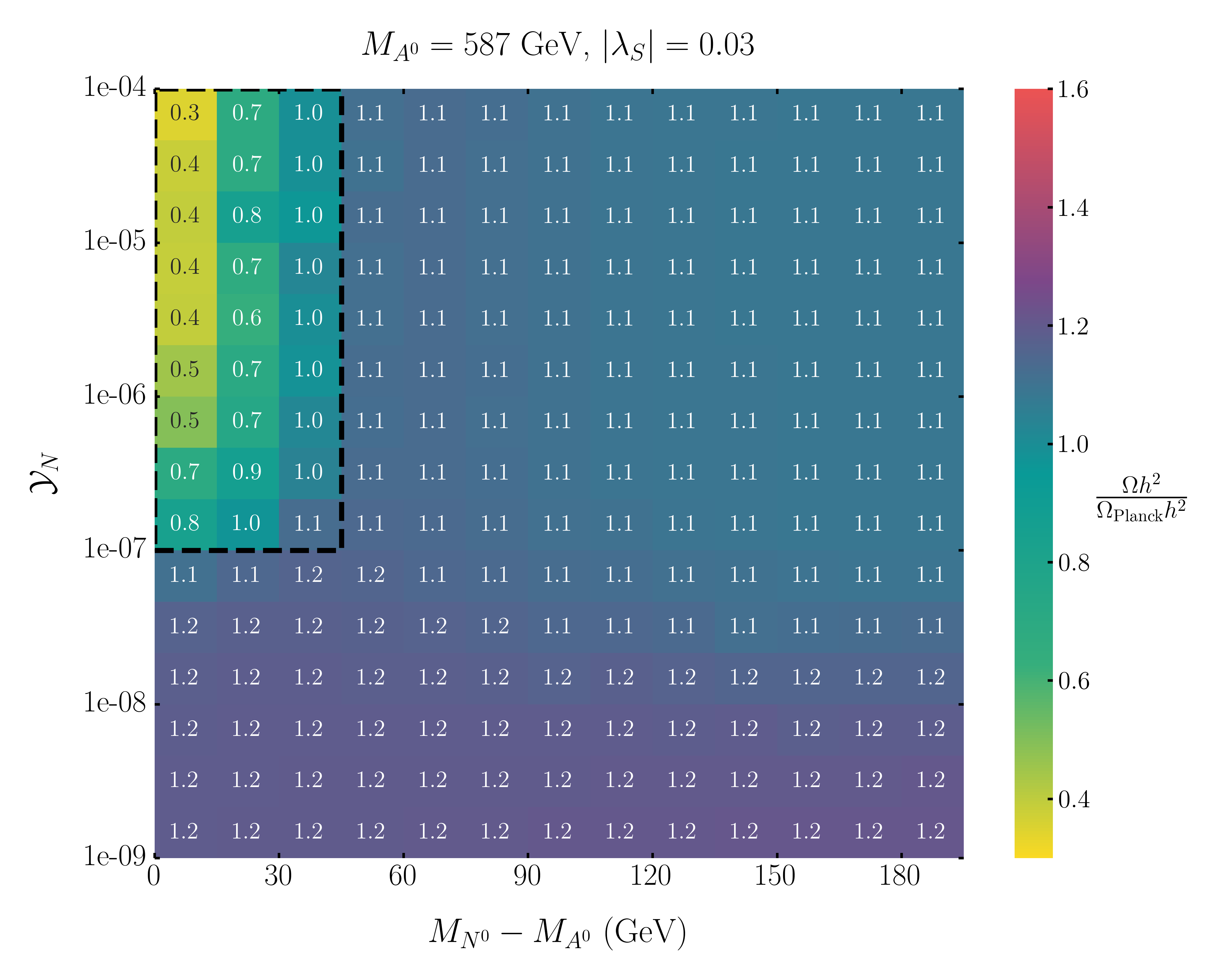}}
	\subfigure[]{\includegraphics[width=0.49\linewidth]{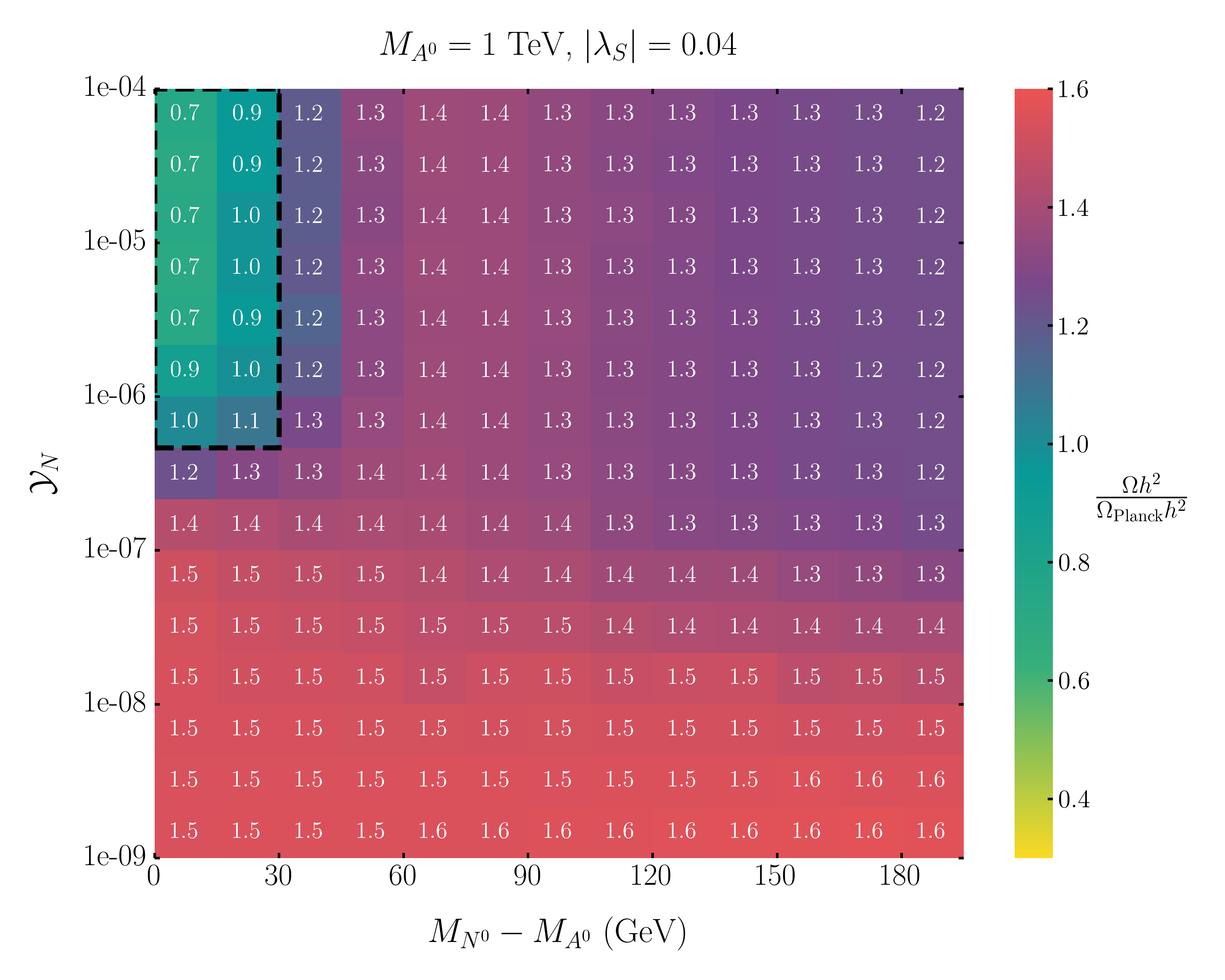}}
	\caption{Heatmap of pseudoscalar DM relic ratios $\frac{\Omega h^2}{\Omega_{\rm Planck} h^2}$, as a function of the mass gap $M_{N^0} - M_{A^0}$, and the Yukawa coupling $\y_N$, fixing the scalar sector mass values in (a) {\bf BP2}, (b) {\bf BP3}. The thick black dashed rectangles enclose the bins that can either satisfy the observed relic, or remain underabundant. Bins with $\frac{\Omega h^2}{\Omega_{\rm Planck} h^2} \geq 1.1$ are overabundant.}
	\label{fig:heatmap}
\end{figure}

It is worthwhile now to take a closer look at the interplay between the pseudoscalar DM species ($A^0$) and fermionic DM species ($N^0$), which is more prominent in the higher mass regions of {\bf BP2} and {\bf BP3}, as seen from \autoref{fig:bpyield}. Fixing the masses of the inert scalars, as well as the quartic couplings, we vary the mass gap $M_{N^0} - M_{A^0}$ between 1-200 GeV, while simultaneously scanning over a range of $\y_N \in [10^{-9},10^{-4}]$, and keep track of the relic density $\Omega h^2$ of each point. We present the ratios of these relic densities with the central value of the Planck 2018 data of observed relic i.e. $\Omega_{\rm Planck} h^2 = 0.1199$, in two-dimensional heatmap plots, as shown in \autoref{fig:heatmap}. 

The primary motivation of \autoref{fig:heatmap} is to study how the feasible space of mass splitting and Yukawa couplings evolve, when one moves up in the DM mass scale. For fixed values of $\lambda_{S}$ at any DM mass, the relic density is determined purely by the interplay, facilitated by this mass gap and $\y_N$. Therefore, the average ratio of $\frac{\Omega h^2}{\Omega_{\rm Planck} h^2}$ within appropriate bins of these two parameters provides a comprehensive picture of the parameter space evolution. \autoref{fig:heatmap}(a) shows the relic ratios $\frac{\Omega h^2}{\Omega_{\rm Planck} h^2}$ obtained from fixing the dark scalar masses as per {\bf BP2}, corresponding to variations in the mass gap $M_{N^0} - M_{A^0}$ (GeV) and the Yukawa coupling $\y_N$. The number on each bin of the plot represents the mean of the relic ratios obtained for the points scanned in the bin. A relic ratio = 1.0 corresponds to the possibility of obtaining the correct relic abundance $\Omega_{\rm Planck} h^2 = 0.1199 \pm 0.0027$, while relic ratios < 1.0 represents under-abundance of the DM. The bins with $\frac{\Omega h^2}{\Omega_{\rm Planck} h^2} \geq$ 1.1 are overabundant, and they are ruled out by the Planck data \cite{Planck:2018vyg}. Clearly, from \autoref{fig:heatmap}(a), the relic-satisfying or underabundant points for the pseudoscalar DM mass of 587 GeV lie in a mass gap range of $(M_{N^0} - M_{A^0}) \in [1-45]$ GeV, with a corresponding Yukawa coupling range of $\y_N \in [10^{-7}, 10^{-4}]$. A black rectangle with thick dashed borders is drawn on the heatmap to enclose this viable region of data points. In \autoref{fig:heatmap}(b), the same result is shown for the scalar mass values fixed as per {\bf BP3}, with a  pseudoscalar DM mass of 1 TeV. Here, the viable region of correct relic or underabundance, enclosed in the black dashed rectangle, has a significant shrink as compared to {\bf BP2}. For this TeV-scale scenario, we are confined to an allowed parameter range of  $(M_{N^0} - M_{A^0}) \in [1-30]$ GeV, and $\y_N \in [5\times10^{-7}, 10^{-4}]$, and beyond that we have an overabundance of DM relic density. In both plots, one sees a saturation in the overabundant regions when the mass gap increases beyond 100 GeV, for $\y_N \gsim 5\times 10^{-8}$ bins, and these regions correspond to the pure IDM yields as discussed in \autoref{fig:bpyield}. However, the highest overabundance region is observed for both cases when one has $\y_N \lsim 10^{-8}$, irrespective of the mass gap. This is accounted for by the fact that such small Yukawa couplings do not affect the co-annihilation as much, but they lead to very late out-of-equilibrium decay of the fermions, which enhances the number density of $A^0$.

This compressed spectrum and delayed decay effect reflect themselves onto the collider phenomenology of our model, leading to some unique and interesting signatures of displaced vertices. In the next section, we discuss the phenomenology of our model at the LHC/FCC-hh.

\section{Decays and production of VLL states}\label{dcprod}

After establishing the benchmark points allowed by all the theoretical and DM experimental bounds, and with a detailed understanding of the dynamics, we now move into their phenomenology at the colliders. The signatures of our model depend on how the VLL states can have displaced decays due to the tiny $\y_N$ values and the compressed mass spectra. Hence in the next subsection we present an overview of these decays.

\subsection{Decays of $N$}\label{dcy}

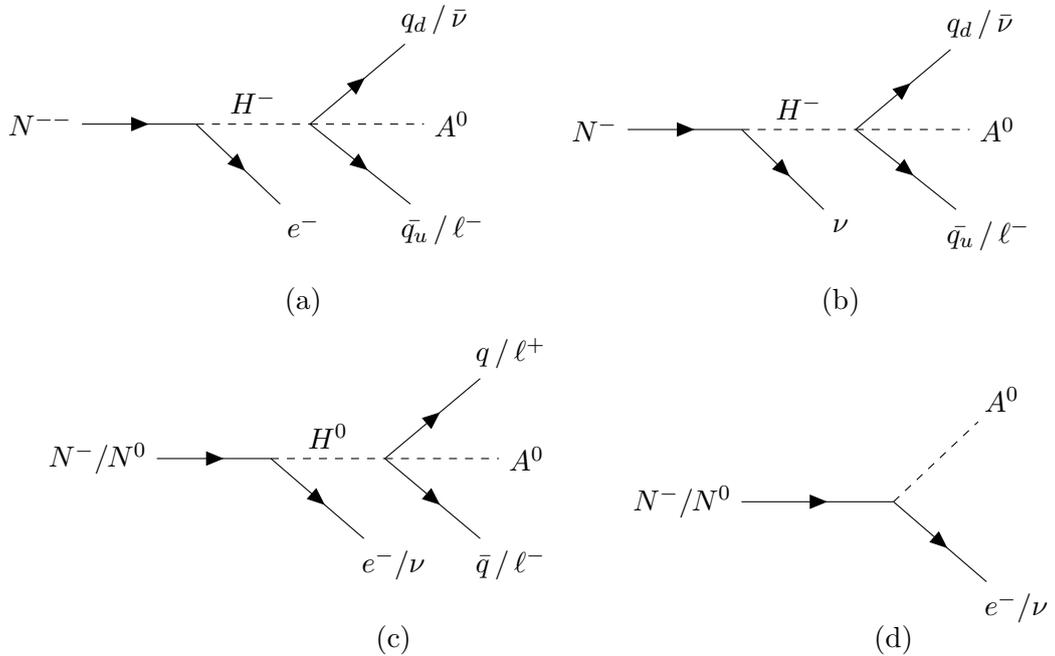
\begin{figure}[ht]
	\centering
	\begin{tikzpicture}
	\begin{feynman}
	\vertex (a1);
	\vertex [left=1.5cm of a1] (i0){$N^{--}$};
	\vertex [below right=1.5cm of a1] (o2){$e^-$};
	\vertex [right=1.5cm of a1] (i1);
	\vertex [above right=1.5cm of i1] (f2){$q_d\,/\, \bar{\nu}$};
	\vertex [right=1.5cm of i1] (i2){$A^0$};
	\vertex [below right=1.5cm of i1] (f1){$\bar{q_u}\,/\, \ell^-$};
	\vertex [below=1cm of o2] (d){(a)};
	\diagram {(i0)--[fermion](a1)--[scalar, edge label=$H^-$](i1), (i1)--[scalar](i2), (a1) --[fermion](o2), (i1)--[fermion](f1), (i1)--[fermion](f2)
	};
	\end{feynman}
	\end{tikzpicture}
	\quad \quad
	\begin{tikzpicture}
	\begin{feynman}
	\vertex (a1);
	\vertex [left=1.5cm of a1] (i0){$N^{-}$};
	\vertex [below right=1.5cm of a1] (o2){$\nu$};
	\vertex [right=1.5cm of a1] (i1);
	\vertex [above right=1.5cm of i1] (f2){$q_d\,/\, \bar{\nu}$};
	\vertex [right=1.5cm of i1] (i2){$A^0$};
	\vertex [below right=1.5cm of i1] (f1){$\bar{q_u}\,/\, \ell^-$};
	\vertex [below=1cm of o2] (d){(b)};
	\diagram {(i0)--[fermion](a1)--[scalar, edge label=$H^-$](i1), (i1)--[scalar](i2), (a1) --[fermion](o2), (i1)--[fermion](f1), (i1)--[fermion](f2)
	};
	\end{feynman}
	\end{tikzpicture}
	
	\quad \quad
	\begin{tikzpicture}
	\begin{feynman}
	\vertex (a1);
	\vertex [left=1.5cm of a1] (i0){$N^{-}/N^0$};
	\vertex [below right=1.5cm of a1] (o2){$e^-/\nu$};
	\vertex [right=1.5cm of a1] (i1);
	\vertex [above right=1.5cm of i1] (f2){$q\,/\, \ell^+$};
	\vertex [right=1.5cm of i1] (i2){$A^0$};
	\vertex [below right=1.5cm of i1] (f1){$\bar{q}\,/\, \ell^-$};
	\vertex [below=1cm of o2] (d){(c)};
	\diagram {(i0)--[fermion](a1)--[scalar, edge label=$H^0$](i1), (i1)--[scalar](i2), (a1) --[fermion](o2), (i1)--[fermion](f1), (i1)--[fermion](f2)
	};
	\end{feynman}
	\end{tikzpicture}
	\quad \quad
	\begin{tikzpicture}
	\begin{feynman}
	\vertex (a1);
	\vertex [left=2.0cm of a1] (i0){$N^{-}/N^{0}$};
	\vertex [below right=1.5cm of a1] (o2){$e^{-}/\nu$};
	\vertex [above right=1.5cm of a1] (i1){$A^0$};
	\vertex [below=1.5cm of a1] (d){(d)};
	\diagram {(i0)--[fermion](a1)--[scalar](i1), (a1) --[fermion](o2)
	};
	\end{feynman}
	\end{tikzpicture}
	\caption{Decay channels of the VLL states.}
	\label{fig:feynnmm}
\end{figure}

For the three benchmark points under  consideration, the masses of the VLL states are higher than those of the IDM scalars. This choice allows $N^{\pm\pm}, \, N^{\pm},$ and $N^0$ to decay into various combinations of $H^0,\, A^0,$ and $H^\pm$, alongside SM fermions. Before going into the discussion of individual decay widths and branching ratios, we present some possible decay chains of the VLL states in \autoref{fig:feynnmm}.

The total decay widths of these VLL states depend upon the values of $\y_N$ and the mass gaps between the two dark sectors. As discussed in \autoref{interplay}, the interplay between VLL and IDM needs lower Yukawa couplings $\mathcal{O}(10^{-7}-10^{-9})$, which also predict displaced decays for the triplet VLL states. This can appear as disappearing charged tracks, and/or displaced productions of leptons and jets. \autoref{tab:VLL_DcyW} portrays the two important pieces of information needed to understand the displaced decay of the VLL states, which are, the total decay width $\Gamma_{\text{tot}}$ (in GeV), and the rest mass decay length $c\tau_0$ (in m) \cite{Bandyopadhyay:2022mej,Bierlich:2022pfr}, respectively. The rest mass decay length can be calculated as $c\tau_0 = (c \hbar)/\Gamma_{\text{tot}}$, which explains the decrease in the decay length with the increase in the total width, as we go from {\bf BP1} to {\bf BP3}. For {\bf BP1}, the decay widths increase slightly as one proceeds from $N^{--}$ to $N^0$ down the rows, because of the drastically different masses of the decay products. Whereas, for {\bf BP2} and {\bf BP3}, the mass gaps remain $\sim 1$ GeV for the scalar decay products, and hence the total widths are governed by the masses of the VLLs only. This is reflected in the decrease in $\Gamma_{\text{tot}}$ as one goes down in mass from $N^{--}$ to $N^0$, for {\bf BP2} and {\bf BP3}. It is also noteworthy that, the smallest value of $\y_N$ in {\bf BP1} ensures the largest decay lengths, of $\mathcal{O}(100)$ meters. While the $\y_N$ are of the same order of magnitude in {\bf BP2} and {\bf BP3}, the lower masses of the VLLs and the smaller mass splitting between the fermionic and scalar dark sectors of {\bf BP2}, allow longer decay lengths of $\mathcal{O}(1)$ m. The VLLs in {\bf BP3} have to settle for the smallest decay lengths, of $\mathcal{O}(0.1)$ m, with the largest total decay widths out of the three BPs.


\begin{table}[ht]
	\centering
	\renewcommand{\arraystretch}{1.2}
	\begin{tabular}{|c|c|c|c|c|c|c|}
		\hline
		\multirow{4}{*}{\makecell{VLL states}} &   \multicolumn{6}{c|}{\makecell{Decay information}} \\
		\cline{2-7}
		& \multicolumn{2}{c|}{\makecell{{\bf BP1} \\ $\y_N = 4.2 \times 10^{-9}$}} & \multicolumn{2}{c|}{\makecell{{\bf BP2} \\ $\y_N = 1.1 \times 10^{-7}$}} & \multicolumn{2}{c|}{\makecell{{\bf BP3} \\ $\y_N = 5.4 \times 10^{-7}$}} \\
		\cline{2-7}
		& $\Gamma_{\text{tot}}$ (GeV) & $c \tau_0$ (m) & $\Gamma_{\text{tot}}$ (GeV) & $c \tau_0$ (m) & $\Gamma_{\text{tot}}$ (GeV) & $c \tau_0$ (m) \\
		\hline
		$N^{\pm \pm}$ & $1.27 \times 10^{-18}$ & 155.42 & $5.92 \times 10^{-17}$ & 3.33 & $1.34 \times 10^{-15}$ & 0.15 \\
		\hline
		$N^{\pm}$ & $1.55 \times 10^{-18}$ & 127.35 & $4.68 \times 10^{-17}$ & 4.22 & $9.09 \times 10^{-16}$ & 0.22 \\
		\hline
		$N^0$ & $1.88 \times 10^{-18}$ & 105.00 & $3.95 \times 10^{-17}$ & 4.99 & $6.24 \times 10^{-16}$ & 0.32 \\
		\hline 
	\end{tabular}
	\caption{Total decay widths ($\Gamma_{\text{tot}}$ in GeV) and rest mass decay lengths ($c \tau_0$ in m) of the VLL states for the three benchmark points. }
	\label{tab:VLL_DcyW}
\end{table}


\begin{figure}[ht]
	\centering
	\subfigure[]{\includegraphics[width=0.4\linewidth]{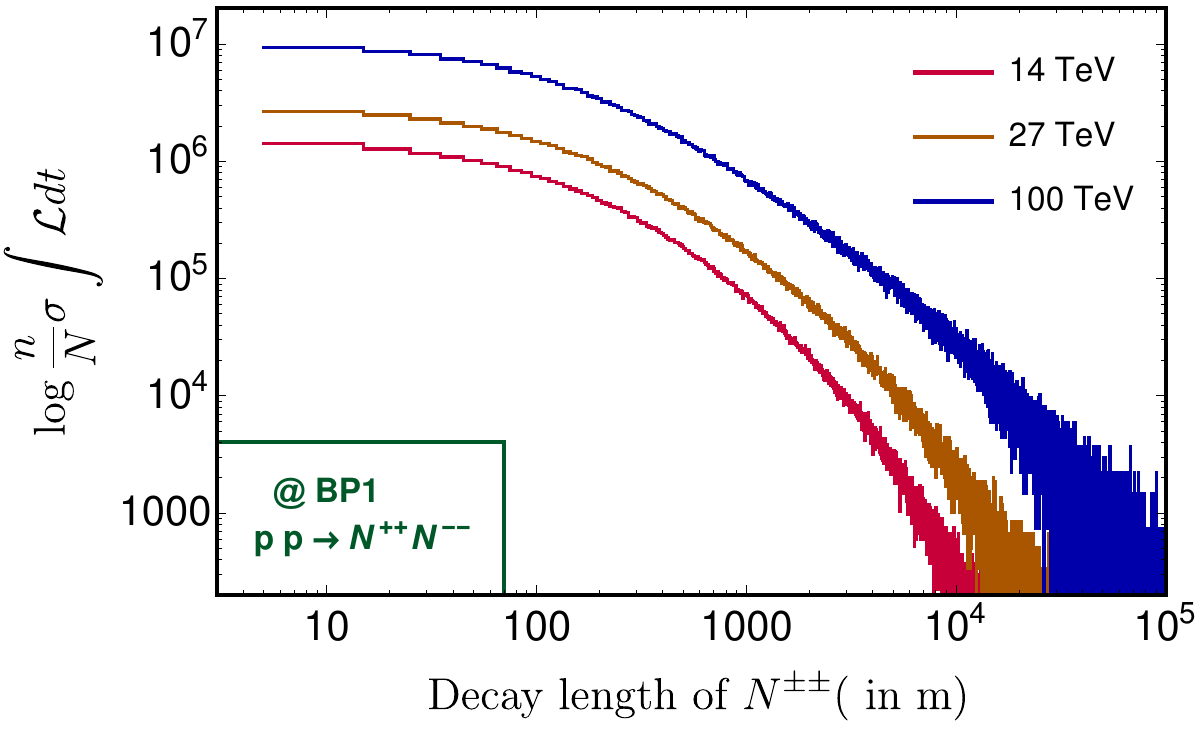}} \quad
	\subfigure[]{\includegraphics[width=0.4\linewidth]{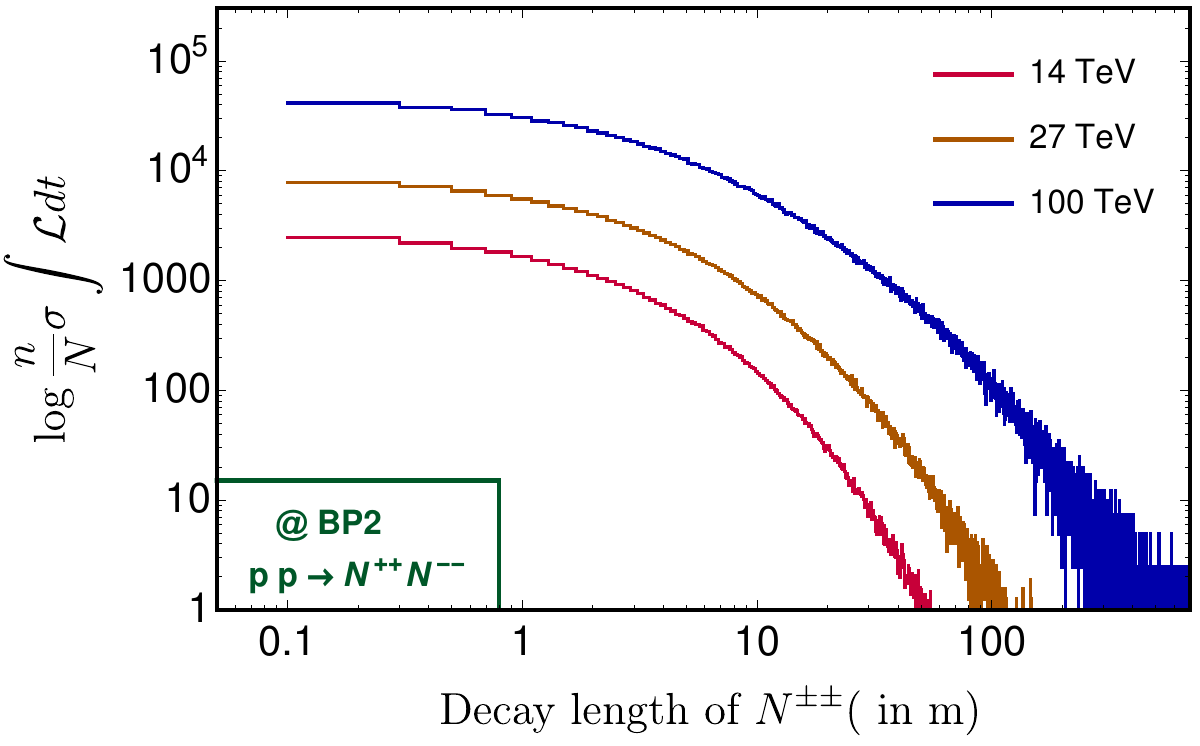}} \quad
	\subfigure[]{\includegraphics[width=0.4\linewidth]{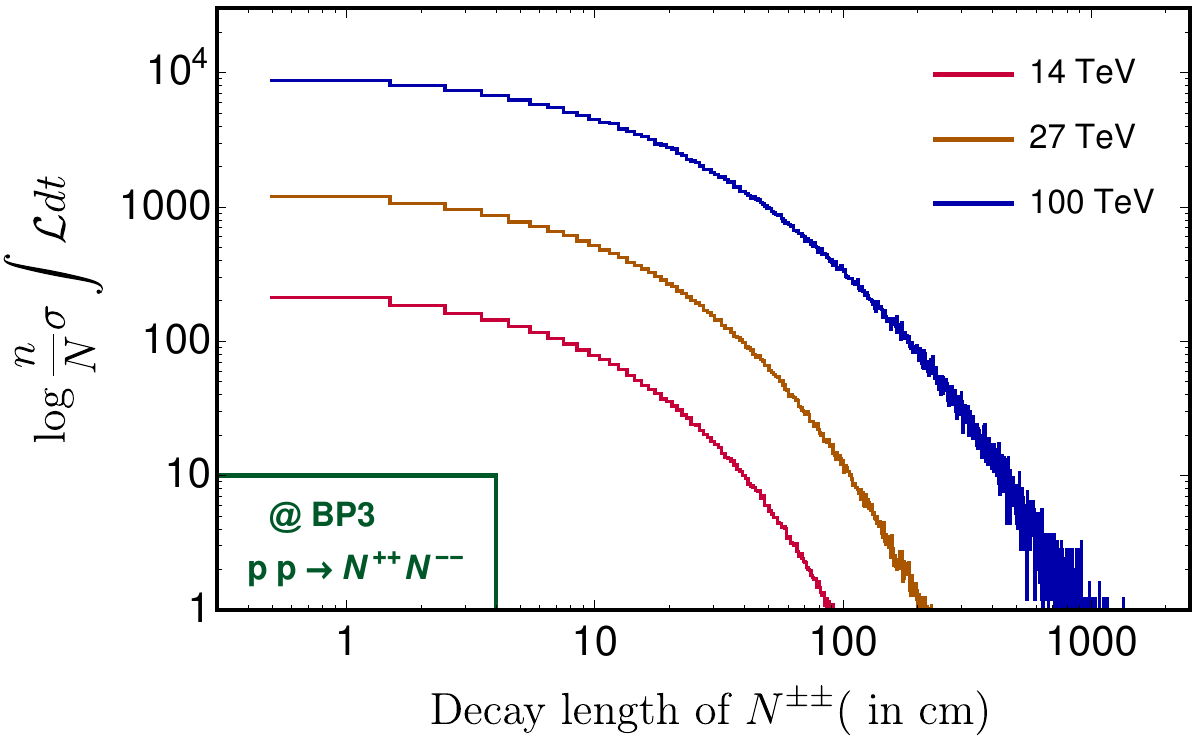}}
	\caption{Displaced decay length distribution of $N^{++}$, coming from the pair production at the LHC/FCC-hh with the centre-of-mass energies of 14, 27 and 100 TeV for (a) {\bf BP1}, (b) {\bf BP2}, and (c) {\bf BP3}. The number of events are normalized with the respective cross-sections and luminosity of 3000 \fbi.}
	\label{fig:dcylNpp}
\end{figure}


The decay lifetime of the produced particle in the lab frame, denoted as $\tau$, follows an exponential distribution. In addition, at a high energy collider, the produced particles also receive a significant amount of boost. Hence, the long-lived particles can travel farther than their rest mass decay length before finally decaying, with the ``decay length" calculated as $\beta \gamma c \tau$ \cite{Bandyopadhyay:2022mej,Bierlich:2022pfr}. \autoref{fig:dcylNpp} illustrates this, showing the distribution of decay lengths of the $N^{\pm\pm}$ from the production mode $p\,p \to N^{++}N^{--}$, at three different centre-of-mass energies of 14, 27, and 100 TeV, for the three BPs. The details of the production modes and the corresponding cross-sections are discussed later in \autoref{prod}. Here, the lowest values of mass allow a sizeable boost effect for {\bf BP1} as seen in \autoref{fig:dcylNpp}(a), where the decay lengths of $N^{\pm\pm}$ can reach up to around 10 km, 30 km, and 100 km at the 14 TeV (red), 27 TeV (dark orange), and 100 TeV (blue) colliders, respectively. {\bf BP2} and {\bf BP3} have significantly shorter decay lengths, with the heavier masses restricting the boost, while the larger $\y_N$ values keep the $c\tau_0$ values lower as well (as per \autoref{tab:VLL_DcyW}). For {\bf BP2}, the maximum obtainable decay lengths reach around 60 m, 150 m, and 700 m, for the three centre-of-mass energies of 14, 27 and 100 TeV, respectively. Even more suppression on the boost effect for the $\sim 1$ TeV mass of $N^{\pm\pm}$ in {\bf BP3} leads to a maximum decay length reach of nearly 90 cm, 110 cm, and 1500 cm, when produced at the colliders with 14, 27, and 100 TeV centre-of-mass energies, respectively.


\begin{table}[ht]
	\centering
	\renewcommand{\arraystretch}{1.4}
	\begin{tabular}{|c|c|c|c|c|}
		\hline
		\multirow{2}{*}{\makecell{BSM particles \\ (Masses for {\bf BP1}, {\bf BP2}, {\bf BP3} respectively in GeV)}} & \multirow{2}{*}{Decay modes} & \multicolumn{3}{c|}{\makecell{Branching ratios (in \%)}} \\
		\cline{3-5}
		& & {\bf BP1} & {\bf BP2} & {\bf BP3} \\
		\hline
		\makecell{$N^{--}$\\ (99.28, 596.8, 1011.9)} & $H^- e^-$ & 100\% & 100\% & 100\%  \\
		\hline
		\multirow{3}{*}{\makecell{$N^{-}$ \\ (98.61, 595.9, 1011.0)}} & $H^- \nu_e$ & 37.9\% & 51.8\% & 62.3\%  \\
		& $A^0 e^-$ & 62.1\% & 29.8\% & 37.7\%  \\
		& $H^0 e^-$ & -- & 18.4\% & --  \\
		\hline
		\multirow{2}{*}{\makecell{$N^0$ \\ (98.25, 595.5, 1010.6)}} & $A^0 \nu_e$ & 100\% & 62.7\% & 100\%  \\
		& $H^0 \nu_e$ & -- & 37.3\% & --  \\
		\hline 
		\multirow{2}{*}{\makecell{$H^+$ \\ (84.76, 588.2, 1001.0)}} & $A^0 q_1 \bar{q_2}$ &  67.2\% & 61.8\% & 63.0\%  \\
		& $A^0 l^+ \nu_{l}$ & 32.8\% & 38.2\% & 37.0\%  \\
		\hline 
		\multirow{5}{*}{\makecell{$H^0$ \\ (117.16, 589.4, 1010.5)}} & $A^0 q_1 \bar{q_1}$ & 45.5\% & 47.6\% & 25.3\%  \\
		& $A^0 \nu_{l} \bar{\nu_{l}}$ & 13.2\% & 21.0\% &  8.5\% \\
		& $A^0 l^+ l^-$ & 6.7\% & 7.1\% & 4.1\%  \\
		& $H^{\pm} q_1 \bar{q_2}$ & 23.0\% & 16.0\% &  42.0\% \\
		& $H^{\pm} l^{\mp} \overset{(-)}{\nu_l}$ & 11.6\% & 8.3\% &  20.1\% \\
		\hline 
		\makecell{$A^0$ \\ (71.57, 587.6, 1000.0)} & -- &--&--&--\\
		\hline
	\end{tabular}
	\caption{Decay modes and branching ratios (in \%) of the BSM particles in our model, for the three BPs. The masses of each particle (in GeV) for {\bf BP1}, {\bf BP2}, {\bf BP3} are written within parentheses in that order. In the table, $l \in \{e, \mu, \tau \}$.}
	\label{tab:branching}
\end{table}

\autoref{tab:branching} details the different decay channels and their respective branching ratios (BR) in percentage, for each of the  particles from VLL and IDM sectors, for the three benchmark points. The masses of each new particle for each BP is also written underneath the particles, within parentheses. Starting off with $N^{--}$, the only possible decay channel is $H^- e^-$, which takes places with 100 \% probability in each BP. This is not the case for $N^-$, where each BP shows different branching ratios. Following the partial decay width expressions as mentioned in \autoref{eq:npdec} and \autoref{eq:npdec2}, one can explain that for {\bf BP1}, the sub-100 GeV masses of $N^-, H^-, A^0$ and the fact that $H^-$ is heavier than $A^0$ by $\sim 14$ GeV, lead to $N^-$ decaying into $A^0 e^-$ with nearly twice the probability compared to $H^- \nu_e$. This behaviour is flipped in {\bf BP3}, where the tiny $\sim 1$ GeV mass splitting of $H^-$ and $A^0$ allows the additional factor of two in \autoref{eq:npdec2} to dominate, leading to almost twice the probability for $N^- \to H^- \nu_e$ than the pseudoscalar mode. In {\bf BP2}, the factor of two dominates again, and $H^0$ being lighter than $N^-$ opens up the decay channel of $N^- \to H^0 e^-$, albeit with lesser probability of $\sim 18\%$ compared to its pseudoscalar counterpart with lower mass. Coming to the decay of $N^0$, {\bf BP1} and {\bf BP3} kinematically allows only the $N^0 \to A^0 \nu_e$ decay channel, whereas in {\bf BP2}, a $\sim37\%$ probability is obtained for the $N^0 \to H^0 \nu_e$ mode as well.
	
In the inert scalar sector, the $Z_2$ symmetry and the heavier fermionic sector only allows three-body decays via an off-shell $W^\pm$ or $Z$-boson exchange, and the percentages of such decays become more nuanced. Taking the case of $H^+ \to A^0 W^{+*}$, we witness how the mass splitting between $H^+$ and $A^0$ governs the probabilities. While for each BP the hadronic decay modes inevitably dominate, the percentage is highest for {\bf BP1} ($\sim 67 \%$), where the large mass splitting of $\sim 14$ GeV allows more phase space for the $ H^+ \to A^0 q_1 \bar{q_2}$ decay. The $\lsim1$ GeV splitting in the other two BPs suppress this mode, and hence the branching ratios are lowered by $\sim 5\%$ in these cases, enhancing the leptonic modes accordingly. The most peculiar decays are witnessed in the case of $H^0$, as it being heavier than both $A^0$ and $H^+$ allows both the $H^0 \to A^0 Z^*$ and $H^0 \to H^\pm W^{\mp*}$ modes. Here, the available phase space is determined by the interplay of two factors: 1) the combined mass of the scalar/pseudoscalar decay product plus the off-shell vector boson counterpart, and 2) the mass splitting between $H^0$ and its two possible inert daughters. In {\bf BP1}, the large mass splittings of $M_{H^0} - M_{A^0} \approx 45$ GeV and $M_{H^0} - M_{H^+} \approx 32$ GeV cause the $H^0 \to A^0 Z^*$ mode to have more phase space, leading to more decay probability compared to $H^0 \to H^\pm W^{\mp*}$. The hadronic and leptonic branching ratios are then subsequently determined by the decay probabilities of the off-shell vector bosons. In {\bf BP2}, the mass spectrum is the most compressed, and hence the slightly larger splitting of $M_{H^0} - M_{A^0} \sim 2$ GeV is favoured by the phase space. The hadronic decay modes ensuing from the $H^0 \to H^\pm W^{\mp*}$ are even more suppressed, with $M_{H^0} - M_{H^+} \sim 1$ GeV. For {\bf BP3}, the behaviour is reversed, as the mass gaps $M_{H^0} - M_{A^0} \approx 10$ GeV and $M_{H^0} - M_{H^+} \approx 9$ GeV are large and almost indistinguishable. Hence, the lower mass of the $W^\pm$-boson ensures the availability of more phase space, with the hadronic and leptonic branching fractions following accordingly. Finally, for the sake of completeness we also mention the DM candidate $A^0$, which does not decay.


\subsection{Production of $N$ and possible final sates  at the LHC/FCC-hh}\label{prod}

\begin{figure}[ht]
	\centering
	\hspace*{-0.6cm}
	\begin{tikzpicture}
	\begin{feynman}
	\vertex (a2);
	\vertex [above left=1.0cm of a2] (a0){\small $q$};
	\vertex [below left=1.0cm of a2] (a1){\small $\bar{q}$};
	\vertex [right=1.0cm of a2] (f1);
	\vertex [above right=1.0cm of f1] (f2){\small $N^{++}$};
	\vertex [below right=1.0cm of f1] (f3){\small $N^{--}$};
	
	\diagram {(a2)--[solid](a1), (a2)--[solid](a0), (a2)--[boson, edge label=\small $Z/ \gamma$](f1), (f1)--[](f2), (f1)--[](f3)
	};
	\end{feynman}
	\end{tikzpicture}
	\quad
	\centering
	\begin{tikzpicture}
	\begin{feynman}
	\vertex (a2);
	\vertex [above left=1.0cm of a2] (a0){\small $q$};
	\vertex [below left=1.0cm of a2] (a1){\small $\bar{q}$};
	\vertex [right=1.0cm of a2] (f1);
	\vertex [above right=1.0cm of f1] (f2){\small $N^{+}$};
	\vertex [below right=1.0cm of f1] (f3){\small $N^{-}$};
	
	\diagram {(a2)--[solid](a1), (a2)--[solid](a0), (a2)--[boson, edge label=\small $Z/ \gamma$](f1), (f1)--[](f2), (f1)--[](f3)
	};
	\end{feynman}
	\end{tikzpicture}
	\quad
	\centering
	\begin{tikzpicture}
	\begin{feynman}
	\vertex (a2);
	\vertex [above left=1.0cm of a2] (a0){\small $q$};
	\vertex [below left=1.0cm of a2] (a1){\small $\bar{q}$};
	\vertex [right=1.0cm of a2] (f1);
	\vertex [above right=1.0cm of f1] (f2){\small $N^{\pm \pm}$};
	\vertex [below right=1.0cm of f1] (f3){\small $N^{\mp}$};
	
	\diagram {(a2)--[solid](a1), (a2)--[solid](a0), (a2)--[boson, edge label=\small $W^{\pm}$](f1), (f1)--[](f2), (f1)--[](f3)
	};
	\end{feynman}
	\end{tikzpicture}
	\quad
	\centering
	\begin{tikzpicture}
	\begin{feynman}
	\vertex (a2);
	\vertex [above left=1.0cm of a2] (a0){\small $q$};
	\vertex [below left=1.0cm of a2] (a1){\small $\bar{q}$};
	\vertex [right=1.0cm of a2] (f1);
	\vertex [above right=1.0cm of f1] (f2){\small $N^{\pm}$};
	\vertex [below right=1.0cm of f1] (f3){\small $N^{0}$};
	
	\diagram {(a2)--[solid](a1), (a2)--[solid](a0), (a2)--[boson, edge label=\small $W^{\pm}$](f1), (f1)--[](f2), (f1)--[](f3)
	};
	\end{feynman}
	\end{tikzpicture}
	\caption{Various production modes of vector-like $SU(2)$ triplet fermions at the LHC.}
	\label{fig:feynProd}
\end{figure}
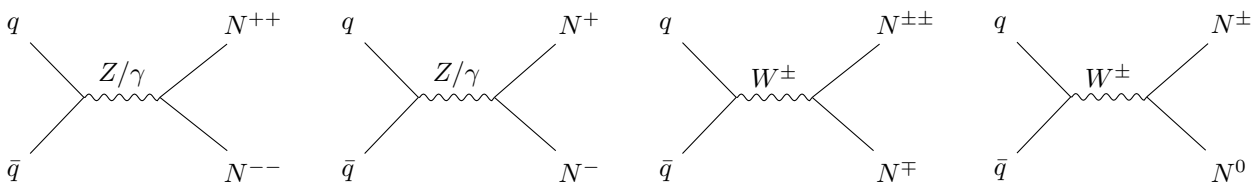

After explaining the various decay modes and the possibility of displaced vertices, we now wish to study how these VLL states can be produced at the LHC/FCC-hh, and how they can lead to salient final states. Here, we  mainly focus on probing  the doubly charged $N^{\pm\pm}$ along with   $N^\pm, \, N^0$ via their displaced  decays at  the LHC/FCC-hh. The compressed mass spectra of the dark sector  and the tiny values of the Yukawa couplings ensure the presence of displaced leptons for the benchmark points, which can be further exploited to establish novel signatures for  such exotic new physics. \autoref{fig:feynProd} depicts  the  relevant Feynman diagrams for the pair and  associated productions of  the VLL particles at the LHC/FCC-hh and in \autoref{tab:cs}  we present these  cross-sections at three different centre-of-mass energies ($E_{CM}$) of 14, 27 and 100 TeV, respectively. As one can see from \autoref{tab:cs}, the two associated production modes of $p\,p\to N^{\pm \pm}N^{\mp}$ and $p\,p\to N^{\pm}N^{0}$ have the highest cross-sections, while $p\,p\to N^{+}N^{-}$ has the lowest, by around one order of magnitude.
Since we are solely interested in the displaced leptonic final states, we will not study the  production  modes of the IDM scalars, signatures of which contain prompt decay products, as discussed in refs. \cite{Jangid:2020qgo, Datta:2016nfz, Belyaev:2016lok,Poulose:2016lvz}.


\begin{table}[ht]
	\centering
	\renewcommand{\arraystretch}{1.2}
	\begin{tabular}{|c|c|c|c|c|}
		\hline
		\multirow{2}{*}{Production modes} & \multirow{2}{*}{\makecell{Benchmark \\ Point}} & \multicolumn{3}{c|}{Cross-section at different $E_{CM}$ (in fb)}\\
		\cline{3-5}
		& & \hspace*{0.5cm} 14 TeV \hspace*{0.5cm} & \hspace*{0.5cm} 27 TeV \hspace*{0.5cm} & \hspace*{0.5cm} 100 TeV \hspace*{0.5cm} \\
		\hline
		\multirow{3}{*}{$p\,p\to N^{++}N^{--}$}& {\bf BP1} & 12337.0 & 27913.0 & 122120.0 \\
		\cline{2-5}
		& {\bf BP2} & 12.7 & 52.0 & 408.0 \\
		\cline{2-5}
		& {\bf BP3} & 0.8 & 5.8 & 66.6 \\
		\hline \hline 
		\multirow{3}{*}{$p\,p\to N^{+}N^{-}$}& {\bf BP1} & 1019.5 & 2283.7 & 9815.4 \\
		\cline{2-5}
		& {\bf BP2} & 1.1 & 4.6 & 35.2 \\
		\cline{2-5}
		& {\bf BP3} & 0.08 & 0.5 & 5.7 \\
		\hline \hline 
		\multirow{3}{*}{$p\,p\to N^{\pm \pm}N^{\mp}$}& {\bf BP1} & 23380.0 & 52306.0 & 224810.0 \\
		\cline{2-5}
		& {\bf BP2} & 23.7 & 93.2  & 710.1 \\
		\cline{2-5}
		& {\bf BP3} & 1.6 & 10.6 & 115.5  \\
		\hline \hline 
		\multirow{3}{*}{$p\,p\to N^{\pm}N^{0}$}& {\bf BP1} & 23826.0 & 53263.0 & 228670.0 \\
		\cline{2-5}
		& {\bf BP2} & 23.8 & 93.6 & 713.3 \\
		\cline{2-5}
		& {\bf BP3} & 1.6 & 10.7 & 115.8 \\
		\hline
	\end{tabular}
	\caption{Pair production cross-sections of different combinations of the triplet vector-like fermions at three different centre-of-mass energies of the LHC/FCC-hh, for the three BPs.}
	\label{tab:cs}
\end{table}

Correlating with the possible decay channels as discussed in \autoref{dcy}, we provide a detailed overview on each of the production modes shown in \autoref{tab:cs} and their subsequent final states, as itemized below.

\begin{enumerate}
	\item \underline{$p\,p\to N^{++}N^{--}$}: 
Due to the doubly charged nature, the pair-production  cross-section for $N^{\pm\pm}$ is relatively  higher  as compared to the singly  charged  pair, and  this process is mediated by $Z/\gamma$ as can be seen from the relevant vertices given in  \autoref{Ndpe}:
\bea\label{Ndpe}
g_{N^{++} N^{--} \gamma} &=& g_1 c_W + g_2 s_W, \quad g_{N^{++} N^{--} Z} = g_2 c_W - g_1 s_W,\\\nn
g_{N^+ N^- \gamma} &=& g_1 c_W,\quad g_{N^+ N^- Z} = - g_1 s_W, 
\eea
where $c_W = \cos{\theta_W}$, $s_W = \sin{\theta_W}$ and $\theta_W$ is the Weinberg angle. 
For this production mode, we can have a maximum of 4 displaced leptons as shown from the decay chain in \autoref{fig:feynnmm}(a). While in principle we can also have displaced jets, the chances of the jets getting clustered is very low, predictably owing to the low momenta of the hadronic decay products from the off-shell $W$-boson. The leptons are also expected to be soft due to the compressed spectrum, however, in a hadronically complicated environment like the LHC/FCC-hh, leptons are considered cleaner signatures than jets for most of the studies. Hence, based on the probability of the leptons getting tagged, we aim to study the following possible  final states from this production process, which will be modified with further requirements and cuts.
	
	\begin{center}
	$\bullet$ 1 displaced lepton + 0 jets + $\ptmiss$ \\
	$\bullet$  2 displaced leptons + 0 jets + $\ptmiss$ \\
	$\bullet$  3 displaced leptons + 0 jets + $\ptmiss$ \\
	$\bullet$  4 displaced leptons + 0 jets + $\ptmiss$\\
	\end{center}
Here $\ptmiss$ is the missing transverse momentum coming from the neutrino and the DM, which is detailed in the next section. 	

\item \underline{$p\,p\to N^{+}N^{-}$}: Due to the singly charged pair, the production cross-section is much smaller  than for the doubly charged pair owing to the factors in the couplings as mentioned in \autoref{Ndpe}.  This  process  also is mediated by $Z/\gamma$.	Following \autoref{fig:feynnmm}(c), a maximum of six displaced leptons are ideally obtainable in this process. Hence, along with the four previously mentioned final states, one can have:

\begin{center}
		$\bullet$ 6 displaced leptons + 0 jets + $\ptmiss$ \\
\end{center}

\item \underline{$p\,p\to N^{\pm \pm}N^{\mp}$}: This mode is mediated by  the $W^\pm$ boson as shown in \autoref{fig:feynProd}.  Due to mass degeneracy the  cross-section  is very  similar to $N^{\pm }N^{0}$ as can be read from \autoref{tab:cs}. This production process, depending on the mass hierarchy, can also lead to the first four final states. However, a decay chain of the type shown in \autoref{fig:feynnmm}(c) can ideally lead to a final state with 5 displaced leptons, albeit with very low probability.

\begin{center}
	$\bullet$ 5 displaced leptons + 0 jets + $\ptmiss$ \\
\end{center}

\item \underline{$p\,p\to N^{\pm}N^{0}$}: When $N^0$ decays to a fully invisible final state of $A^0 \nu_e$, this process can lead to a maximum of three displaced leptons. However, $N^0 \to H^0 \nu_e$ can lead to two additional displaced leptons. Hence, ideally this process can contribute to the final states with one to five displaced leptons.

\end{enumerate}
\section{Collider Simulation at the LHC/FCC-hh}\label{csim}
To simulate the phenomenology of our model at the LHC/FCC-hh, we obtain the input files for \chep \cite{Belyaev:2012qa} from the model files written in \sarah \cite{Staub:2013tta,Staub:2015kfa}, and use them to generate events at the leading order for the four production processes mentioned in \autoref{tab:cs}, using the parton distribution function (PDF) of \texttt{NNPDF30\_lo\_as\_0130\_QED} \cite{NNPDF:2014otw}. The event \texttt{lhe} files are then showered with \py \cite{Sjostrand:2014zea}, and the subsequent jets are clustered with \fj \cite{Cacciari:2011ma} using the \texttt{anti-KT} algorithm \cite{Cacciari:2008gp}, keeping the radius parameter $R = 0.4$. It is important to note that, the generation-level cuts on the kinematic variables and calorimeter coverage are different for the 14/27 TeV LHC compared to the 100 TeV FCC-hh machine, governed by the existing plans for CMS/ATLAS \cite{CMS:2007sch,ATLAS:1999vwa}, as well as the proposed detector upgrade of FCC-hh \cite{FCC:2018vvp}. The choices are presented in detail as follows:

\begin{itemize}
	\item The calorimeter coverage is taken as $\abs{\eta} \leq 4.5 (6)$ for the 14, 27 (100) TeV collider energies.
	
	\item The minimum transverse momentum of the jets are taken to be $p_{T, min}^{jet} = 20$ GeV for all three energies, and are ordered in $p_T$.
	
	\item The maximum pseudorapidity for identifying the charged leptons is taken as $\abs{\eta_{max}^{\ell^\pm}} = 2.5 (4.8)$ for the centre-of-mass energies of 14,27 (100) TeV, where $\ell^\pm$ stands for $e^\pm, \mu^\pm$. 
	
	\item Owing to the compressed spectrum and the probable emergence of majority soft leptons, we collect the leptons with $p_T^{\ell^\pm} \geq 5 (3)$ GeV, for the  14, 27 (100) TeV collider energies.

	\item The leptons are hadronically cleaned, demanding the total hadronic transverse momenta,  within a cone of $\Delta R = 0.3$ with the leptonic axis, to be $\leq 0.15\, p_T^{\ell^\pm}$ .
	
	\item The charged leptons are isolated from the jets with a cone radius of $\Delta R_{\ell j} \geq 0.2$, and the same value of $\Delta R_{\ell \ell} \geq 0.2$ is used to isolate the leptons from each-other. 
	
	\item Since we are interested to probe the displaced vertex signature of our model, we demand the production vertex of the isolated leptons to be at least 1\,mm away from the interaction point \cite{CMS:2021kdm, ATLAS:2020wjh}.
\end{itemize}

Additionally, the missing transverse momentum $\ptmiss$ is calculated as the vector sum of the visible transverse momenta i.e. those of the isolated leptons and clustered jets, to account for the missing energy carried by the undetectable $A^0$, and the neutrinos.

\subsection{Kinematic distributions}\label{kinem}
Below we discuss the  different kinematical distributions before the detailed simulation at the LHC/FCC-hh for the benchmark points. 

\begin{itemize}
	\item {\bf Multiplicity of displaced leptons:}

\begin{figure}[ht]
	\centering
	\subfigure[]{\includegraphics[width=0.48\linewidth]{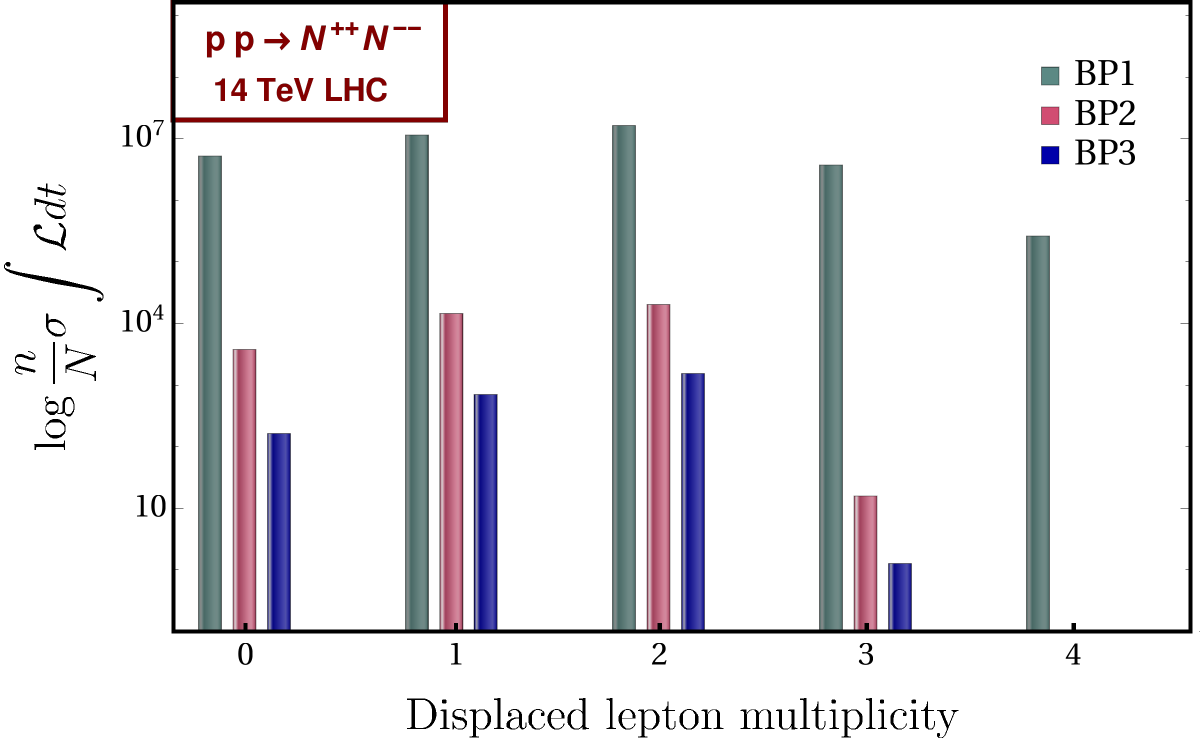}} \quad
	\subfigure[]{\includegraphics[width=0.48\linewidth]{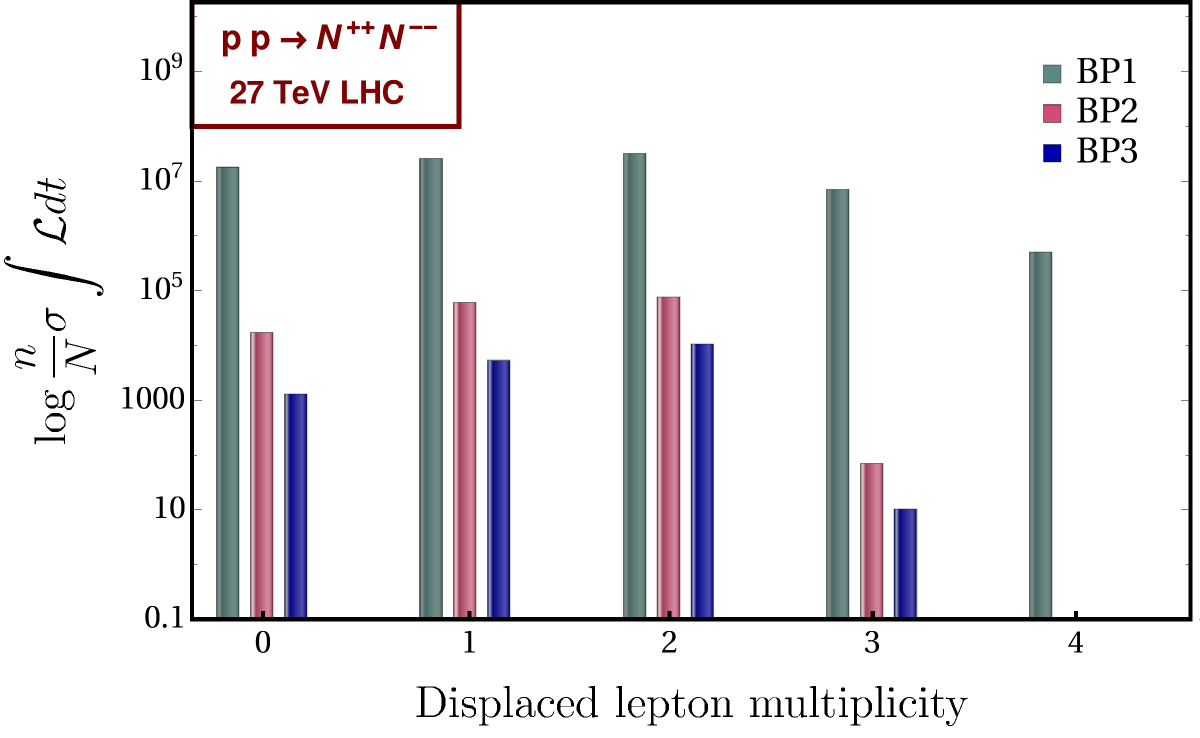}} \quad
	\subfigure[]{\includegraphics[width=0.48\linewidth]{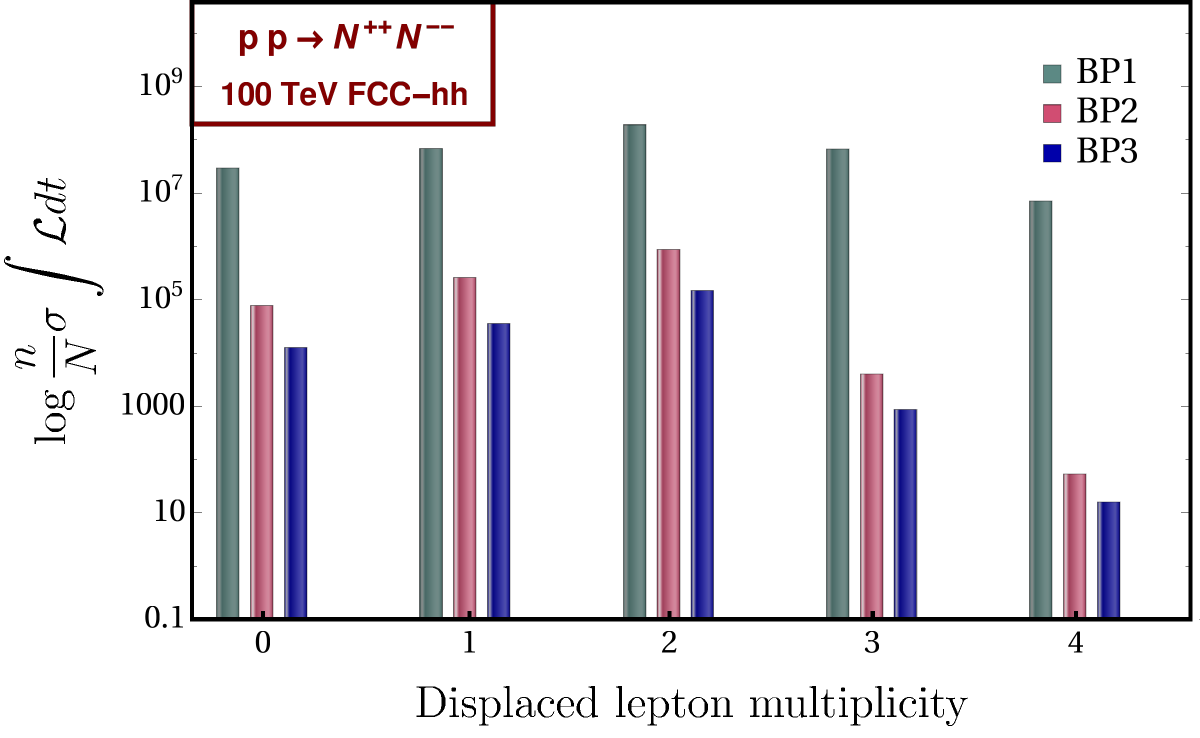}}
	\caption{Multiplicity of displaced leptons for the process $p p \to N^{++}N^{--}$  at the centre-of-mass energies of (a) 14 TeV, (b) 27 TeV and (c) 100 TeV. The number of events are normalized with respective cross-sections and luminosity of 3000\,fb$^{-1}$.}
	\label{fig:lepmul}
\end{figure}
\autoref{fig:lepmul} shows the multiplicity of displaced leptons for the three benchmark points of our model, from the production process $p\,p\to N^{++}N^{--}$ at the 14, 27 TeV LHC and 100 TeV FCC-hh. The low mass gaps between the dark sector particles keep the $p_T$ of the leptons very soft, and hence we put a cut of $p_T \geq 5$ GeV while collecting the displaced leptons. 
In each BP, the first and second displaced leptons originate mainly from the decay of $N^{\pm\pm} \to H^\pm e^\pm$, and the $\sim10$ GeV mass splitting between $N^{\pm\pm}$ and $H^\pm$ across the three BPs keep the distribution shape more-or-less the same upto the second lepton. The effect of the mass splitting between $H^{\pm}$ and $A^0$ is clearly reflected in the multiplicity distribution, when it comes to the third and the fourth displaced leptons getting tagged. The $\sim 15$ GeV splitting in {\bf BP1} allows the third and fourth displaced leptons to have more momenta compared to the $\sim1$ GeV splitting in {\bf BP2} and {\bf BP3}. Hence, {\bf BP1} shows the highest number of events for three- and four-displaced leptons, while less than 10 three-lepton events can be seen for {\bf BP2} and {\bf BP3}, with no events being recorded with four displaced leptons. The situation improves for {\bf BP2} and {\bf BP3} when we move to the 27 TeV LHC, where the increase in boost slightly enhances the number of three-lepton events. The four-lepton events for {\bf BP2} and {\bf BP3} only show up at the FCC-hh energy of 100 TeV, due to the larger boost to the otherwise soft leptons from $H^\pm$ decays.

	\item {\bf $p_T$ of displaced leptons: }
	
	\begin{figure}[ht]
	\centering
	\subfigure[]{\includegraphics[width=0.48\linewidth]{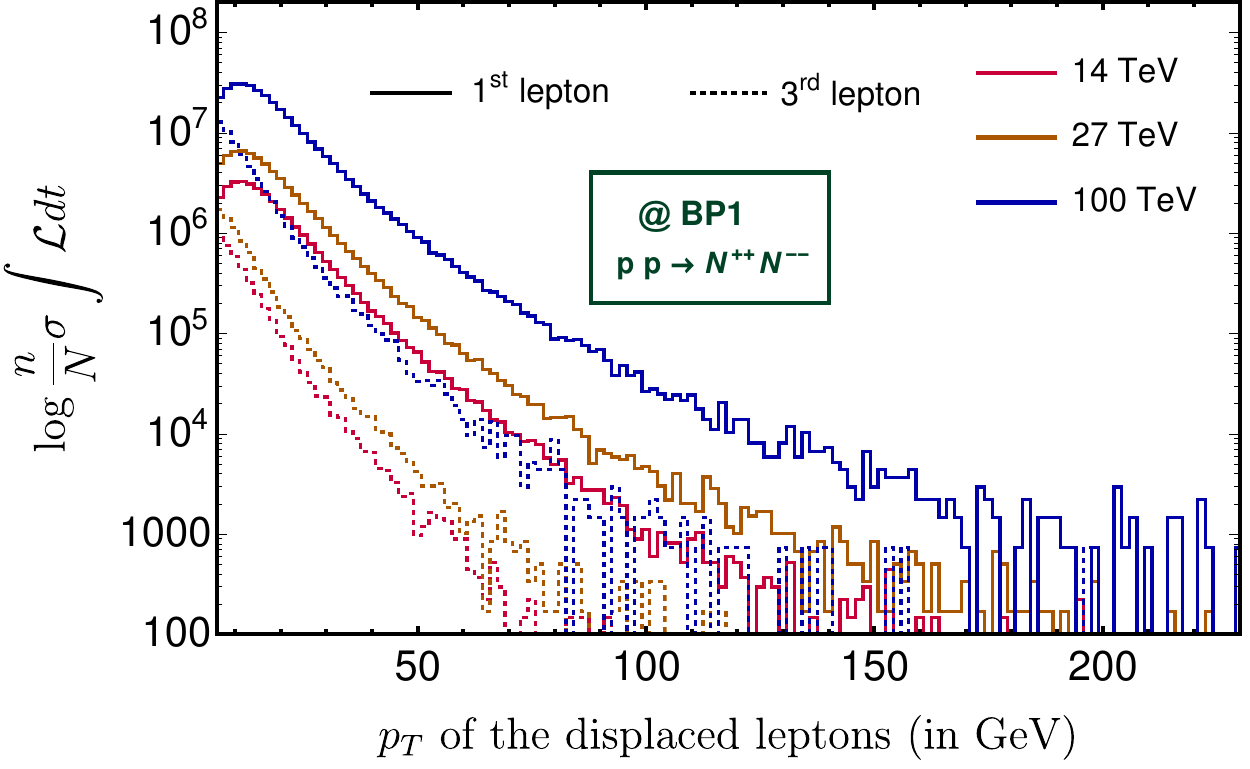}} \quad
	\subfigure[]{\includegraphics[width=0.48\linewidth]{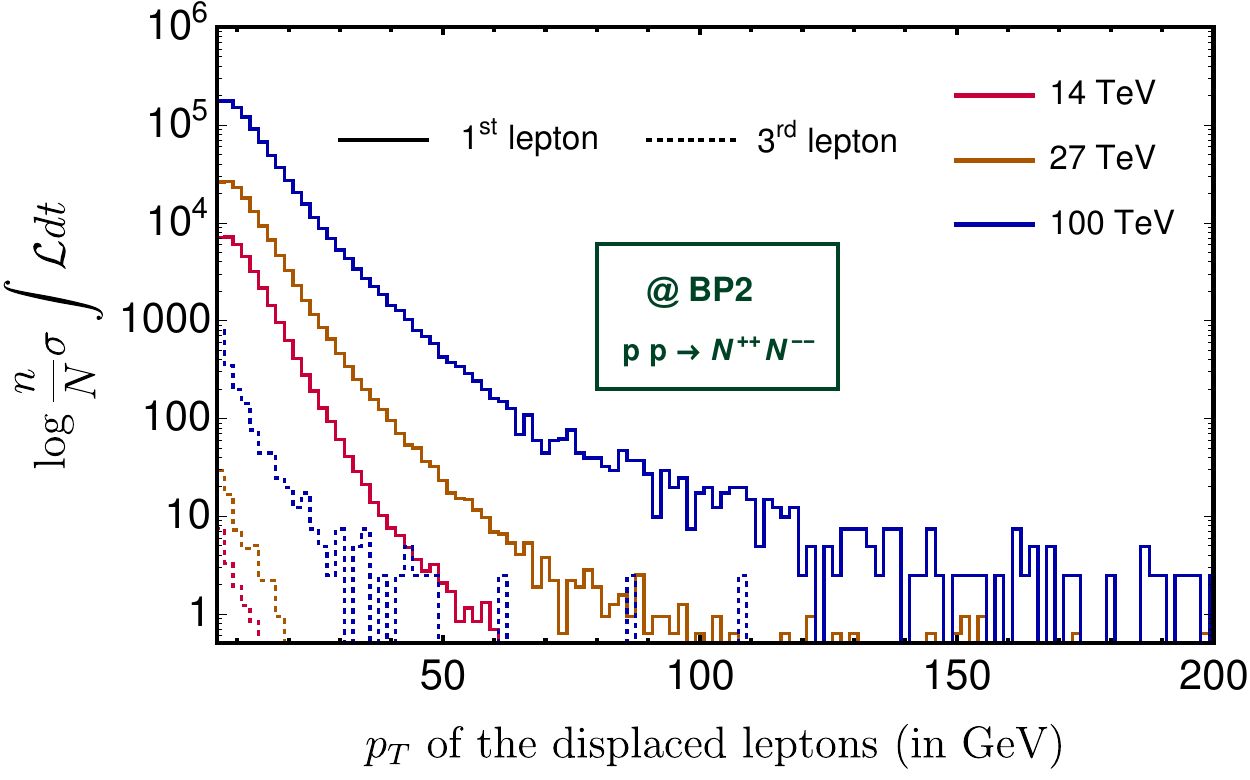}} \quad
	\subfigure[]{\includegraphics[width=0.48\linewidth]{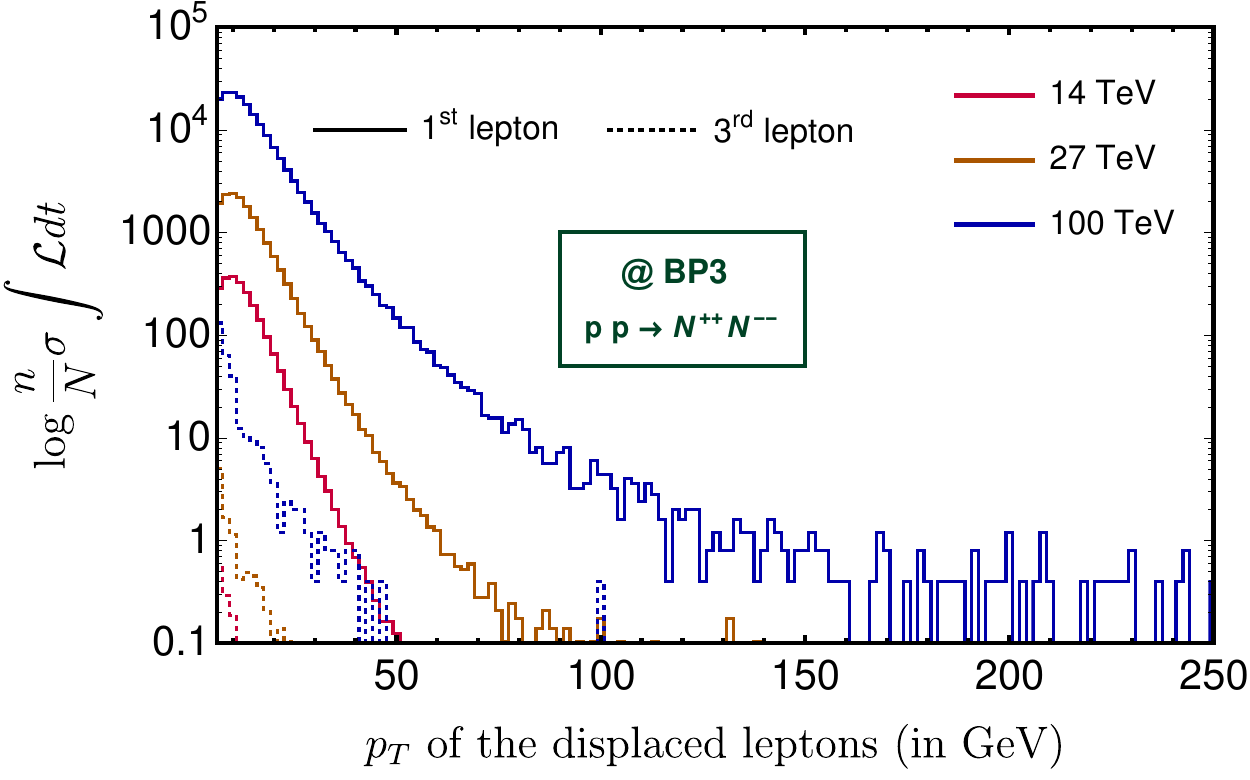}}
	\caption{$p_T$ of the first (solid line) and the third (dotted line) displaced leptons for the process $p\,p\to N^{++}N^{--}$ at the centre-of-mass energies of 14, 27 and 100 TeV for (a) {\bf BP1}, (b) {\bf BP2} and (c) {\bf BP3}. The number of events are normalized with respective cross-sections and luminosity of 3000\,fb$^{-1}$.}
	\label{fig:leppt}
\end{figure}

As mentioned previously, the displaced leptons that emerge from such a compressed mass spectrum are expected to be very soft. \autoref{fig:leppt} is presented to fortify this claim, for the process $p\,p\to N^{++}N^{--}$. Here, panels (a), (b), (c) correspond to {\bf BP1}, {\bf BP2}, and {\bf BP3}, respectively. For each BP, the plot shows the $p_T$ distribution of the first displaced lepton (in solid line) and the third displaced lepton (in dotted line), for the three centre-of-mass energies of 14 TeV (red), 27 TeV (darker orange), and 100 TeV (blue). As the first two leptons emerge from the decays of $N^{\pm\pm}$, which have similar $p_T$ spectrum, and the third and fourth lepton emerge from $H^\pm$ decays, we have chosen only to show the first and third lepton momenta in the plot, to exemplify the mass splitting effects. For {\bf BP1}, the $\sim15$ GeV mass gap between $N^{\pm\pm}$ and $H^\pm$ leads to the first lepton $p_T$ distribution peaking at the 15-20 GeV bin, while the off-shell three-body decay of $H^\pm$ can only provide a peak of <10 GeV for the third lepton. The length of the tails of these distributions increases as one move up along the collider energies of 27 TeV and 100 TeV. Especially at the 100 TeV FCC-hh energy, the typically softer third lepton can also get boosted to a tail end of $p_T \sim 200$ GeV. For {\bf BP2}, a <10 GeV mass splitting in both the scalar and fermionic dark sector leads to both the first and third lepton $p_T$ distributions peaking in the 5-10 GeV bin. As seen from the multiplicity plots in \autoref{fig:lepmul}, the small number of three-lepton events mean that the $p_T$ distributions of the third lepton do not cover as much area of the histogram compared to {\bf BP1}, and even the 100 TeV FCC-hh is not enough to boost their momenta above $\sim100$ GeV. Similar behaviour is also observed for {\bf BP3}, with the only difference being the first lepton $p_T$ distributions peaking at the 10-15 GeV bin, due to the $\sim10$ GeV mass gap between $N^{\pm\pm}$ and $H^\pm$.

	\item {\bf Missing transverse momentum ($\ptmiss$):}

\begin{figure}[ht]
	\centering
	\subfigure[]{\includegraphics[width=0.48\linewidth]{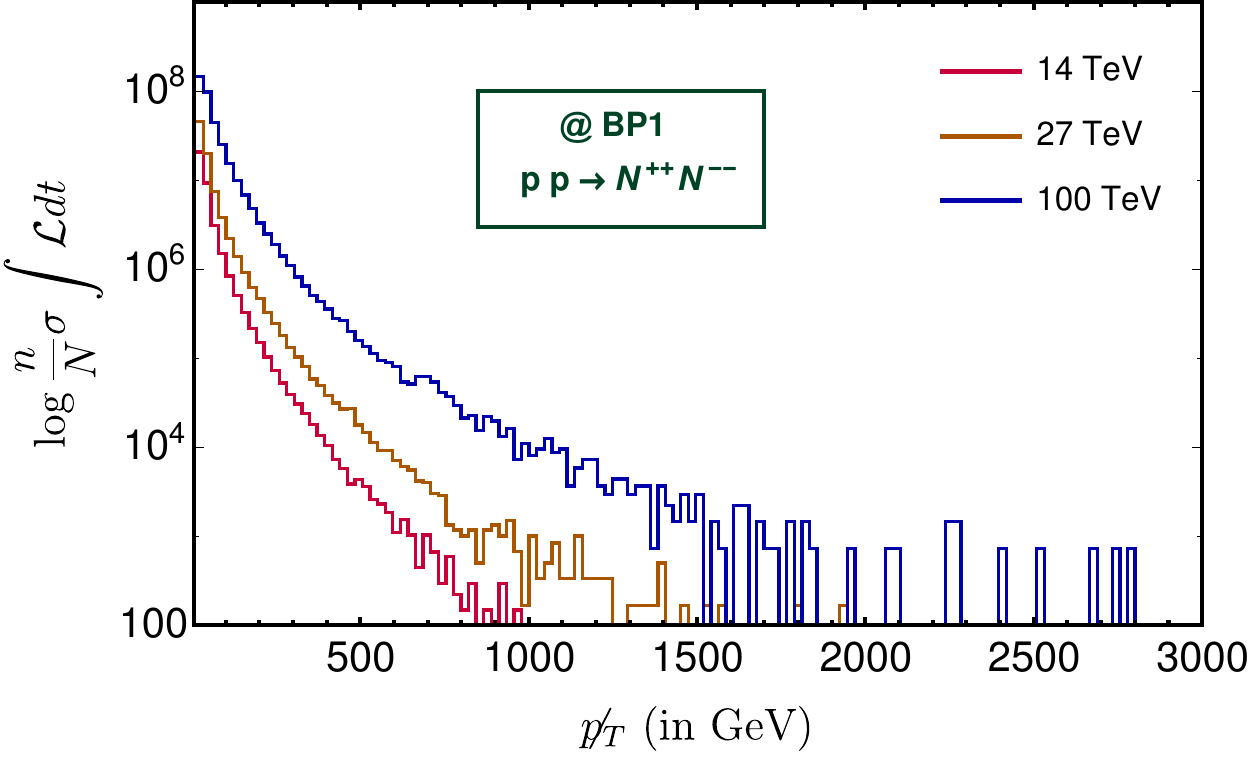}}\quad
	\subfigure[]{\includegraphics[width=0.48\linewidth]{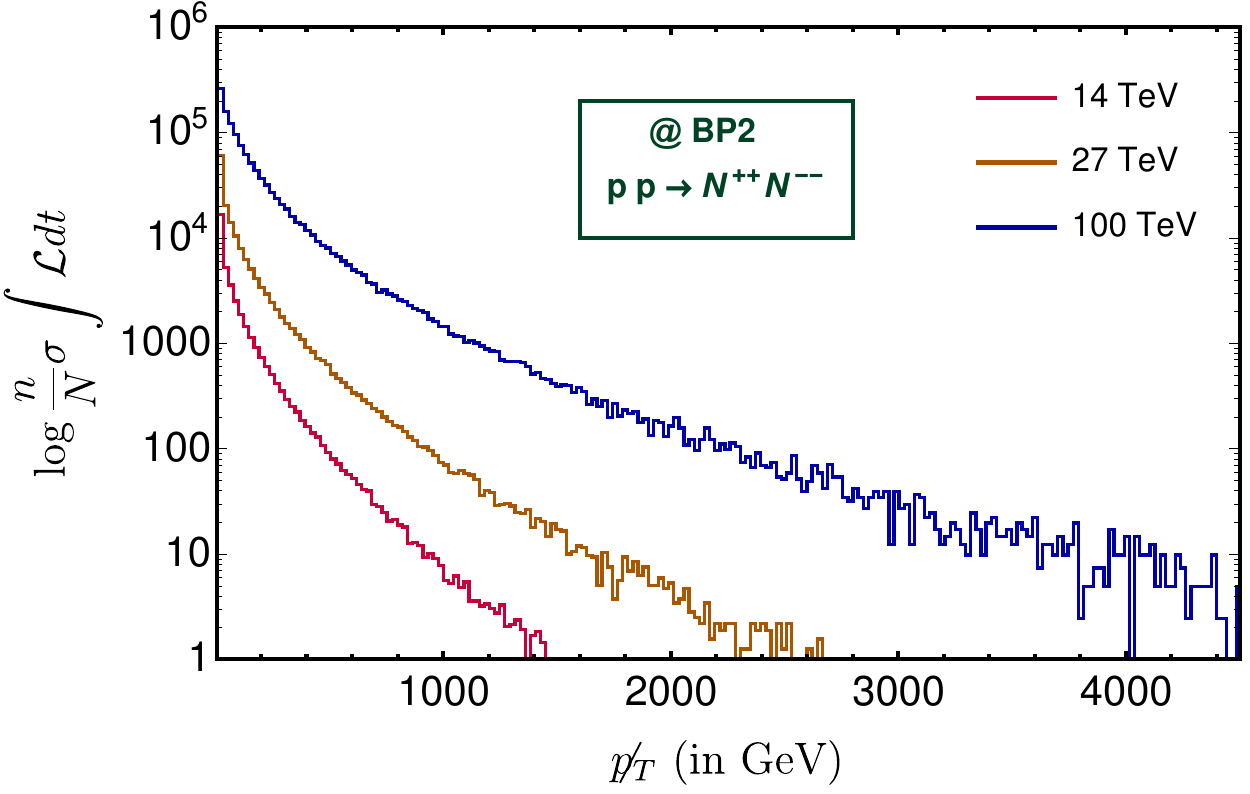}}
	\quad
	\subfigure[]{\includegraphics[width=0.48\linewidth]{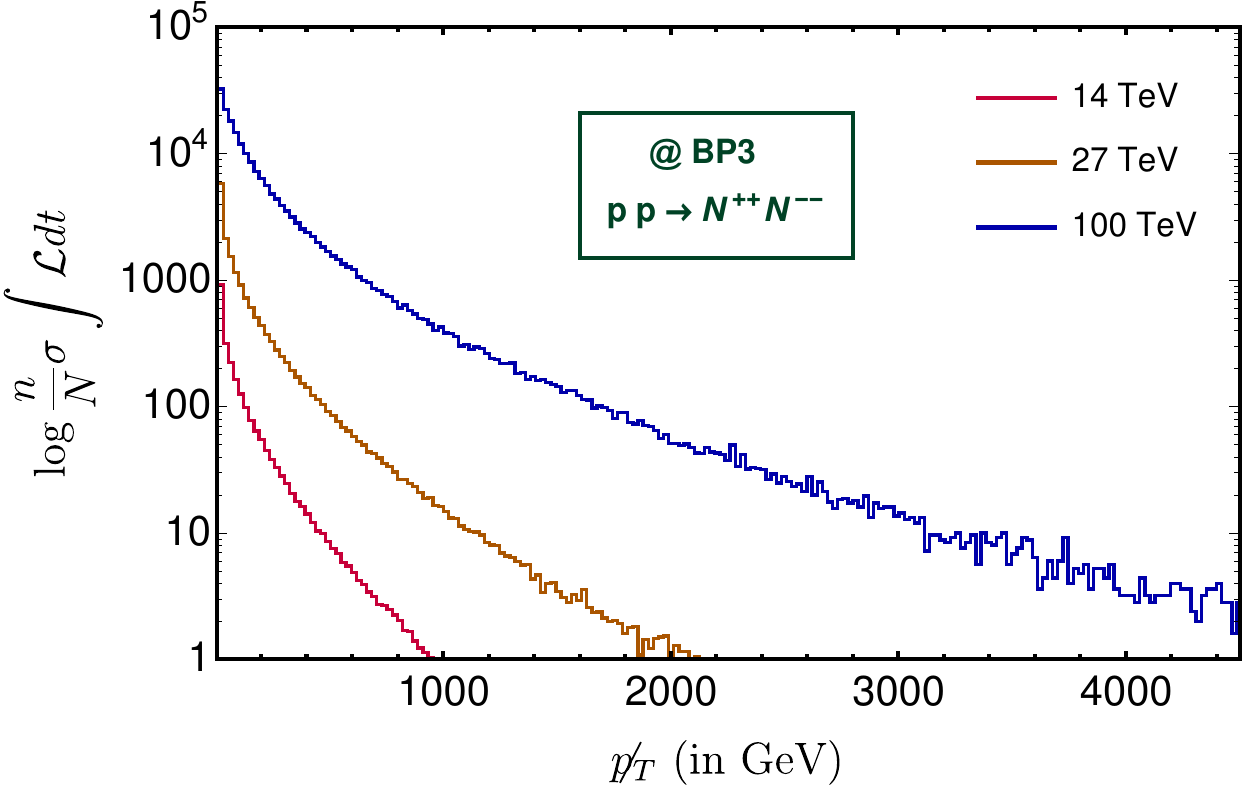}}
	\caption{Distribution of the missing transverse momentum ($\ptmiss$) for the process $p\,p\to N^{++}N^{--}$ at the centre-of-mass energies of 14, 27 and 100 TeV for (a) {\bf BP1}, (b) {\bf BP2} and (c) {\bf BP3}. The number of events are normalized with respective cross-sections and luminosity of 3000\,fb$^{-1}$.}
	\label{fig:met}
\end{figure}

In our model, the triplet VLLs decay to the pseudoscalar dark matter $A^0$ and SM neutrinos, following the multiple decay chains shown in \autoref{fig:feynnmm}. In a pair production, such as $p\,p\to N^{++}N^{--}$ the transverse momentum imbalance can be seen as net missing momentum ($\ptmiss$), which can be experimentally estimated as $\ptmiss= \sqrt{\left(\sum p^{\rm jet}_x + \sum p^{\ell}_x \right)^2 + \left(\sum p^{\rm jet}_y + \sum p^{\ell}_y \right)^2}$.  In \autoref{fig:met}(a), (b), (c), we plot the distribution of the missing transverse momentum for the process $p\,p\to N^{++}N^{--}$, for {\bf BP1}, {\bf BP2}, and {\bf BP3} respectively. Each plot contains the distributions for $\ptmiss$ at the three collider energies of 14 TeV (red), 27 TeV (darker orange), and 100 TeV (blue). The back-to-back pair production of such heavy particles leads to cancellation in the missing transverse momentum, and hence in each of the three BPs, the $\ptmiss$ peaks at the 0-20 GeV bin. However, availability of larger boost for the visible sector, as we move up the collider energies, lead to increasingly longer tails in the $\ptmiss$ distributions. Additionally, the lengths of the tails gradually enhance with the increase in the masses of the particles in {\bf BP2} and {\bf BP3}, pertaining to the events becoming more and more central, and the visible sector having more momenta to recoil against.

	%
	
\end{itemize}

\section{Results}\label{results}
In this section we summarize the results after simulation at the CMS, ATLAS and  LHC/FCC-hh for the centre of energies of 14, 27 and 100 TeV, respectively and at MATHUSLA.  As can be  seen from \autoref{tab:bp}, the {\bf BP1} mass values are relatively lower, thus the kinematics and cuts suitable for this analysis  are different than the  other benchmark points. Thus we present the analysis of {\bf BP1} separately in \autoref{anaBP1}, whereas the results for {\bf BP2} and {\bf BP3} are summarized in \autoref{anaBP23}.

\subsection{Analysis of events from the low-mass {\bf BP1}}\label{anaBP1}


After discussing the kinematical distributions of our model, we are now equipped enough to begin the analysis of the displaced multi-lepton final states. It is noteworthy that, the difference in mass scales of our chosen benchmark points prompts us to use two different sets of cuts: one for the sub-100 GeV masses in {\bf BP1}, and another for the $\sim 500-1000$ GeV mass scale of {\bf BP2} and {\bf BP3}. We will begin by presenting the analysis for {\bf BP1}, at three different centre-of-mass energies of 14, 27, and 100 TeV, respectively, including a dedicated analysis at the proposed MATHUSLA detector for long-lived particles (LLP). We  mention here that, the existing searches using displaced leptons at the LHC experiments tend to look for pairs of muons that originate from the same displaced vertex, in order to reconstruct any possible neutral resonances \cite{CMS:2022qej, CMS:2021sch, CMS:2021juv}. The searches involving electrons look for displaced di-leptons with high cuts on their $p_T$ \cite{CMS:2021kdm, ATLAS:2020wjh}. While the three- and four-lepton searches for BSM particles demand prompt leptons \cite{CMS:2019lwf}, these strategies are not applicable to our model signatures, dominated by soft displaced electrons from the charged VLL decays. A combination of hard and soft displaced lepton final states in the context of  lepton flavour violation are relevant in the context of supersymmetric decays \cite{Bandyopadhyay:2015iij, Bandyopadhyay:2011qm, Bandyopadhyay:2014sma}.

Inside the CMS and ATLAS detectors of the existing LHC, an electron is identified by reconstructing tracker hits and matching them with corresponding electromagnetic calorimeter (ECal) hits. Hence, electrons that come from an LLP that decays beyond the ECal are missed by the detection system. Nonetheless, the probabilistic nature of the boosted decay lengths allow a decent number of events, where the electrons from our VLL decays are produced before the ECal ends enabling their detection. The same behaviour is also expected from the proposed FCC-hh detector, whose dimensions we will be using for the 100 TeV FCC-hh analysis. In the first part of this study, we present the numbers of events with one to four displaced leptons in hadronically quiet final states, which are produced with a displacement of at least 1 mm from the interaction point, with an upper limit on the production vertex provided by the reach of the ECals. A simplified approximation of the ECal limit has been taken for this analysis, keeping in mind the variations in the radii of the barrel and endcap regions. For CMS and ATLAS, the upper limit of the displacement is taken as 1.7 meters and 2.3 meters respectively \cite{CMS:2006myw,CMS:2007sch,ATLAS:1999uwa,ATLAS:1999vwa}, while for the proposed FCC-hh detector, an optimistic ECal limit of 5 meters is considered \cite{FCC:2018vvp}. In \autoref{tab:nBP1}, we present the number of events inside these detector regions, for each of the displaced $n$-lepton final states, from all possible contributing production processes, at the three centre-of-mass energies of 14, 27, and 100 TeV, respectively. The integrated luminosity for quoting the numbers is considered to be 300 \fbi.


\begin{table}[ht]
	\centering
	\renewcommand{\arraystretch}{1.2}
	\begin{tabular}{|c|c|c|c||c|c||c|}
		\hline
		\multirow{3}{*}{\makecell{Final States\\for {\bf BP1}}} & \multirow{3}{*}{\makecell{production \\ modes}} &  \multicolumn{5}{c|}{\makecell{centre-of-mass energies }} \\
		\cline{3-7}
		&  & \multicolumn{2}{c||}{14 TeV} & \multicolumn{2}{c||}{27 TeV} & 100 TeV \\
		\cline{3-7}
		& & CMS & ATLAS & CMS & ATLAS & FCC-hh detector \\
		\hline\hline
	\multirow{1.5}{*}{	FS1:  1 displaced lepton} & $N^{++}N^{--}$ & 274.1 & 400.0 & 352.6 & 688.4 & 2877.4  \\
		\multirow{2.0}{*}{with $p_{T_\ell} \geq 30$\,GeV}& $N^{\pm \pm}N^{\mp}$ & 2401.7 & 3904.5 & 5469.2 & 8392.4 & 24343.9 \\
		\multirow{2.5}{*}{+ 0 jet}& $N^+N^-$ & 389.7 & 581.5 & 689.3 & 1070.2 & 5958.8 \\
		& $N^{\pm}N^{0}$ & 11149.5 & 16989.0 & 20862.9 & 31326.4 & 173654.0 \\
		\cline{2-7}
		& Total & 14215.0 & 21875.0 & 27374.0 & 41477.4 & 206834.1 \\
		\hline \hline
		\multirow{1.5}{*}{FS2: 2 displaced leptons} & $N^{++}N^{--}$ & 140.8 & 214.9 & 386.2 & 520.5 & 3014.5 \\
		\multirow{2.0}{*}{with $p_{T_{\ell_{1}}} \geq 20 $\,GeV } & $N^{\pm \pm}N^{\mp}$ & 927.0 & 1811.8 & 1634.5 & 3111.8 & 35190.2 \\
		\multirow{2.0}{*}{and $p_{T_{\ell_{2}}} \geq 10 $\,GeV}& $N^+N^-$ & 40.4 & 85.8 & 86.3 & 167.2 & 1957.5 \\
		\multirow{2.5}{*}{+ 0 jet}& $N^{\pm}N^{0}$ & 0.0 & 0.0 & 0.0 & 32.0 & 0.0 \\
		\cline{2-7}
		& Total & 1108.2 & 2112.5 & 2107.0 & 3831.5 & 40162.2 \\
		\hline \hline
		\multirow{1.5}{*}{FS3: 3 displaced leptons} & $N^{++}N^{--}$ & 44.4 & 140.8 & 134.3 & 302.2 & 5480.8 \\
		\multirow{2.0}{*}{with $p_{T_{\ell_{1}}} \geq 15$\,GeV} & $N^{\pm \pm}N^{\mp}$ & 126.4 & 323.0 & 408.6 & 785.8 & 13256.6 \\
		\multirow{2.0}{*}{and $p_{T_{\ell_{2,3}}} \geq 5$\,GeV}& $N^+N^-$ & 0.6 & 0.6 & 0.0 & 0.0 & 0.0 \\
		\multirow{2.5}{*}{+ 0 jet}& $N^{\pm}N^{0}$ & 0.0 & 0.0 & 0.0 & 0.0 & 0.0 \\
		\cline{2-7}
		& Total & 171.4 & 464.4 & 542.9 & 1088.0 & 18737.4  \\
		\hline \hline
		\multirow{1.5}{*}{FS4: 4 displaced leptons} & $N^{++}N^{--}$ & 14.8 & 37.0 & 16.8 & 33.6 & 1096.2 \\
		\multirow{2.0}{*}{with $p_{T_{\ell_{1,2,3,4}}} \geq 5$\,GeV}& $N^{\pm \pm}N^{\mp}$ & 0.0 & 0.0 & 0.0 & 0.0 & 0.0 \\
		\multirow{2.5}{*}{+ 0 jet}& $N^+N^-$ & 0.0 & 0.0 & 0.0 & 0.0 & 0.0 \\
		& $N^{\pm}N^{0}$ & 0.0 & 0.0 & 0.0 & 0.0 & 0.0 \\
		\cline{2-7}
		& Total & 14.8 & 37.0 & 16.8 & 33.6 & 1096.2  \\
		\hline
	\end{tabular}
	\caption{Number of displaced lepton events at CMS, ATLAS for 14 and 27 TeV centre-of-mass energies and at proposed FCC-hh detector for 100 TeV centre-of-mass energy for {\bf BP1} with integrated luminosity of 300 \fbi.}
	\label{tab:nBP1}
\end{table}


From the first column of \autoref{tab:nBP1}, one can see that the transverse momentum cut on the displaced leptons vary, depending on their multiplicity in the demanded final state, as well as the requirement to establish triggers in the absence of any jets.  For the first final state we choose  FS1:  1 displaced lepton  with $p_{T_\ell} \geq 30$\,GeV plus zero jets, to effectively trigger the event. The most number of events for this minimalistic final state comes from the production mode of $N^\pm N^0$, where the fully invisible decay mode of $N^0$ ensures that one gets a majority of one-lepton events from $N^\pm \to A^0 e^\pm$ decay mode. The next largest contribution comes from the $N^{\pm\pm} N^\mp$ production, having a larger cross-section, as well as comparatively harder electrons coming from the $N^\pm \to A^0 e^\pm$ decay, due to the larger mass splitting between $N^\pm$ and $A^0$ compared to $N^{\pm\pm}$ and $H^\pm$. The same effect keeps the number of events higher in the production mode of $N^+ N^-$, despite having lower cross-section than its counterpart, the $N^{++} N^{--}$ mode. In total, even at the 14 TeV LHC with 300 \fbi of data, one can ideally see $\sim14000$ events at CMS, and $\sim 22000$ events at ATLAS for the hadronically  quiet displaced mono-lepton final state. At the 27 TeV upgrade, the numbers rise to $\sim 27000 (41000)$ for CMS (ATLAS), while the FCC-hh detector at the 100 TeV collider can detect $\sim 200000$ such events with 300 \fbi of data.
	
Next we consider FS2: the hadronically quiet displaced di-lepton final state with $p_{T_{\ell_{1}}} \geq 20 $\,GeV  and $p_{T_{\ell_{2}}} \geq 10 $\,GeV, and the numbers are presented in the second row of \autoref{tab:nBP1}. Expectedly, the numbers of events decrease sharply compared to the mono-lepton case, as it is less probable to find two leptons within the ECal regions for particles with $c\tau_0 \geq 100$ meters. No events are obtained from the $N^\pm N^0$ mode, as only one lepton can possibly come in {\bf BP1}, from their respective decays shown in \autoref{tab:branching}. In fact, the contribution from this process remains zero for the three- and four-lepton final states as well. The highest contribution is obtained via the production mode $N^{\pm\pm} N^\mp$ for the two displaced leptons, owing to a majority of two-lepton events emerging from the subsequent decays, and the harder leptons from the $N^\pm$ decay. While both $N^{++}N^{--}$ and $N^+ N^-$ modes promise two displaced leptons, the higher cross-section of the former contributes more number of events. In total, one obtains $\sim 1100 (2100)$ events at CMS (ATLAS) with the 14 TeV LHC, which increases to $\sim2100 (3800)$ moving to the 27 TeV upgrade. The FCC-hh detector with the higher cross-sections at 100 TeV can see $\sim 40000$ of such events.

In FS3,  three  hadronically  quiet displaced leptons are demanded, one of them to be harder with $p_T \geq$ 15 GeV to trigger the event, while the remaining two are rather soft with $p_T \geq 5$ GeV. As expected, no contribution is obtained from the $N^+ N^-$ and $N^\pm N^0$ production modes, and the majority of events come from $N^{\pm\pm} N^\mp$ owing to its higher cross-section.  With 300 \fbi of integrated luminosity, $\sim 170 (460)$ events are predicted at CMS (ATLAS) at the 14 TeV LHC, which increase to $\sim 540 (1000)$ at the 27 TeV LHC. At the FCC-hh detector of the 100 TeV collider, $\sim 18000$ events can still be obtained for this case. 

Finally in  FS4,  as events with four displaced leptons are a rarity,  we keep the minimum $p_T$ requirement for this final state as 5 GeV only. As can be seen from \autoref{tab:branching}, only the $N^{++} N^{--}$ production mode can contribute to this final state. At the 14 TeV LHC with 300 \fbi of integrated luminosity, this mode leads to $\sim 15 (37)$ events at CMS (ATLAS). While one expects to see more events at the 27 TeV LHC due to the increased cross-section, the higher boost makes the isolation of the displaced leptons harder, and hence only $\sim 17(34)$ events are predicted at CMS (ATLAS). At the 100 TeV FCC-hh however, the overwhelmingly large cross-section leads to $\sim 1100$ observable events. It is noteworthy that, the leptons coming  from SM processes are prompt ones and with the aforementioned $p_T$ cuts on the displaced leptons, it is difficult to  mimic our signals.


\begin{table}[ht]
	\centering
	\renewcommand{\arraystretch}{1.2}
	\begin{tabular}{|c|c|c|c|c|}
		\hline
		\multirow{2}{*}{\makecell{Final States for {\bf BP1}  \\ at MATHUSLA}} & \multirow{2}{*}{\makecell{production \\ modes}} &  \multicolumn{3}{c|}{\makecell{centre-of-mass energies}} \\
		\cline{3-5}
		&  & 14 TeV & 27 TeV & 100 TeV \\
		\hline\hline
		\multirow{3}{*}{FS5: 1 displaced lepton} & $N^{++}N^{--}$ & 445.1 & 976.9 & 1710.4 \\
		\multirow{3.5}{*}{with $p_{T_\ell} \geq 20$\,GeV}& $N^{\pm \pm}N^{\mp}$ & 1505.5 & 2995.8 & 7164.3 \\
		& $N^+N^-$ & 129.8 & 243.1 & 681.0 \\
		& $N^{\pm}N^{0}$ & 3677.0 & 7110.4 & 24231.4 \\
		\cline{2-5}
		& Total & 5757.4 & 11326.2 & 33787.1 \\
		\hline \hline
		\multirow{3}{*}{FS6: 2 displaced leptons} & $N^{++}N^{--}$ & 249.1 & 507.7 & 1690.6 \\
		\multirow{3.5}{*}{with $p_{T_{\ell_{1,2}}} \geq 10$\,GeV}& $N^{\pm \pm}N^{\mp}$ & 259.8 & 623.8 & 2114.7 \\
		& $N^+N^-$ & 1.7 & 1.8 & 10.3 \\
		& $N^{\pm}N^{0}$ & 0.0 & 0.0 & 0.0 \\
		\cline{2-5}
		& Total & 510.6 & 1133.3 & 3815.6 \\
		\hline
	\end{tabular}
	\caption{Number of displaced lepton events at MATHUSLA for {\bf BP1} with integrated luminosity of 300 \fbi at the 14, 27 and 100 TeV LHC/FCC-hh.}
	\label{tab:mathusla1}
\end{table}


From \autoref{fig:dcylNpp}, one can see that a significant number of displaced lepton events can be obtained at the proposed MATHUSLA detector \cite{Curtin:2018mvb,MATHUSLA:2022sze,Curtin:2023skh} for {\bf BP1}. This dedicated LLP detector is proposed to be placed on the surface of earth at a distance of 68 meters in the longitudinal direction from the CMS interaction point, with a transverse distance of 60 meters, providing a $25 \times 100 \times 100$ m$^3$ of decay volume, covering a fraction amounting to 0.27 of the total azimuthal angle coverage. Additionally, MATHULSA is proposed to be situated above the upper one of the two halves of the CMS cylinder, and hence it can maximally detect two displaced leptons coming from our processes, namely from the $N^{++} \to H^+ e^+ \to (A^0 \ell^+ \nu) e^+$ decay chain. Hence, in \autoref{tab:mathusla1} we present the number of observable events with one and two displaced leptons produced inside the decay volume at MATHUSLA as FS5 and FS6, for the three collider energies of 14, 27, and 100 TeV, with an integrated luminosity of 300 \fbi. The one-lepton final state clearly shows more number of events, as any one of the two produced VLLs can decay to an electron inside MATHUSLA. For the one displaced lepton case, a cut of $p_T \geq 20$ GeV is put on the displaced lepton to trigger the event. As the $N^\pm \to A^0 e^\pm$ decay provides a comparatively harder electron, the most number of events is seen from the $N^\pm N^0$ mode, closely followed by the $N^{\pm\pm} N^\mp$ mode. The $N^+ N^-$ mode provides the least contribution due to the lowest cross-section. With 300 \fbi of integrated luminosity, one is able to see $\sim 5700,\, \sim 11000$, and $\sim 34000$ mono-lepton events at MATHUSLA, for the centre-of-mass energies of 14, 27 and 100 TeV, respectively. When it comes to two displaced leptons, the contribution becomes negligible from the two production modes without $N^{\pm\pm}$, failing to provide the adequate decay chain. The cut on the lepton $p_T$ is reduced to 10 GeV, and as in both cases the two leptons only come from the $N^{\pm\pm}$ decay, similar number of events are observed for both the $N^{++}N^{--}$ and $N^{\pm\pm} N^\mp$ modes. In total, with 300 \fbi of luminosity, one can observe $\sim 500, \, \sim 1100$, and $\sim 3800$ events with two displaced leptons at MATHUSLA, at the 14, 27, and 100 TeV machines. Emanating from a doubly charged VLL, {\it  i.e.} $N^{\pm\pm}$, these two leptons are ideally of same-sign, which can be a definitive signature of this model coming from the same vertex.


\subsection{Analysis of events for {\bf BP2} and {\bf BP3}}\label{anaBP23}

The low mass, high boost, and larger mass gaps allowed us to work with stronger cuts on the displaced leptons in case of {\bf BP1}, in order to trigger events with comparative ease at the LHC/FCC-hh. For the case of {\bf BP2} and {\bf BP3}, the heavier masses of the dark sector particles reduce the available boost, while the smaller mass gaps don't help the leptons to achieve sufficient momenta either. Hence, we analyze the displaced lepton signatures for {\bf BP2} and {\bf BP3} with a separate set of cuts, proposing softer triggers at the detectors, and quoting the numbers of events with 3000 \fbi of luminosity as opposed to the 300 \fbi for {\bf BP1}. Also it is  noteworthy  that, the boost effect is not enough to have sufficient events at MATHUSLA with this luminosity, and hence we stick to the events within the CMS/ATLAS and the proposed FCC-hh detector ECals, as mentioned in the previous subsection.


\begin{table}[ht]
	\centering
	\renewcommand{\arraystretch}{1.2}
	\begin{tabular}{|c|c|c|c||c|c||c|}
		\hline
		\multirow{3}{*}{\makecell{Final States\\for {\bf BP2}}} & \multirow{3}{*}{\makecell{production \\ modes}} &  \multicolumn{5}{c|}{\makecell{centre-of-mass energies }} \\
		\cline{3-7}
		&  & \multicolumn{2}{c||}{14 TeV} & \multicolumn{2}{c||}{27 TeV} & 100 TeV \\
		\cline{3-7}
		& & CMS & ATLAS & CMS & ATLAS & FCC-hh detector \\
		\hline\hline
		\multirow{1.5}{*}{FS1a: 1 displaced lepton} & $N^{++}N^{--}$ & 223.9 & 296.6 & 1022.9 & 1390.0 & 4608.9 \\
		\multirow{2.0}{*}{with $p_{T_\ell} \geq 15$\,GeV}& $N^{\pm \pm}N^{\mp}$ & 655.9 & 897.4 & 2778.1 & 3792.5 & 21584.6 \\
		\multirow{2.5}{*}{+ 0 jet}& $N^+N^-$ & 14.8 & 20.6 & 65.7 & 91.0 & 527.4 \\
		& $N^{\pm}N^{0}$ & 238.7 & 339.9 & 1047.8 & 1494.1 & 9309.3 \\
		\cline{2-7}
		& Total & 1133.3 & 1554.5 & 4914.5 & 6767.6 & 36030.2 \\
		\hline \hline
		\multirow{1.5}{*}{FS2a: 2 displaced leptons} & $N^{++}N^{--}$ & 410.3 & 666.5 & 1464.9 & 2446.0 & 17982.2 \\
		\multirow{2.0}{*}{with $p_{T_{\ell_{1,2}}} \geq 10 $\,GeV } & $N^{\pm \pm}N^{\mp}$ & 233.6 & 395.9 & 822.4 & 1457.3 & 10232.8 \\
		\multirow{2.5}{*}{+ 0 jet}& $N^+N^-$ & 3.3 & 6.0 & 12.4 & 23.5 & 165.0 \\
		& $N^{\pm}N^{0}$ & 0.1 & 0.1 & 0.5 & 0.6 & 8.6 \\
		\cline{2-7}
		& Total & 647.3 & 1068.5 & 2300.2 & 3927.4 & 28388.6 \\
		\hline \hline
		\multirow{1.5}{*}{FS3a: 3 displaced leptons} & $N^{++}N^{--}$ & 1.7 & 2.1 & 4.7 & 7.5 & 104.9 \\
		\multirow{2.0}{*}{with $p_{T_{\ell_{1,2,3}}} \geq 5$\,GeV} & $N^{\pm \pm}N^{\mp}$ & 1.0 & 1.1 & 1.7 & 4.4 & 124.9 \\
		\multirow{2.5}{*}{+ 0 jet}& $N^+N^-$ & 0.0 & 0.0 & 0.0 & 0.1 & 2.8 \\
		& $N^{\pm}N^{0}$ & 0.0 & 0.0 & 0.0 & 0.0 & 4.7 \\
		\cline{2-7}
		& Total & 2.7 & 3.2 & 6.4 & 12.0 & 237.3  \\
		\hline 
	\end{tabular}
	\caption{Number of displaced lepton events at CMS, ATLAS for 14 and 27 TeV centre-of-mass energies and at proposed FCC-hh detector for 100 TeV centre-of-mass energy for {\bf BP2} with integrated luminosity of 3000 \fbi.}
	\label{tab:nBP2}
\end{table}


\autoref{tab:nBP2} displays the number of events obtained from the four aforementioned production modes for {\bf BP2}, with one, two or three displaced leptons designated as FS1a, FS2a and FS3a respectively within the ECal of CMS and ATLAS for the 14, 27 TeV LHC, and of the proposed FCC-hh detector for the 100 TeV FCC-hh, considering the integrated luminosity of 3000 \fbi. For the final state with one displaced lepton and no jets (FS1a), we put a softer cut of $p_{T_\ell} \geq 15$ GeV to trigger the events. The highest contribution to this final state in {\bf BP2} comes from the mode $N^{\pm\pm}N^\mp$, as the mass gaps between the fermions and the scalars remain $\sim 7-8$ GeV, and hence the displaced electrons coming from the decays $N^{\pm\pm} \to H^\pm e^\pm$ and $N^\pm \to A^0/H^0 e^\pm$ have similar probabilities of obtaining sufficient boost and getting tagged. While the cross-section of the $N^\pm N^0$ mode is almost twice than that of the $N^{++}N^{--}$ mode, the contributions from both are similar, owing to the fact that any one of the two electrons coming from the doubly charged VLLs have the probability of getting tagged in the latter case, with the second electron losing out due to the applied cuts. Low cross-sections keep the contribution of the $N^+ N^-$ production mode substantially small. All-in-all, the 14 TeV LHC can lead to $\sim 1100 (1500)$ such soft displaced mono-lepton events to be detected at CMS (ATLAS), which increase to $\sim 4900 (6800)$ events at the 27 TeV LHC. The FCC-hh detector, operating with the 100 TeV collider, can detect a very promising $\sim 36000$ events for this final state.
	
Coming to the final state with two displaced soft leptons and no jets in FS2a, we reduce the minimum transverse momentum demand to be $p_{T_{\ell}} \geq 10$ GeV for both the leptons. Here, the dominant contribution comes from the $N^{++}N^{--}$ mode, with a probability of having at least two of the four possible displaced leptons getting tagged. Out of the remaining production modes, only $N^{\pm\pm}N^\mp$ contributes significantly. The total number of events are reduced compared to the previous final state, with $\sim 650 (1100)$ observable events at the CMS (ATLAS) with the 14 TeV LHC, enhancing to $\sim 2300 (3900)$ with the 27 TeV upgrade. The 100 TeV FCC-hh however, enables detection of $\sim 28000$ such events, due to the higher production rates.
	
Finally, as is clear from the multiplicity plots in \autoref{fig:lepmul}, the number of events with three displaced leptons are negligible for {\bf BP2} at the 14 and 27 TeV energies. Hence even with a very soft cut of $p_{T_\ell} \geq 5$ GeV, one can see only $\sim 3$ events at both the detectors with the 14 TeV LHC, and $\sim 6 (12)$ events at CMS (ATLAS) when one moves to the 27 TeV LHC. However, at the 100 TeV FCC-hh, the third lepton coming from the $H^\pm$ or $H^0$ decay receives more boost, leading to it being tagged and $\sim 240$ events being observed at the FCC-hh detector. Unlike {\bf BP1}, we do not have any events here with four displaced leptons, and hence we restrict our analysis to these three final states only.


\begin{table}[ht]
	\centering
	\renewcommand{\arraystretch}{1.2}
	\begin{tabular}{|c|c|c||c||c|}
		\hline
		\multirow{3}{*}{\makecell{Final States\\for {\bf BP3}}} & \multirow{3}{*}{\makecell{production \\ modes}} &  \multicolumn{3}{c|}{\makecell{centre-of-mass energies }} \\
		\cline{3-5}
		&  & \multicolumn{1}{c||}{14 TeV} & \multicolumn{1}{c||}{27 TeV} & 100 TeV \\
		\cline{3-5}
		& & CMS/ ATLAS & CMS/ ATLAS & FCC-hh detector \\
		\hline\hline
		\multirow{1.5}{*}{FS1a: 1 displaced lepton} & $N^{++}N^{--}$ & 45.8 & 360.4 & 1120.3  \\
		\multirow{2.0}{*}{with $p_{T_\ell} \geq 15$\,GeV}& $N^{\pm \pm}N^{\mp}$ & 209.5 & 1479.0 & 7846.0 \\
		\multirow{2.5}{*}{+ 0 jet}& $N^+N^-$ & 6.9 & 51.1 & 269.0 \\
		& $N^{\pm}N^{0}$ & 115.0 & 796.0 & 4123.8 \\
		\cline{2-5}
		& Total & 377.2 & 2686.5 & 13359.1  \\
		\hline \hline
		\multirow{1.5}{*}{FS2a: 2 displaced leptons} & $N^{++}N^{--}$ & 182.5 & 1206.3 & 6700.1 \\
		\multirow{2.0}{*}{with $p_{T_{\ell_{1,2}}} \geq 10 $\,GeV } & $N^{\pm \pm}N^{\mp}$ & 144.6 & 890.1 & 4621.1 \\
		\multirow{2.5}{*}{+ 0 jet}& $N^+N^-$ & 2.6 & 16.9 & 79.6 \\
		& $N^{\pm}N^{0}$ & 0.0 & 0.2 & 0.7 \\
		\cline{2-5}
		& Total & 329.7 & 2113.5 & 11401.5 \\
		\hline \hline
		\multirow{1.5}{*}{FS3a: 3 displaced leptons} & $N^{++}N^{--}$ & 0.2 & 1.4 & 39.9 \\
		\multirow{2.0}{*}{with $p_{T_{\ell_{1,2,3}}} \geq 5$\,GeV} & $N^{\pm \pm}N^{\mp}$ & 0.2 & 1.0 & 17.1 \\
		\multirow{2.5}{*}{+ 0 jet}& $N^+N^-$ & 0.0 & 0.0 & 0.2 \\
		& $N^{\pm}N^{0}$ & 0.0 & 0.0 &  0.0 \\
		\cline{2-5}
		& Total & 0.4 & 2.4 & 57.2  \\
		\hline 
	\end{tabular}
	\caption{Number of displaced lepton events at CMS, ATLAS for 14 and 27 TeV centre-of-mass energies and at proposed FCC-hh detector for 100 TeV centre-of-mass energy for {\bf BP3} with integrated luminosity of 3000 \fbi.}
	\label{tab:nBP3}
\end{table}


In \autoref{tab:nBP3}, the number of events in the same three final states are presented for {\bf BP3}, at the three collider energies of 14, 27 and 100 TeV with an integrated luminosity of 3000 \fbi. Notably, from \autoref{fig:dcylNpp}, the maximum decay lengths of the VLL states are 0.9(1.1)\,m for the 14(27)\,TeV LHC and hence, almost all the displaced leptons here are produced within the CMS ECal. Therefore, the final state numbers show only minute differences for CMS and ATLAS at these energies, allowing us to present one number for both the detectors. Much like {\bf BP2}, the highest number of events with a single displaced lepton is observed from the $N^{\pm\pm}N^\mp$ for the same reason as mentioned above. The difference with {\bf BP2} comes in the number of events from the  $N^\pm N^0$ and $N^{++}N^{--}$ modes, which in the previous case were nearly the same. Here, the $N^\pm N^0$ process promises a majority of one-lepton events and has twice the cross-section to that of the latter process. On the other hand, the larger mass splitting of $\sim10$ GeV between $N^{\pm\pm}$ and $H^\pm$ allows more two-lepton events to be tagged than one-lepton events from the $N^{++}N^{--}$ mode, and combined with the lower cross-section, it contributes a much smaller number of events than the $N^\pm N^0$ mode. However, the smallest contribution remains from the $N^+N^-$ mode, because of the lowest cross-sections. All contributions combined, $\sim 380 (2700)$ events with only one displaced lepton can be detected by CMS/ATLAS at the 14 (27) TeV LHC, while at the 100 TeV FCC-hh, the proposed detector can see $\sim 13300$ such events. The behaviour of contributions remain the same as {\bf BP2} for the two and three displaced lepton final states in case of {\bf BP3} as well. Receiving the highest contribution from the $N^{++}N^{--}$ mode, $\sim 330 (2100)$ events are detectable with the CMS/ATLAS at the 14 (27) TeV LHC, with the numbers increasing to $\sim 11400$ at the FCC-hh detector operating with the 100 TeV collider. The numbers of events with three displaced leptons remain negligibly small at the 14 and 27 TeV LHC, and only the high cross-sections at the 100 TeV FCC-hh allow $\sim 57$ such events to be seen by the proposed detector.

As an alternative to these displaced multilepton signatures, one can invoke the idea of heavy stable charged particles (HSCP). The HSCP are defined massive long-lived charged particles that traverse the detector layers without decaying, leaving behind anomalously high ionizations in the tracker layers, and arriving significantly late at the muon spectrometers of CMS/ATLAS \cite{CMS:2013czn, CMS:2016ybj, ATLAS:2022pib, ATLAS:2023zxo}. In the context of our benchmarks, the VLL states in \textbf{BP1} possess large enough $c\tau_0$ and boost, to completely decay outside the CMS/ATLAS detectors, which in principle can leave behind HSCP-like tracks. However, the $\sim100$ GeV masses allow them to receive a significant boost even at the 14 TeV LHC, with the majority of events containing $\beta \geq 0.9$, contradicting the optimal requirement of $\beta < 0.9$ for HSCP reconstruction. In contrast, the displaced leptons at MATHUSLA favours the larger boost, and hence comes out on top as the optimal search strategy for this scenario. Again from  \autoref{fig:dcylNpp}(c), for \textbf{BP3}, the decay lengths are contained entirely within CMS/ATLAS or the FCC-hh detector , which is again not desirable for HSCP reconstruction. \textbf{BP2} however, with its $\sim 600$ GeV mass and $\mathcal{O}(10)$m of maximum decay length, can in principle be viable for HSCP searches, especially with the doubly-charged VLL, as per the recent ATLAS studies \cite{ATLAS:2023zxo}. However, the demand for the VLL to traverse through the muon spectrometer, combined with the stringent selection criteria, cuts on the boost, and the low efficiency, significantly reduces the signal events. At the 14 TeV LHC with 3000 \fbi of luminosity, the event counts for at least one doubly-charged HSCP are two orders of magnitude less compared to the FS1a in \autoref{tab:nBP2}, and one order of magnitude lesser than those of FS2a. Nonetheless, improvement in reconstruction efficency in the future may allow HSCP signatures to be a complementary probe of a \textbf{BP2}-like scenario.

\section{Discussion and conclusion}\label{concl}

In this article we have considered a scenario where the SM is augmented by a $Z_2$-odd inert Higgs doublet and a $Y=-1$  vector-like lepton triplet, to study their interplay in attaining the  correct DM relic. Due to its non-zero hypercharge, the neutral component of the VLL cannot serve as the cold DM candidate, due to its large vector coupling to the $Z$-boson. However, the same VLL can play a crucial role, through its sufficiently late decay and co-annihilation, to the relic abundance of the pseudoscalar dark matter $A^0$ coming from the IDM. This specific behaviour predicts certain values of the Yukawa coupling $\y_N$, and of the mass splitting  between the two dark sectors. Typically, the Yukawa coupling ranges from $10^{-9}$ to $10^{-7}$,  as we study  the  mass spectrum of $\mathcal{O}(100)$ GeV to  $\mathcal{O}(1)$ TeV, which is perfect for the searches of displaced decays at the colliders. Due to the non-zero hypercharge of the VLL,  the doubly charged lepton $N^{\pm \pm}$ can travel from $\mathcal{O}(10)$ cm to $\mathcal{O}(1)$ km before it decays, while possible decays of $N^\pm, N^0$ are also of the same order.  Eventually, their decay products can produce a maximum of four displaced charged leptons which can be detected by CMS and ATLAS detectors of the LHC. The $\mathcal{O}(100)$ m decay  length in one quarter-sphere of the beam axis can be detected by the proposed detector MATHUSLA, which in principle can detect the $N^{\pm \pm}$ coming from only one leg  of the  decay chain, a smoking gun signature of this model. Due to the asymmetric position of MATHUSLA, it cannot detect all four leptons coming from the $N^{\pm\pm}N^{\mp\mp}$ pair. However, these can be detected by the ECal of CMS and ATLAS detectors. Based on the constraints obtained from the DM experiments, we establish three benchmark points with VLL masses of $\sim 100$ GeV, $\sim 600$ GeV, and $\sim 1$ TeV respectively, and perform a detailed study of displaced lepton signatures from various symmetric and asymmetric pair productions at the 14 and 27 TeV LHC, as well as the 100 TeV FCC-hh. The events numbers for hadronically quiet final states with displaced $4\ell,\, 3\ell, \, 2\ell$ and $1\ell$ are presented at the CMS and ATLAS for all the three BPs at the aforementioned $E_{CM}$ values. The first benchmark, {\bf BP1}, with the lowest values of mass ($\sim 100$ GeV) and Yukawa coupling ($\y_N  \sim 10^{-9}$ ) can also provide a significant number of events with a maximum of two displaced leptons at the MATHUSLA detector, which are mostly same-sign, coming from the decays of $N^{\pm\pm}$. While {\bf BP2} can theoretically lead to a few displaced lepton events at the MATHUSLA, the cuts on the kinematics and the limited azimuthal angle coverage of MATHUSLA render them undetected. Additionally, the compressed mass spectrum of our model keeps the displaced leptons really soft, and hence our kinematical cuts and vetoes differ from those of the existing LHC searches, to care for the softness. 

It is important to discuss the novelty of our signatures as compared to existing models that involve similar searches. One can have a $Y=-1$ VLL triplet without a $Z_2$ symmetry, which allows the doubly-charged component to decay into $W^\pm \ell^\pm$, provided sufficient mixing with the SM leptonic sector \cite{Delgado:2011iz, Ma:2013tda}. While this mode can lead to a  $4\ell + \ptmiss$ signature, the leptons here are prompt and hence they do not mimic our signals. Insufficient mixing and a compressed spectrum in this sort of a scenario leads to displaced pions instead of SM leptons, which is not exhibited by the VLLs in our setup. 

Next, we discuss the different seesaw scenarios and other charged scalar/fermion models with displaced vertex signatures.  Considering the existing studies of the type-II seesaw model with a long-lived doubly-charged scalar \cite{BhupalDev:2018tox, Antusch:2018svb}, one can mimic the final state with four displaced leptons, enabled via very tiny Yukawa couplings for the decay mode involving the charged lepton pair. However, such final states do not involve a DM particle, and hence no subsequent missing transverse momenta is present. Additionally, the dual impact of the compressed mass spectrum and the small $\y_N$ values allow much a larger displacement of the leptons in our doubly-charged VLL decays, which is not achievable with the same order of couplings from a type-II seesaw. A doubly-charged scalar decaying in a $W^\pm W^\pm$ channel can lead to multi-lepton signatures with some $\ptmiss$ (see e.g. \cite{Ghosh:2017pxl, Ashanujjaman:2022ofg, Dey:2022whc}), but such leptons are prompt, and hence they also differ from our signatures. 

Coming to the case of singly-charged LLPs, a real \cite{Cirelli:2005uq}  triplet scalar DM model with $Y=0$, which always involves a compressed spectrum, can also in principle provide a displaced $2\ell + \ptmiss$ signal. However, this final state is heavily suppressed by the $\sim 2\%$ branching ratio, with the majority of the decay leading to soft displaced pions, which are not present in our setup. In case of a  $Y=0$  complex scalar triplet \cite{Bandyopadhyay:2020otm}, there is one additional charged Higgs which decays promptly. 
The type-III fermionic triplet with $Y=0$ possesses quite a bit of similarity with our signals \cite{Jana:2020qzn, Sen:2021fha,Das:2020uer}, however for higher than di-lepton final  states, the other leptons come from the decays of $Z, W^\pm$ bosons. Generically these leptons are thus harder in $p_T$ and for higher multiplicity the di-lepton invariant mass peaks around $Z$ boson mass. In contrast, for our scenario  none of these leptons come from the SM gauge bosons and thus no such invariant mass peak  is possible. Additionally, we have a relatively larger missing energy coming from the inert doublet dark matter in contrast to type-III case, where the missing energy consists of SM neutrinos only. Moreover, our model features a doubly charged lepton as $N^{\pm \pm}$, which is absent in the type-III case, giving rise to a doubly charged track. For lower Yukawa coupling  such excitations can lead to same sign dilepton events at the distance of  MATHUSLA, which is  not possible for  type-III case, as for lower Yukawa the charged type-III fermion decay to charged pion is enhanced \cite{Sen:2021fha}.  
For $Y=0$ supersymmetric triplet scenario the triplinos can have the displaced singly-charged/neutral decays to leptons due to either compressed spectrum or/and smaller mixing with the doublet \cite{Bandyopadhyay:2010wp,Bandyopadhyay:2014raa}, and can be distinguished from our case using the properties discussed above.

To summarize, a vector-like lepton triplet with $Y=-1$ in tandem with the much-studied inert doublet model enables significant interplay in the dark sectors during their cosmological evolution. Such interplay manifests itself as displaced decays of the VLLs, which can provide novel signatures at the LHC/FCC-hh, with hadronically quiet soft displaced leptons being observable at the CMS/ATLAS detectors (or their proposed counterpart for the FCC-hh), as well as the dedicated MATHUSLA detector for long-lived particles.

\section*{Acknowledgements}
PB wants to thank SERB's MTR/2020/000668 grant for support during project, and also Concordia University for the arrangements of the collaborative visits. PB also acknowledges Kiyoharu Kawana for some useful discussion. The work of MF has been partly supported by NSERC
through the grant number SAP105354. CS would like to thank the MoE, Government of India for supporting her research via SRF. SP acknowledges the Council of Scientific and Industrial Research (CSIR), India for funding his research (File no: 09/1001(0082)/2020-EMR-I). CS and SP would also like to thank Abhishek Roy for helpful discussion regarding the dark matter studies. The authors offer their gratitude to P. Poulose for discussions and valuable inputs.
\vspace{2cm}
\bibliography{idmvll_refs}

\end{document}